\documentclass[usenames,dvipsnames,twoside,11pt]{article}
\pdfoutput = 1

\usepackage[table,dvipsnames]{xcolor}

\usepackage[inline,nomargin,marginclue,index,draft]{fixme}
\makeatletter
\renewcommand*\FXLayoutInline[3]{%
  {\@fxuseface{inline} \ignorespaces[#3 \fxnotename{#1}: #2]}}
\makeatother
\fxsetup{theme=color,mode=multiuser} \FXRegisterAuthor{Javi}{ame}{\color{Blue}Javi}
\fxsetup{theme=color,mode=multiuser} \FXRegisterAuthor{Julia}{am}{\color{green}Julia}
\fxsetup{theme=color,mode=multiuser} \FXRegisterAuthor{Alex}{a}{\color{red}Alex}

\usepackage{tikz}
\usetikzlibrary{patterns}
\usetikzlibrary{plotmarks}
\usetikzlibrary{external}
\usepackage{soul}
\usepackage{booktabs}
\usepackage{xfrac}
\usepackage{bm}
\usepackage{rotating}

\usepackage{cite}
\usepackage[colorlinks=true,linkcolor=blue,urlcolor=magenta,citecolor=blue,pagebackref=true]{hyperref}
\usepackage{graphicx}
\usepackage{epsfig}
\usepackage{amssymb}
\usepackage{amsmath}
\usepackage{epstopdf}
\usepackage[latin1]{inputenc}
\usepackage{url}
\usepackage{slashed}
\usepackage{mathtools}
\usepackage[makeroom]{cancel}
\usepackage{dsfont}

\DeclareMathAlphabet{\mathpzc}{OT1}{pzc}{m}{it}
\usepackage[weather]{ifsym}
\usepackage{authblk}
\usepackage{pdflscape}

\numberwithin{equation}{section}

\newcommand{\additionalinfo}[1]{}
\newcommand{\exclude}[1]{}

\newcommand{\mpl}{m_{\rm P}}

\newcommand{\A}{A}

\newcommand{\mPl}{m_{\rm P}}
\newcommand\veca[1]{{\bf #1}}

\newcommand{\n}{{\mathpzc{n}}}

        \newcommand{\ct}{{\z}}
        \newcommand{\z}{{\tau}}
        \newcommand{\ctc}{{\ct_c}}
        
        \newcommand{\x}{x}
        
        \newcommand{\cf}{{\Phi}}
        
        \newcommand{\lp}{{\lambda_\phi}}
        \newcommand{\m}{{|\phi|}}
        
        \newcommand{\ctheta}{{\psi}}
        \newcommand{\cmass}{{m_\ctheta}}
        \newcommand{\cmassTO}{{m^2_\ctheta}}
        \newcommand{\cctheta}{{\psi}}
        \newcommand{\nrc}{{\Upsilon}}
        
        \newcommand{\ms}{m_s}
        \newcommand{\prs}{{\rm PRS}}
        \newcommand{\sa}{{\sigma_\star}}
        \newcommand{\axit}{\circledcirc}

	\newcommand{\osc}{{\rm osc}}
	\newcommand{\losc}{L_\osc}

\renewcommand\({\left(}
\renewcommand\){\right)}
\renewcommand\[{\left[}
\renewcommand\]{\right]}

\newcommand{\be}{\begin{equation}}
\newcommand{\ee}{\end{equation}}
\newcommand{\bea}{\begin{eqnarray}}
\newcommand{\eea}{\end{eqnarray}}

\usepackage[normalem]{ulem}

\title{Early seeds of axion miniclusters}
\author[1]{Alejandro Vaquero}
\author[2,3]{Javier Redondo}
\author[4]{Julia Stadler}
\affil[1]{\normalsize Department of Physics and Astronomy, University of Utah, Salt Lake City, Utah, 84112, USA}
\affil[2]{\normalsize Departamento de F\'{i}sica Te\'{o}rica,  
Universidad de Zaragoza, 50009 Zaragoza}
\affil[3]{\normalsize  Max-Planck-Institut f\"{u}r Physik,  
80805 M\"{u}nchen, Germany} 
\affil[4]{\normalsize Institute for Particle Physics Phenomenology, Durham University, South Road, Durham, DH1 3LE, UK}

\date{\today}

\begin{document}

\noindent
\hspace{\fill} MPP-2018-238

\begingroup
\let\newpage\relax
\maketitle
\endgroup

\begin{abstract}

We study the small scale structure of axion dark matter in the post-inflationary scenario, which predicts the formation of low-mass, high density clumps of gravitationally bound axions called axion miniclusters. To this end we follow numerically the cosmological evolution of the axion field and the network of strings and domain walls until the density contrast is frozen. Our simulations, comprising up to $8192^3$ points, are the largest studies of the axion field evolution in the non-linear regime presented so far. Axitons, pseudo-breathers of the Klein-Gordon equation, are observed to form in our simulation at late times. Studying their properties analytically and numerically, we observe that in particular the earliest axitons contribute to density perturbations at the typical length scale of miniclusters. We analyse the small scale structure of the density field, giving the correlation length, power spectrum and the distribution of high density regions that will collapse into axion miniclusters. The final density field of our simulations can be used to calculate the minicluster mass fraction in simulations including gravity. In particular, we find that typical minicluster progenitors are smaller than previously thought and only of moderate, $\mathcal{O}(1)$ overdensity. We expect these miniclusters to have a rich sub-structure, emerging from small-scale fluctuations produced in the collapse of the string-wall network and from axitons.

\end{abstract}

\section{Introduction}

\subsection{Axion solution to the strong CP problem}

The axion is a hypothetical particle predicted in the Peccei-Quinn mechanism~\cite{Peccei:1977ur,Peccei:1977hh,Wilczek:1977pj,Weinberg:1977ma} to solve the so-called strong CP problem~\cite{Peccei:2006as}. The essence of the mechanism consists of promoting the $\theta$-angle of quantum-chromodynamics (QCD) into a full-fledged dynamical field, $\theta\to \theta(x^\mu)$. Non-perturbative effects (instantons) imply a different QCD-vacuum energy density as a function of $\theta$, $V_{\rm QCD}(\theta)$, which is minimised at the glorious CP-conserving value $\theta=0$. Was the field given enough time to relax to its lowest, it would easily accommodate the most stringent experimental constraints $|\theta|<1.3\times 10^{-10}$ by a wide berth. The relaxation time depends on the central and essentially only parameter of the Peccei-Quinn mechanism, the axion decay constant $f_A$. This new energy scale defines the canonically-normalised $\theta$-field as the \emph{axion} field $A(x^\mu)=\theta(x^\mu) \times f_A$, but most importantly, it gives a characteristic time-scale for the relaxation, $t_A\sim 1/m_A$ where $m_A= V''_{\rm QCD}|_{\rm \theta=0}/f_A$  is the {\rm axion-mass}. Axion models with values of $f_A\ll 10^8$ GeV are ruled out because many stellar systems would cool faster than observed by the emission of axions (see~\cite{Raffelt:1996wa} and the recent revisions \cite{Viaux:2013lha,Bertolami:2014wua,Ayala:2014pea,Chang:2018rso}). For such really large values of $f_A$, quantum effects are suppressed~\cite{Dvali:2017ruz} and the cosmological relaxation of $\theta(x)$ towards its minimum can be taken as that of a classical field suffering only the usual Hubble friction due to the expansion of the Universe. 

It was soon realised that the axion field relaxes as damped harmonic oscillations that behave like a coherent state of particles, effectively a gas of very cold dark matter (CDM) \cite{Preskill:1982cy,Abbott:1982af,Dine:1982ah}. Using $m_A\simeq 57\, \mu{\rm eV} (f_A/10^{10} {\rm GeV})^{-1}$ from the latest calculations of the QCD topological susceptibility \cite{diCortona:2015ldu,Borsanyi:2016ksw} with initial conditions of $\theta(t \sim 0)\sim \mathcal{O}(1)$ the typical oscillation amplitude today would be $\theta_{\rm today}\sim 10^{-21} (60\, \mu{\rm eV}/m_A)^{0.6}$, a wide berth indeed. The appeal of the axion as a solution of the strong CP problem is therefore manyfold. Not only do QCD effects take good care of cancelling the CP violation by sending the axion field to a CP conserving minimum, but also by doing it so imperfectly due to a finite lifetime of the Universe, they easily account for all the observed CDM of the Universe. Moreover, despite the minute values of the axion DM field today, there are experimental ideas galore to detect axions as DM, see \cite{Irastorza:2018dyq} for a recent review. The general trick to beat the smallness of the predicted signals is to couple the axion field with a resonant detector~\cite{Sikivie:1983ip}, but since the axion frequency $\nu_A\simeq m_A/2\pi$ is unknown, this haystack-needle hunting reveals as a painstakingly long endeavour.

The darkest side of this story is that the abundance of axion DM \emph{crucially depends on its initial conditions} and this prevents us in general from making the most desirable statement: ``the axion mass
should be ... to account for all the observed DM of the Universe'', which would allow a more focused search. However, a small hope glows in the so-called post-inflationary Peccei-Quinn scenario. This entails a broad set of  \emph{cosmological histories} where the axion field appears as an effective angular field (like a pseudo-Goldstone boson, for instance) after a phase transition at a very high temperature (usually $T_c\sim f_A$),   when it takes random initial values in the range $\theta \in (-\pi,\pi)$ in causally disconnected regions. Since there are zillions of those in today's horizon size, the DM abundance can be computed as a statistical average as a function of $m_A$ (equivalently $f_A$), the only unknown. This would pinpoint a clear experimental target! Moreover, if axion DM is experimentally found at that precise mass, not only the strong CP and the DM problems can be solved, but we will have another (very needed) handle on the very early  Universe. As a side remark, the opposite (pre-inflation) scenario consists on having cosmic inflation after the axion took its initial conditions. The axion field in a small patch is stretched to today's Universe size so the power of statistic inference is lost. Although anthropic arguments can be invoked to avoid overproduction of DM, there is no strongly favoured axion DM mass in such a
case.

\subsection{Phenomenological implications of post-inflation PQ symmetry breaking}

Besides the axion DM mass, the post-inflation scenario has another most relevant prediction: the distribution of DM will be highly inhomogeneous at very small scales, $\losc \sim 0.1$ pc --- the comoving size of the Universe at time $\sim t_A$. Regions with ${\cal O}(1)$ overdensities tend to collapse gravitationally very soon, even prior to average matter-radiation equality (redshift $z_{\rm eq}\sim 4000$), into small DM halos, usually called axion {\rm miniclusters}~\cite{Hogan:1988mp,Kolb:1993zz,Kolb:1993hw,Zurek:2006sy}. These regions can contain a large fraction of the total axion DM mass but are quite compact, a typical radius being $r_{\rm mc} \lesssim \losc/z_{\rm eq}\sim 10^{12}$ cm, and a typical mass $M_{\rm mc} \sim 10^{-12} M_\odot$~\cite{Kolb:1993zz,Kolb:1993hw,Kolb:1994fi}, with some dependence on the cosmological history prior to big-bang nucleosynthesis \cite{Nelson:2018via,Visinelli:2018wza}. Encounters with the Earth would enhance enormously the DM signal but are mighty rare. Depending on the minicluster fraction, it could be more advantageous to give up resonant detection techniques and focus on broadband experiments like a dielectric haloscope~\cite{TheMADMAXWorkingGroup:2016hpc,Millar:2016cjp} or even a dish-antenna~\cite{Horns:2012jf}. On the other hand, such compact objects can have interesting phenomenological consequences that can lead to an indirect detection or exclusion, like fast-radio bursts~\cite{Tkachev:2014dpa,Pshirkov:2016bjr}. Most discussed lately is the issue of dilute axion-stars
\cite{Kolb:1993zz}--- gravitationally bound solitons~\cite{Ruffini:1969qy,Kaup:1968zz,Chavanis:2011zi,Chavanis:2011zm,Chavanis:2016dab,Chavanis:2017loo} of the axion field which can easily appear in this scenario (for instance in the cores of miniclusters~\cite{Levkov:2018kau,Veltmaat:2018dfz}, but not only). More compact objects can lead to more pronounced~\cite{Bai:2016wpg,Bai:2017feq,Eby:2017xaw} and even coherent effects, see~\cite{Hertzberg:2018zte}. Recently, a new branch of ``dense" axion stars (energy density  $\sim V_{\rm QCD}$) was suggested~\cite{Braaten:2015eeu}, with even more spectacular consequences
\cite{Iwazaki:2014wka,Raby:2016deh}. However these are highly unstable objects associated with pseudo-breathers, oscillons~\cite{Visinelli:2017ooc}, and axitons in the context of axions~\cite{Kolb:1993hw}, which are only expected to appear on time scales $\sim t_A$ in the collapse of dilute axion stars \cite{Eby:2016cnq,Eby:2017xrr} before bursting into semi-relativistic axions~\cite{Levkov:2016rkk}.

Axitons themselves will appear extensively later on in our discussion. They correspond to quasi-stable oscillons of the Sine-Gordon equation, which can only exist for a brief epoch after $t_A$. At these times the axion field is dominated by non-relativistic axions while the axion mass growths as a fast power law. At later times, once the axion mass has settled to its zero-temperature value, axitons become unstable and diffuse away. While axitons contribute to small-scale inhomogeneities in the axion energy density, and hence require a careful treatment in the study of miniclusters, they are temporary objects whose remains will be scrambled by the gravitational evolution of the axion field.

Most importantly, miniclusters can be searched with femto-, pico-\cite{Kolb:1995bu} and micro-lensing~\cite{Fairbairn:2017sil,Fairbairn:2017dmf} techniques. While the former technique is still not very
sensitive~\cite{Katz:2018zrn}, the latter was recently pointed out as particularly suitable for the task.
In fact, the preliminary analysis~\cite{Fairbairn:2017sil,Fairbairn:2017dmf} of the HSC data on a long exposure tracking of M31 stars~\cite{Niikura:2017zjd} seems to already disfavour a minicluster mass
fraction above $10\%$~\cite{Fairbairn:2017sil,Fairbairn:2017dmf}\footnote{Recently, it has been pointed out that the geometric optics approximation used in~\cite{Niikura:2017zjd} is not completely justified at
the lowest black-hole masses excluded $\lesssim 10^{-11}M_\odot$~\cite{Inomata:2017vxo}, which will also affect the results of~\cite{Fairbairn:2017sil,Fairbairn:2017dmf}.}.
In reference~\cite{Fairbairn:2017sil,Fairbairn:2017dmf}, most reasonable assumptions about the minicluster mass and concentration distribution are taken but it is clear that something better than a very nice
and educated guess will be required to make any stronger claims.
Indeed, before proceeding to rule it out entirely, it seems reasonable to study in detail the axion distribution in the particularly predictive post-inflationary scenario. 
As miniclusters are expected to be generated copiously in the post-inflationary scenario, but \emph{not} in the pre-inflationary scenario, the detection or exclusion of a significant minicluster population with the relevant predicted properties will help distinguishing between the scenarios, providing also a handle on the energy-scale of inflation and other elusive aspects of particle cosmology.

In this paper we aim at shedding some light on the birth of axion density perturbations in the early Universe. In a further publication, we will study the collapse and evolution of the axion miniclusters
themselves, but it seems sensible to have a firm ground on the origin of the DM fluctuations first. The task requires performing cosmological numerical simulations of the axion field around $t_A$ with random
initial conditions. For the values of $f_A$ of interest, $t_A$ corresponds to shortly before the QCD phase transition. The immediate problem with simulating a random angular field is the formation of a network
of global strings by the Kibble mechanism~\cite{Kibble:1980mv}. Around an axion string the values of $\theta$ wind a factor of  $2\pi$, which topologically traps a region where $\theta$ is undefined. This calls
for an UV completion of the axion field, which is normally taken to be a complex scalar field $\phi=\rho e^{i\theta}$ with a Mexican-hat potential $V_\phi\sim \lambda(|\phi|^2 -f_A^2)^2$, like in the original
KSVZ model~\cite{Kim:1979if,Shifman:1979if}. In regions where $\theta\sim \pi$, the axion potential is larger resulting in a domain wall attached to the cosmic strings. The domain wall's tension helps in destroying the
axion string network shortly after $t_A$.
The first numerical studies of the axion field in this scenario\cite{Kolb:1993hw,Zurek:2006sy,Hardy:2016mns} artificially limited the amplitude of axion initial conditions to avoid strings and domain walls. Thus they could not  be reliably used to compute the distribution of axion fluctuations, but just to characterise the typical effects to be encountered. They emphasised the role of axion attractive self-interactions
in enhancing the overdensities~\cite{Kolb:1993zz,Kolb:1994fi} and the occurrence of axitons.
Recent simulations of the cosmic strings network have been used to compute the total axion DM density, either directly~\cite{Fleury:2015aca,Klaer:2017ond} or by studying the energy loss from the axion string
network, see~\cite{Hiramatsu:2012gg,Kawasaki:2014sqa} and the particularly recent~\cite{Kawasaki:2018aa,Gorghetto:2018myk}.  These works predict different values for the axion DM mass in this scenario, namely
\begin{itemize}
  \item $m_A= 115  \pm 25  \mu {\rm eV}$~\cite{Kawasaki:2014sqa},
  \item $m_A= 26.2 \pm 3.4 \mu {\rm eV}$~\cite{Klaer:2017ond},  
\end{itemize}
which are not in agreement by more than 3$\sigma$. But they are of a similar order of magnitude, enough for most purposes. One of the most recent papers emphasises the uncertainties in the extrapolation of
small-tension results and the need for a deeper understanding of the string dynamics~\cite{Gorghetto:2018myk}. 

These studies have not, however, discussed the DM distribution. This entails a series of problems on its own, like resolving the axiton core and understanding its role in the power spectrum.  The spatial
resolution of the grid presents itself as the biggest challenge in these simulations. On the one hand, cosmic strings need to be extremely thin to have enough tension, stand the pull of the domain walls
and not be destroyed unphysically. On the other hand, the axiton core size shrinks as $1/m_A$ in a period when the axion mass is increasing at a fast pace due to QCD thermal effects. The larger our grids are
the higher string tension we can simulate and the longer we can keep track of axitons.
Therefore, as the fundamental tool for our study, we have developed our own numerical code based on MPI/OpenMP parallelism and prepared to run efficiently in large computer clusters\footnote{The code is public
and can be found in \url{https://github.com/veintemillas/jaxions}}. Thanks to this code, we have performed simulations of strings, walls and axitons with the most refined grids ever achieved, up to $8192^3$
grid points.

The rest of this paper is organised as follows. In Sec.~\ref{simulations} we describe the physical time-scales and our simulations.  In Sec.~\ref{spectrum} we discuss the simulation results, in particular the
spectrum of axions obtained after the string network has decayed and its connection with the literature. In Sec.~\ref{density} we analyse the spectrum of density fluctuations and how we disentangle it from the
all-pervading axitons, which receive their own analysis in Sec.~\ref{sectionaxitons}. The statistics of minicluster seeds are discussed in Sec.~\ref{minicluster}. A summary and discussion of our findings is
presented in Sec.~\ref{conclusions} where we also present our conclusions.

\section{Simulations of the axion DM field}\label{simulations}

In the following section we introduce the equation of motion for the axion field implemented in our simulation. Cosmic strings necessitate a UV completion of the theory, for which we will assume a Mexican-hat potential, motivated e.g. by the original KSVZ model. A major challenge to this kind of simulations is the large separation of scales between the core size of a cosmic string and the cosmological horizon. We discuss this issue and its implications at length and advocate the viewpoint that the physical quantities we aim to investigate, namely the axion DM yield and its spatial distribution, will not be strongly affected by the string tension. In this context we also rely on the Press-Ryden-Spergel (PRS) trick~\cite{Press:1989yh,Moore:2001px}, keeping the size of the string core constant in comoving coordinates. Section \ref{simulations} is completed by a discussion of all restrictions on the simulation's parameters, which are necessary to capture all physical effects as accurately as possible. In this context, a conflict emerges between resolving the axiton cores and arriving at a fully non-relativistic axion population. Possible solutions are to prematurely cut the power-law growth of the axion mass or to employ an analytical late-time approximation to the evolution equations (c.f. Eq.~\ref{solumode}). We choose the latter option.

\subsection{Evolution equations of the axion field}

The axion field evolution in the early Universe is described by the Klein-Gordon equation in an expanding Friedman-Robertson-Walker metric,
\be
\label{eom1}
\ddot \theta +3 H \theta-\frac{1}{R^2}\nabla^2\theta+\frac{1}{f^2_A}\,\partial_{\theta}V_{\rm QCD}(T,\theta) = 0\,,
\ee
where $T$ is the temperature of the universe, $R(t)$ is the scale factor and $H=\dot R/R$. The temperature dependence of the axion potential is due to the fact that instantons are suppressed at high
temperatures as QCD becomes less non-perturbative. At sufficiently large $T\gg T_c\sim 157$ MeV the potential can be written in the dilute instanton gas approximation as\footnote{In the low $T$ regime the
potential away from the minimum is not well-described by \eqref{potential}. However, it seems to be a good approximation above $T\sim 2T_c$ or, at least, the quartic coupling computed in~\cite{Bonati:2015vqz}
does. The fact that at these temperatures the axion field is \emph{mostly} very close to $\theta\sim 0$ renders the inaccuracy of small importance to us in this paper, but, as we will see, it is not entirely
irrelevant.} 
\be
\label{potential}
V_{\rm QCD} \simeq \chi(T)(1-\cos\theta), 
\ee 
and the axion mass inherits a temperature dependence from $\chi(T)$, the QCD's topological susceptibility, since $m_\A^2f^2_\A=\chi(T)$. Recently, $\chi(T)$ has been calculated in lattice QCD up to
$T\sim 2$~GeV \cite{Borsanyi:2016ksw}. The axion mass increases very abruptly as the universe's temperature decreases and saturates at a value $\chi(0) = \chi_0 = (75.5(5){\rm MeV})^4$ below $T_c$
\cite{diCortona:2015ldu,Borsanyi:2016ksw}, which allows to define the zero-temperature axion mass,
\be
\label{axionmassT0}
m_A(T=0) = \frac{\sqrt{\chi_0}}{f_A} = 57 ~ \mu{\rm eV}~\left( \frac{10^{11}\rm GeV}{f_a} \right).
\ee 

At early times, the axion potential is irrelevant and Eq.~\eqref{eom1} is a simple relativistic wave equation. When the potential term $\partial_{\theta}V_{\rm QCD}/f_A^2 = m_A^2\sin\theta$ becomes relevant,
the axion field will tend to roll down the potential and oscillate around $\theta= 0$ with an effective equation of state like a non-relativistic particle gas, i.e. dark matter. We define the characteristic
time-scale for this transition as\footnote{Definitions in the literature differ slightly, for instance Sikivie and Kawasaki use $3H_1=m_A$ as the defining scale. Fortunately for comparison's sake, this has only
mild effect on the value of $H_1$ as the temperature dependence of $m_A$ is quite strong.}
\be
\label{Hm1}
H(T_1) = m_\A(T_1) \equiv H_1.
\ee
We will assume that the Universe is already radiation dominated at the temperatures of interest. Alternative cosmological histories give in general different miniclusters~\cite{Visinelli:2018wza} and will be
treated in a separate publication. The Hubble expansion parameter is given by 
\be
H^2 = \frac{8\pi}{3 \mpl^2} \(\frac{\pi^2}{30}g_*(T) T^4\)\,,
\ee
where $\mPl$ is the Planck mass and $g_*(T)$ the effective number of relativistic degrees of freedom of the plasma. Using the very complete lattice QCD output for axion cosmology of~\cite{Borsanyi:2016ksw} we
obtain numerically in the region of interest, 
\bea
\label{Tm1}
T_1   &\simeq& 1.694\, { \rm GeV}\(\frac{m_a}{50\,\mu\rm eV}\)^{0.1638},  \\
H_1   &\simeq& 3.45\times 10^{-3} {\mu\rm eV} \(\frac{m_a}{50\,\mu\rm eV}\)^{0.338},   \\
1+z_1 &\simeq& R^{-1}_1 =1.956\times 10^{13} \(\frac{m_a}{50\,\mu\rm eV}\)^{0.1712}, \\
L_1   &\equiv& \frac{1}{H_1 R_1} \simeq 
1.116 \times 10^{17} {\rm cm}\(\frac{50\,\mu\rm eV}{m_a}\)^{0.167}
=  0.0362 \,{\rm pc}  \(\frac{50\,\mu\rm eV}{m_a}\)^{0.167}\,,
\label{Tm1-end}
\eea
where we have given the redshift at which Eq.~\eqref{Hm1} is satisfied as well as the physical and comoving size of the horizon at that time.
In that region of temperatures, one can express $\chi(T)\simeq \chi(T_1)\({T_1/T}\)^{\n'} \simeq \chi(T_1)\({R}/{R_1}\)^\n $ where $\n\sim 7.3$. The value of $\n  \sim 7.3$ is smaller than $\n'\sim 8.2$ due to
the changing number of degrees of freedom. Indeed, assuming quasi-thermal equilibrium conditions during the expansion of the SM plasma, the entropy in a comoving volume $\propto g_{*S}T^3 R^3$ is conserved,
leading to $T\sim 1/g_{*S}^{1/3}R$. Since the number of entropy degrees of freedom $g_{*S}\sim g_{*}$ is decreasing at those temperatures~\cite{Borsanyi:2016ksw} $T$ decreases a bit slower than $1/R$.

It is advantageous to use conformal time, $d\eta = dt/R$, normalised to $\eta_1=L_1$, rescaled coordinates and a conformally rescaled axion field
\be
\ct = \frac{\eta}{L_1} \quad , \quad  \veca \x =  \frac{\vec x}{L_1} \quad, \quad \ctheta = \ct \theta.
\ee
In these coordinates Eq.~\eqref{eom1} takes the very simple form
\be
\label{eom11}
\ctheta_{\ct\ct} -\nabla^2 \ctheta + \ct^{\n+3}\sin\(\frac{\ctheta}{\ct}\) - \frac{R_{\ct\ct}}{R}\ctheta= 0,
\ee
where $\ctheta_\ct = d\ctheta/d\ct$. These conformal quantities, made dimensionless with appropriated powers of $H_1$ (energy) and $L_1$ (length), will be referred to as to be in ADM units.

The term  $R_{\ct\ct}/R$ is much smaller than the rest because $\ct \simeq R/R_1$ with very good precision. In fact, $\tau$ and $R/R_1$ scale differently only due to the change of the effective number of
degrees of freedom, 
\be
d\ct = R_1 H_1 d\eta = \frac{R_1 H_1dR}{H R^2} = \frac{R_1 H_1dR}{ H_1\(\frac{g(T)}{g_1}\)\(\frac{T}{T_1}\)^2 
R_1^2 \( \frac{g_1T^3_1}{g(T)T^3} \)^{2/3} } = 
\(\frac{g(T)}{g_1}\)^{1/6}  \frac{dR}{R_1}\,,
\ee
and only to a very mild power of it. In the following we neglect the term $R_{\ct\ct}/R$ but keep implicitly the effects of the changing $g(T)$ in the axion mass time-dependence coefficient, $\n$. 

\subsection{Complex scalar field as UV completion}
Endowing the axion field with random initial conditions $\theta\in[-\pi,\pi)$ in causally disconnected regions produces a network of cosmic strings~\cite{Kibble:1980mv}, which requires an UV completion of the
axion model to regularise the string-core energy and enable its dynamics. As many before us, we use the simplest of such completions, a complex scalar field $\phi=\m e^{i\theta}$ with the following Lagrangian
density, 
\be
\label{lagrangian}
{\cal L} = \int d^3x\,dt\,R^{3}\(\frac{1}{2}\left|\frac{d\phi}{dt}\right|^2-\frac{1}{2R^2}|\nabla \phi|^2 - V(\phi)\). 
\ee
The full potential,
\be
\label{PQ1QC1D1}
V(\phi) = V_{\rm PQ}(|\phi|)+V_{\rm QCD}(\theta) = 
\frac{\lp}{8}\(|\phi|^2-f_A^2\)^2+\chi(T)\(1-\frac{{\rm Re}(\phi)}{f_A}\), \\
\ee
is composed of the QCD potential $V_{\rm QCD}$ and the saxion potential $V_{\rm PQ}$. The QCD potential reproduces $(1-\cos\theta)$ when $\lambda_\phi f_A^4\gg \chi$ and the radial mode is at its minimum $\m=f_A$. Henceforth, we will denote the radial mode as ``saxion", as it is usually called in the context of supersymmetry, see for example~\cite{Asaka:1998xa}. However, we will not assume that our saxion has any of the characteristics derived from the SUSY context. It is rather the new scalar singlet introduced in the KSVZ axion models~\cite{Kim:1979if,Shifman:1979if}.

As with the axion field, it is advantageous to define a properly normalised conformal complex field as 
\be
\cf=\ct \frac{\phi}{f_A}. 
\ee
The equations of motion for $\phi$ are extremely simple in ADM units\footnote{In this formula we again neglect the difference between $\ct$ and $R/R_1$ except in $\n$. }, 
\be
\label{eom2}
\cf_{\z\z}-\nabla^2\cf+\frac{\lambda}{2}\cf(|\cf|^2-\ct^2) - \ct^{\n+3}=0 ,  
\ee
having only one dimensionless parameter $\lambda$. In physical units, $\lp$ determines the the saxion mass, $\ms = \sqrt{\lp} f_A$. In ADM units, $\lambda$ reflects the ratio of the saxion mass to the Hubble
rate at $\ct=1$, which, by definition, is also the axion mass at that relevant time $\ct=1$,
\be
\lambda = \lp \frac{f^2_A}{H_1^2} = \frac{\ms^2}{H_1^2} = \frac{\ms^2}{m_A^2(T_1)}. 
\ee 
Realistic parameters like $f_A\sim 10^{11}$ GeV and $\lp\lesssim 1$ require simulations with $\lambda\sim 10^{57}$, which we cannot perform at the moment. The reason is that the saxion field regularises the
divergence of the axion gradient energy density at the core of cosmic strings at a length scale related to the inverse saxion mass, but at the same time we need to simulate distances longer than the causal
horizon $\sim 1/H$. With current memory constraints we can simulate lattice grids of $\sim 512^3$ in a laptop and $\sim 8192^3$ in a supercomputer, therefore we can only have a hierarchy of scales
$\ms/H \sim {\cal O}(10^3-10^4)$ and $\lambda\sim 10^7$. The assumption will be that the physics, in this case the axion DM yield and its spatial distribution does not strongly depend on this parameter. Barring
the unphysical destruction of domain walls pointed out in~\cite{Fleury:2015aca}, this seems not to be entirely unreasonable as the energetics of the problem suggest that $\lambda$ enters into the system only as
$\sim\log\lambda$.

\subsubsection{Strings}

A straight cosmic string is the static and minimum energy solution of the Hamiltonian derived from \eqref{lagrangian} where the axion field wraps linearly a factor of $2\pi$ around a straight line. In
cylindrical coordinates $(\rho, \varphi, z)$ these solutions are found with the ansatz $\theta(\varphi) = \varphi$ and $\m=\m(\rho)$. The solution for $\m$ is $\m\sim  0.5302\ms \rho + ...$ at the core, while
$\m\to f_A$ as $\rho\to \infty$. An approximate fit gives,
\be
\m (\rho) \simeq f_A\frac{0.43 \rho'+0.164 \rho'^2+0.036 \rho'^4}{1+0.39 \rho'+0.2 \rho'^2+0.036 \rho'^4} 
\quad , \quad \rho' = \ms \rho, 
\ee
so $\m$ is half way to $f_A$ around $\ms \rho\sim 1.4$. 

The tension (energy per unit length) of such straight string is, 
\be
\label{tension}
\mu_{\rm st} = 2\pi\int \rho\, d\rho\(\frac{1}{2}\(\frac{d\m}{d\rho}\)^2
+V_{\rm PQ}+\frac{1}{2}\(\frac{\m}{\rho}\)^2\) =  f_A^2\(4.5 +\pi \log\(\frac{\ms \rho_{\rm cut}}{4}\) \) \,.
\ee
The first term involves the saxion gradient and potential terms while the second comes from the axion gradient and  diverges at $\rho\to \infty$, so we have cut the integration along the radial coordinate at a
distance $\rho_{\rm cut}$ from the centre. Physically,  the role of this cut-off will be played by the distance to nearby strings or the  radius of a string loop. The right-hand side of formula~\eqref{tension}
is a good approximation for $\ms \rho_{\rm cut}\gtrsim 2$. The energy inside $\ms \rho_{\rm cut}\simeq 4$ is shared in equal amounts between the axion and saxion fields, each giving $\sim 2.25 f_A^2$. From
$\ms \rho_{\rm cut}\simeq 4$, essentially all energy is in the form of the axion field gradients.

In numerical simulations of cosmic strings one encounters the problem that the Universe expands while the string cores remain physically constant with radial dimensions $\rho\sim 1/\ms$.
The Press-Ryden-Spergel (PRS) trick~\cite{Press:1989yh,Moore:2001px} consists in simulating with a quartic coupling that decreases on time, effectively
\be
\lambda \to \frac{\lambda_{\prs}}{\ct^2}\,,
\ee
with a constant $\lambda_\prs$ so that the saxion mass decreases as $\ms\propto 1/\ct$ and the string core radius stays constant in comoving coordinates along the Universe evolution. Physically this corresponds
to the strings ``fattening" as the Universe expands, so these simulations were called ``fat"-string simulations~\cite{Moore:2001px}. In our opinion this name makes poor justice to the trick. In a simulation,
the lattice spacing limits the value of $\lambda$ at the \emph{latest times} (when strings are thinner in comoving coordinates). Fixing the string thickness to be acceptable at late times, $\prs$-strings are
\emph{thinner} than standard-strings throughout the whole simulation! Instead of referring these strings as ``fat'', we will call them $\prs$, which transmits the notion of ``compressed" strings.

\subsubsection{Small digression on energetics and the axion yield}
\label{sec-digression-axion-yied}

The string network is thought to evolve following a sort of \emph{scaling solution} where the normalised average length of strings per Hubble patch volume, defined as
\be
\xi = \frac{\ell_{{\cal V}} t^2}{{\cal V}},
\ee
is  a  constant of the order of 1~\cite{Yamaguchi:1998gx,Yamaguchi:1999yp}. Here $\ell_{\cal V}$ is the total length of the strings in a given {physical volume $\cal V$ greater than the causal horizon $\sim t^3$}. 
The reason why $\ell_{\cal V}/{\cal V} \sim 1/t^2$ is the following: string loops of size $\ell$ take at least a time of order $t\sim \ell$ to collapse and string bends of a radius of curvature $\sim \ell$ take
a similar time to straighten up, so that after a time $t$ all the structures of length $\lesssim t$ had time to relax and only relatively straight strings, that stretch over more than 1 horizon-volume, can
survive. Therefore $\ell_{\cal V}$ cannot be much larger than $t$ in a ${\cal V}\sim t^3$ volume if the string network relaxation is efficient. Note that the constancy of $\xi$ has been challenged recently by a
number of works, which show a soft (logarithmic) increase with simulation time \cite{Fleury:2015aca,Gorghetto:2018myk,Kawasaki:2018aa}, also visible in our simulations.

The average energy density stored in strings and their fluffy axion-gradients is therefore $\varrho_{\rm st}= \xi(t) \mu_{\rm eff}/t^2$, where $\mu_{\rm eff}$ is given by Eq.~\eqref{tension} with an adequate cut-off,
usually taken to be $\rho_{\rm cut}= 2t/\sqrt{\xi}=1/H\sqrt{\xi}$ as the typical inter-distance between strings. 
The string energy, highly localised around the string cores, is dissipated into the axion field as the string-loops collapse and the network relaxes\footnote{A small part is radiated into saxions as well
\cite{Gorghetto:2018myk}. The energy released in gravitational waves is much smaller~\cite{Davis:1985pt}. }. This energy is stored as kinetic and gradient energy in the axion field between the strings and can be understood as relativistic axion waves. Later, around $\ct\sim 1$ the axion potential becomes important, and domain walls attached to the axion strings build up tension, dragging the remaining strings to collapse. At similar times, the lowest momentum axion modes start to oscillate and
become non-relativistic axions, i.e. dark matter. With increasing times the relativistic axions become non-relativistic too. The energy density of the string network $\varrho_{\rm st}$, that of relativistic
axions $\varrho_{A}$, and the energy stored in the QCD potential $\varrho_{\rm mis}$ are parametrically of the same order $\sim f_A^2 H_1^2$ at the key time, $t_1$
\be
\label{tworhos}
\varrho_{\rm st} :: \varrho_{A} :: \varrho_{\rm mis} = 
\frac{4 \pi \kappa }{\ct^4} :: \frac{4 \pi \kappa}{\ct^4}\frac{\kappa}{3} :: \frac{1}{2}\ct^\n , 
\ee 
see~\cite{Gorghetto:2018myk}, but the string and axion energy densities are enhanced by the large log of the tension,
\be
\kappa = \frac{\mu_{\rm st}}{\pi f_A^2} = \log\(\frac{\ms}{H}\),
\ee
which is of order $\sim 70$ for standard values of $\lambda\sim \mathcal{O}(1)$.
The good news is that the gigantic energy scales involved in string cores do not appear in the energy-balance as terms proportional to $\lambda$ but only as a $\log\sqrt{\lambda}$. Most of the energy of the
strings is in a small region around the core, while the already radiated axions and the QCD energy fill almost all space. The bad news is that the energy density is not the key parameter to compute the final
dark matter density, the axion \emph{number} is an adiabatic invariant in the further evolution.

As benchmark for the axion DM production,  one usually introduces the ``naive" misalignment contribution as the comoving axion number that follows from the evolution equation \eqref{eom1} when neglecting the
laplacian and averaging over initial conditions on $\theta(t=0)\in[-\pi,\pi)$.
For $\n=7$ the numerically obtained number of axions $N_A$ per comoving volume $V$ is
\be
\label{misalignment}
\frac{N_A}{V} =  16.00 \times H_1 f_A^2 R_1^3,   \quad \text{(misalignment-only, }\n=7) \text{\quad \cite{Fleury:2015aca}}\,.
\ee
\additionalinfo{Normalising it to the comoving entropy $S=2\pi^2 g_{*S}(T) T^3 R^3/45$ and using $\n=7$ we  have~\cite{Fleury:2016xrz}
\be
\frac{N_A^{\rm mis}}{S} =  16.85 \frac{H_1 f_A^2}{\frac{2\pi^2}{45}g_{*S}(T_1) T_1^3},
\ee}
The fact that this is a relatively large number compared to the naive estimate $\sim H_1 f_A^2 R_1^3$ comes in part from the non-harmonic delay of the axion field oscillations for initial conditions close to
$\theta\simeq \pi$ and in part from the steep increase of the axion mass $\propto \ct^{\n/2}$. From the comoving axion number and relations \eqref{Tm1} to \eqref{Tm1-end}, the relic abundance as
fraction of the critical density today $\varrho_c$ and the reduced Hubble parameter $h$ is
\be
\Omega_c h^2 = \frac{m_A(T=0)\frac{N_A}{V}}{\varrho_c}h^2=0.0037 \(\frac{50\,\mu\rm eV}{m_a}\)^{1.176}\frac{N_A/V}{H_1 f_A^2 R_1^3}.
\ee

Coming back to the energy ratios given in Eq.~\eqref{tworhos}, the large factor of $\kappa$ implies that relativistic axions dominate the axion energy at $t_1$. However, in order to know whether they also dominate
the axion number we need to know the \emph{average axion energy}. There is a long-standing controversy regarding how efficiently string energy is converted into axion number. The situation has been recently
exquisitely reviewed in~\cite{Gorghetto:2018myk}. In their own numerical study, which reaches values of $\kappa\simeq 6$, the authors conclude that the energy of strings goes into a few relativistic axions
instead of many low energy ones. They study standard and \prs~strings. The system simulated is essentially the one we have discussed here, neglecting the axion mass. The new results of~\cite{Gorghetto:2018myk}
seem to agree with previous simulations of~\cite{Fleury:2015aca} for $\prs$ strings. Reference~\cite{Fleury:2015aca} did not discuss the conversion of string energy into axions, but it simply counted the axions
at the end of their simulation, without discussing the spectrum. In doing so they obtained a surprising result: for $\kappa\simeq 6$, the axion yield is half the ``naive" misalignment value! Clearly, this
invites to interpret that strings are not efficient in radiating axions.

On the other hand, the interpretation of the results of~\cite{Fleury:2015aca,Gorghetto:2018myk} is in opposition with the previous state of the art\cite{Hiramatsu:2010yu,Hiramatsu:2012gg}, which has been very
recently updated in~\cite{Kawasaki:2018aa}. These authors simulate ``standard'' strings and conclude that the mean momentum of radiated axions is soft and that the axion number production is dominated by the
string emission. They attribute part of the possible misunderstanding to the usage of $\prs$ strings. However, we note that~\cite{Gorghetto:2018myk} and \cite{Kawasaki:2018aa} disagree on the
\emph{mean radiated axion energy} when they simulate ``standard'' strings. Even if the latest simulations of \cite{Kawasaki:2018aa} have more dynamical range thanks to larger boxes, the analysis of
\cite{Gorghetto:2018myk} and in particular their extrapolation study is much more transparent. It appears to us that the different studies might be using a different definition of \emph{mean radiated energy}
\cite{RedondoPatras2018} and certainly a different extrapolation technique. We will put up our explanation in a forthcoming publication.

A most interesting and complementary piece of information comes from a recently proposed new way of simulating high-tension global strings~\cite{Klaer:2017qhr}. Using two scalar fields charged under a $U(1)$
gauge field, the remaining Goldstone direction can become the axion with an effective value of $f_A$, which is suppressed by an easily controllable factor. Each global string becomes attached to a local string
produced by the $U(1)$ field. The local string can enhance the global string tension without disturbing the long-range string dynamics~\cite{Vilenkin:2000jqa}. In this sense, large values of
$\kappa$, up to the physical value $\kappa \sim 70$, can be simulated! In these simulations, the networks become denser $\xi\sim 3-4$ in agreement with the idea that strings of higher tension take more time to be accelerated
(have more inertia). However, in \cite{Klaer:2017qhr} the axion yield increases only little with respect to the $\xi\sim 1, \kappa\sim 6$ result, and the extrapolation to the physically motivated values stays
below the naive misalignment estimate. This is perfectly consistent with the idea that the string energy goes into few relativistic axions and the final axion number is dominated by the latest produced axions,
which cannot be easily distinguished from those usually attributed to misalignment. Certainly, we are still in need of further simulations that investigate the mechanism of axion production with the strings
and put into context all the previous works, but as of now it seems that the results of the different groups clearly point in the same direction. We plan to attack this issue in detail in a further publication.

In this work we want to shed some light on an aspect that has received little attention until recently and that will have very important consequences: the resulting density inhomogeneities of the axion field. For the moment, we will take the relatively optimistic view, that seems to emerge from the recent simulations, i.e. that axions are not efficiently radiated from strings, so that using high or ultra-high tension does not amount to a large change in the properties of the simulations. Further understanding on the properties of the scaling solution can change this conclusion dramatically~\cite{Gorghetto:2018myk}, a circumstance under which we could be forced to revisit most of the conclusions presented here. 

\subsection{Simulations}

We have performed numerical simulations of the complex scalar and the axion field (c.f. Eq.~\eqref{eom2} and Eq.~\eqref{eom11}) around the relevant times when axions become dark matter $\ct\sim 1$. We have evolved the fields on
cartesian discretised grids with up to $8192^3$ lattice points. Our simulations are defined by three central physical parameters,
\be
\n  , \quad \quad  L,\quad  \quad \lambda_\prs.
\ee
The first controls how fast the axion mass increases around $\tau\sim 1$, the second is the physical length of our simulation box and the third determines the saxion mass/string tension, which needs to be taken
as large as possible. In this section, all quantities are in ADM units: energies are normalised to $H_1$ and comoving lengths to $L_1$. We define $a=L/N$ as our comoving  lattice spacing. The two parameters
controlling discretisation effects are $a$ for space and $d\ct/a$ for the time.

We will report on simulations with ${\prs}$ strings because they are thinner than physical strings and so have a much larger tension throughout the whole simulation. Indeed, physical strings are deemed to be a
factor $\sim \ct_f/\ct_i$ wider than $\prs$ strings at initial time. This makes the string density smaller than it should have been at that time in the scaling solution. The alternative would be to start the
simulation later, but then we are more sensitive to our initial conditions. Since we are not interested in details of the scaling solutions we opt for the most physical option, which, funnily enough, seems to
us the use of the ``unphysical'' $\prs$ trick.

The restrictions on our type of simulations have been discussed at length in different references, e.g.~\cite{Yamaguchi:1998gx,Yamaguchi:1999yp,Hiramatsu:2010yu,Fleury:2015aca}. We summarise them in the following. 

\subsubsection{Resolving string-core}
Resolving the string cores with a few points imposes a lower limit on  $\ms a$. Note that we only present results for $\prs$ strings, for which $\ms a$ is constant throughout the simulation.
References \cite{Fleury:2015aca,Gorghetto:2018myk} study the dependence of network quantities like $\xi$ as a function of $\ms a$ and conclude that values below $\ms a\lesssim 1.5$ are acceptable. However,
reference~\cite{Gorghetto:2018myk} reports that the axion spectrum from string radiation requires $\ms a\lesssim 1.0$. In the interval $\ms a\in(1.0,1.6)$ their instantaneous spectrum hardens by a factor of
$\sim k^{0.11}$. The main features of our results are not very sensitive to $\ms a$, so we have tended to use the highest reasonable tensions $\ms a\sim 1.5$, although we also present results with values in the
range $\sim 0.75-1.5$.

\subsubsection{Time scales of the simulations}

At early times the network of cosmic strings evolves by collapsing loops, smoothing bends and reconnecting long strings so that the typical distance between strings is of the size of the horizon,
$\ell_H\sim t$. The axion field becomes increasingly homogeneous, with fluctuations (relativistic axions) on top of a relatively smooth distribution between strings.

In our ADM units the axion field starts responding to its potential at $\ct\sim 1$ (c.f. Eq.~\eqref{Hm1}). At that time the surfaces of $\theta\sim \pi$ that connect strings start behaving like domain
walls with a surface tension given by $8 m_Af^2_A$. The wall tension builds up at a very fast rate (due to the fast increase of $m_A$ as temperature drops) and pulls the strings to collapse around the regions
where $\theta\sim \pi$. We define $\ct_2$ as the time when the wall tension starts dominating over the string tension, analogous to \cite{Hiramatsu:2012gg}
 
\be
8m_Af_A^2 t_2  = \mu_{\rm st} \to \ct_2 \sim \(\frac{\pi \kappa}{4}\)^\frac{2}{\n+4}.
\ee
The collapse of the network is limited essentially only by causality and therefore at $\sim 2\ct_2$ the walls have had time to drag the strings a distance $\ell_H(\ct_2)$ and collapse them. Note that the time required for the collapse would
reduce if $\xi$ was sizeable larger than $1$. Thus our simulations require approximately a time span of $2\ct_2$ to reach over the collapse of the string-wall network. For the parameters we can simulate $\kappa\sim 6- 8$ and $\n=7$,   
\be
2\ct_2 \sim 2.8. 
\ee

By that time the bulk of the axions are already non-relativistic. However, we want to ensure that by the end of our simulations even the highest momentum axions have become non-relativistic. The largest momentum
an axion can have along each direction of the discretised simulation grid is the Nyqvist momentum or UV cut-off, $k_{\rm Ny} = \pi/a$. Imposing that even this mode is non-relativistic, $p_{\rm Ny} = k_{\rm Ny}/R \ll m_A$, translates into a lower limit for the simulation time,
\be
\label{timeNyqvist}
\ct \gg \ct_{\rm Ny} \equiv  \(\frac{\pi}{a/L_1}\)^\frac{2}{\n+2}.
\ee

The last time-scale to consider is the time corresponding to the critical temperature of the QCD phase transition. After this time the axion mass acquires its $T\sim 0$ value \eqref{axionmassT0}. 
Since we normalise $\ct=1$ at $m_A=H_1$ and the critical temperature $T_c$ is fixed, $\ctc$ depends on the axion mass, although only slightly in the region of interest, that is for $m_A\sim 50\, \mu$eV.
Here we choose $T_c=140$ MeV because it is the temperature at which the axion mass growth becomes $\n\simeq 2$, above which axitons become unstable (see Sec.~\ref{sectionaxitons}). We find 
\be
\label{ctc}
\ctc \simeq 16 \(\frac{m_A}{50\mu \rm eV}\)^{0.17},  
\ee
where we have used Eq.~\eqref{Tm1} and taken into account the fast decrease of $g_{*S}$ around $T_c$.

\subsubsection{Unphysical DW destruction and shift correction}

The QCD potential \eqref{PQ1QC1D1} also contributes to the saxion potential. One can easily show that the value of $\m$ that minimises the potential \eqref{PQ1QC1D1} for a given value of $\theta$
is\footnote{For this formula, evaluate the arctan between 0 and $\pi$.}
\be
\m_{\rm min}(\theta) = \frac{2 f_A}{\sqrt{3}} \cos\(\frac{1}{3}\arctan\(\frac{\sqrt{1-27 \varepsilon^2\cos^2\theta}}{3\sqrt{3}\varepsilon\cos\theta}\)\) \,,
\ee
where the ``small" $\varepsilon$ parameter controls the QCD correction,
\be
\varepsilon = \frac{\chi}{\lp f_A^4} =  \frac{m_A^2(t)}{\ms^2}
\to \frac{\ct^\n}{\lambda} \to \frac{\ct^{\n+2}}{\lambda_\prs}.
\ee
For physical values $\lambda\sim 10^{60}$ these corrections are irrelevant. But for the limited computational resources available, $\varepsilon$ can easily grow sizeable at times $\ct \gtrsim 1$, and indeed it
would eventually beat the saxion mass if the $\ct^\n$ scaling of the axion mass was maintained. We can correct our physics at the first order in $\varepsilon$ by noting that 
\be
\m_{\rm min}/f_A = 1 + \varepsilon \cos\theta + ... \,,
\ee
so the valley of the Mexican hat stops being a circle in a strict sense. However, close to $\theta=0,\pi$ the valley is still locally a circle around a centre that has shifted, from $\phi=0$ to
$\phi_\varepsilon \simeq \varepsilon(1+0i)$. The axion field close to its minimum can then be redefined as $\arg(\phi-\phi_\varepsilon)$ and still keeps its expected mass, although there are small corrections
to the cosine potential $\propto \varepsilon$.

However, when $|\varepsilon|\sim 1/\sqrt{27}\sim 0.2$ the saxion stops having a minimum at $|\theta|=\pi$ (this is the first value to fail but others follow). At this point the QCD potential has tilted the Mexican hat so much, that the axion field will not roll down its potential through the surrounding valley, but go unimpeded over the hat's top!

In~\cite{Fleury:2015aca}, the authors showed that even for smaller values of $\varepsilon$ this process happens in axionic domain walls, allowing the trapped axion field to relax to zero in an ``unphysical"
manner. They computed the critical value of $\varepsilon$ to be $\sim 1/40$, smaller than the $0.2$ obtained naively, because in the domain walls the gradient energy of the axion field  pushes the
$\theta\sim\pi$ region with even more force than the QCD potential only. To avoid the unphysical destruction of domain walls by this \emph{roll-over-the-top mechanism}, we must keep $\varepsilon$ below that
critical value as long as we simulate the complex PQ field (c.f. Eq.~\eqref{eom2}). Imposing the condition at least \emph{until the time where strings and walls have collapsed}, $\ct\sim \beta\ct_2$ with
$\beta\simeq 2$ as argued before, restricts the saxion mass to $\ms^2<40 m_A^2(\beta\ct_2)$.

For $\prs$ strings this translates into a restriction on the tension $\lambda_\prs$,
\be
\lambda_\prs > 40\times \beta^{\n+2}\(\frac{\pi\kappa}{4}\)^\frac{2\n+4}{\n+4} \quad \(\mathrm{for}\quad\varepsilon(2\ct_2)<\frac{1}{40}\) \,,
\ee 
or, equivalently, to a constraint on the lattice spacing $a$
\be
\label{spacing}
a  < \frac{\ms a}{\sqrt{40}\times \beta^{\n/2+1}}\(\frac{4}{\pi\kappa}\)^\frac{\n+2}{\n+4}.
\ee
We show the upper bound this condition imposes on the physical box size $L$ in Fig.~\ref{simugraph} for different values of $N=L/a$. Equation \eqref{spacing} has the undesirable property that depends quite
strongly on $\beta$. We find that $\beta\sim 1.9$ is sufficient to make sure that there is no unphysical destruction of domain walls. However, the limit is quite subtle because at the final stage of their
evolution loops are already very small and detecting an increase in string length requires loads of statistics. For $\n=7,\beta=1.9$ and $\ms a=1$, we require $L/L_1\lesssim 2.2, 4.4, 6.6, 8.8,17.6$ for
the number of points per dimension $N=1024,2048,3072,4096,8192$, respectively. As we will see, $L/L_1<6$ is prone to have some finite volume effects, so in principle we need at least $(N \sim 3072)^3$ boxes.
The condition can be typically fulfilled in the large boxes we can simulate, although not by a huge margin.

As $\n$ is large, the distortion parameter $\varepsilon$ increases very fast, and distortions of the potential would become very severe quickly after the criterion $\varepsilon=1/40$  is met. To keep the
axion physics as close as possible to the physical picture, we simulate the complex scalar field (c.f. Eq.~\eqref{eom2}) until strings have disappeared and then switch to the evolution of $\theta$ (i.e.
Eq.~\eqref{eom11}) so that uncontrolled distortions of the potential and unphysical roll-over-the-top behaviours cannot happen. In the change of variables, we parametrise the axion as the angle around the
\emph{shifted} centre of the Mexican hat, where the shift is computed as $\m_{\rm min}(\theta=0)-f_A$. More details are given below, in subsection~\ref{transition}.

\subsubsection{Box size and finite volume effects}
Relativistic waves travel at the speed of light, which in our conformal time and coordinates corresponds to 1. We use periodic boundary conditions to maximise the usable statistics. In a periodic grid of size $L$ a spherical relativistic wave generated at a time $\ct_0$ interferes with the waves emitted from the 6 nearby copies at a time $\ct_0+L/2$. To avoid this ``finite volume'' effect would naively require to limit the the simulation time to $\ct<L/2$ (using $\ct_0\sim 0$). Combined with the objective to evolve our simulation at least to the time where the destruction of the string-wall network has completed this
translates into a lower limit on the box size.

\begin{figure}[tbp]
\begin{center}
\includegraphics[width=0.8\textwidth]{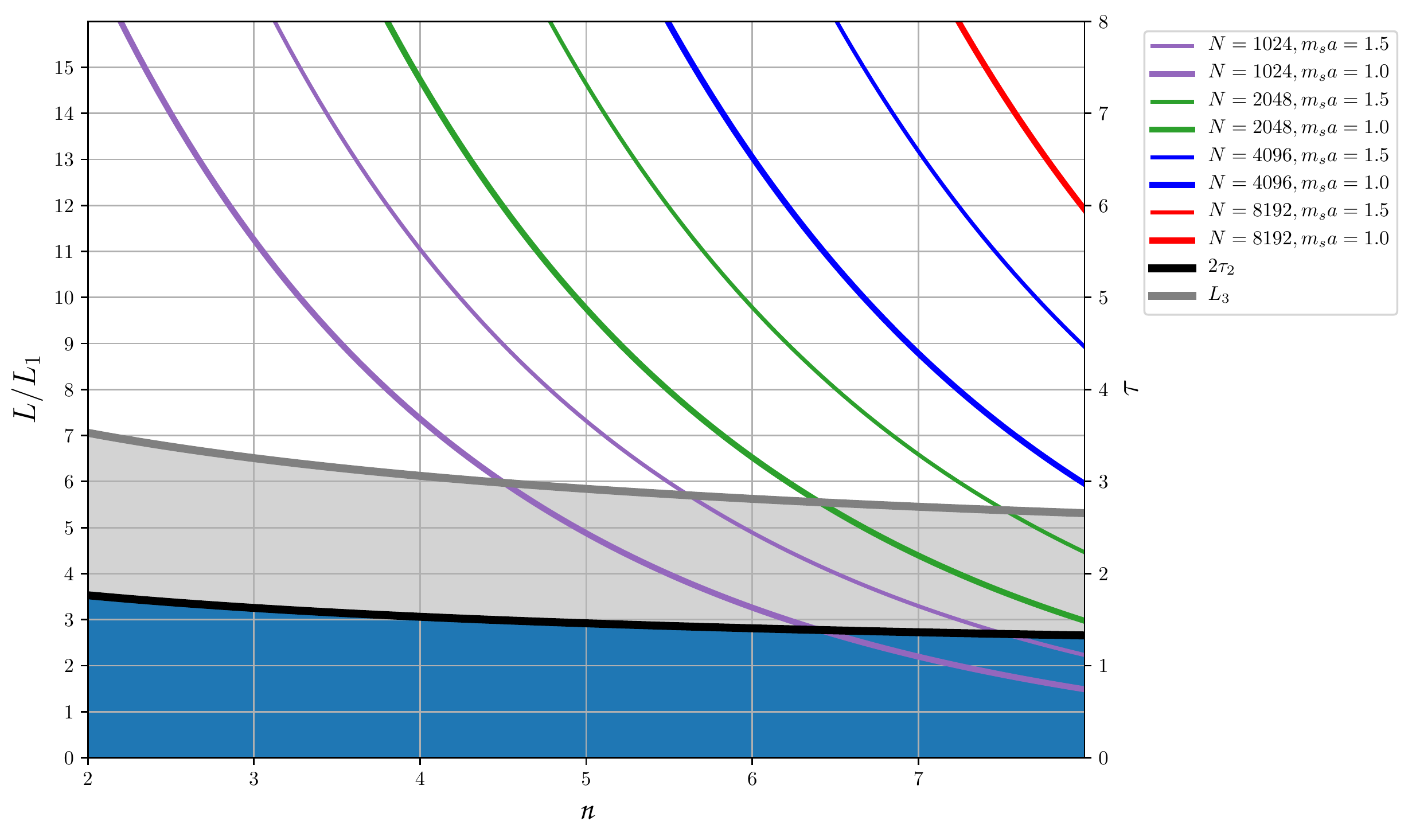}
\caption{Minimum and maximum required lengths $L$ for a given number of points $N$ per linear dimension and string core resolution $\ms a$ to ensure that no volume effects and no unphysical destruction of the
domain walls occur. Coloured lines show the upper limit on the length of the simulation box to avoid unphysical destruction of domain walls at $\ct=2\ct_2$ computed from~\eqref{spacing}. We show this upper
limit as a function of the index of the axion mass growth $\n$. The blue/gray regions mark the box sizes where simulations will suffer from volume effects at $\ct_2$
and $2\ct_2$ respectively. Simulations should have a length above the gray region and below the pertinent coloured line.}
\label{simugraph}
\end{center}
\end{figure}

Moreover, we have argued that the string network scaling has on average one string per horizon, which is $\sim \ct$ in conformal space-time. At $\ct \sim L/2$ the network will start to depart from this scaling \cite{Fleury:2015aca} and there is the danger that strings annihilate with their periodic copies instead of different anti-strings if the box size is smaller than $\ct_2$. In Fig.~\ref{simugraph} we show the
minimal lengths to \emph{ensure} no volume effects at $\ct_2$ (black) and $2\ct_2$ (grey) when the string-wall network starts and finishes its final hecatomb.

Coming back to free-streaming, neither axions nor saxion waves travel at speed $v=1$ because of their non-zero mass. Axions with comoving momentum $k$ would free-stream a comoving length
\be
\label{freesteam}
\lambda_{\rm free-stream} = \int\frac{dt}{R}v
= L_1 \int^\ct_0 d\ct \(\frac{1}{{1 + \frac{\ct^{\n+2}}{(k L_1)^2}}}\)^{1/2}
= L_1\times \left\{ \begin{array}{lr}  \ct & (\ct\ll 1)  \\ 
\sim 	\frac{1.55+\n}{\n} \(k L_1\)^\frac{2}{\n+2} & (\ct\gg 1)
\end{array}\,.
\right.
\ee
For $\n\sim 7.0$, and the highest momenta we can simulate $kL_1\sim \pi N/L$ we obtain $\lambda_{\rm free-stream}\sim 7 L_1$. However, as we will see below, the relevant momenta are at least a factor of 10
smaller ($k L_1\lesssim 100$), for which we get $\lambda_{\rm free-stream}\sim 3.3$. Imposing $L>2\lambda_{\rm free-stream}$ the criterion is typically only a bit more stringent than the gray line in Fig.
\ref{simugraph}.
  
\subsubsection{Transition to axion-only simulations}\label{transition}

In our simulations we switch variables from the PQ complex field to the conformal axion field, $\ctheta$, when the length of strings has been zero for a few time steps,  
\bea
\label{totheta}
\ctheta &=&   \ct \theta =  \ct \arg(\cf)\,, \\ \nonumber
\partial_\ct\ctheta  &=& \ct \partial_\ct\theta  + \theta = 
\ct{\rm Im}\left\{\frac{\partial_\ct \cf}{\cf}\right\} + \theta\,.
\eea
Indeed, this corresponds to the limit $\ms\to \infty$ and makes the simulation as physical as it gets, preventing unphysical roll-over-the-top events and uncontrolled distortions of the potential for the final
part of the simulation. As a bonus, this reduces the memory requirements and number of calculations per time-step, speeding up the computations. 

The naive translation \eqref{totheta} entails a problem related to the multivariate nature of an angle. Although the topology of the axion field should be trivial when there are no strings in the simulation,
there is the possibility to find \emph{spherical domain walls} at the cores of axitons, discussed below and in more detail in Sec.~\ref{sectionaxitons}. We have implemented a simple routine that adds factors of
$2\pi$ whenever a jump from $-\pi$ to $\pi$ is found between two points, repeating iteratively until the number of ``mends'' is zero. Spherical DWs tend to appear prominently in small simulations with
unphysical small volumes, but typically no mending is required in our large simulations with $L>6 L_1$.

A related type of spherical domain walls are the so-called axion nuggets~\cite{Zhitnitsky:2018mav,Liang:2016tqc,Ge:2017idw}, which can be rendered stable if a large baryon number is stored inside in the
colour-superconducting phase. Most intriguing is the claim that they constitute viable dark matter candidates and store the missing baryons to explain the baryon asymmetry of the Universe and the similar amount
of baryon and dark matter. It would be very interesting to study the influence of the quark plasma on the dynamics of axitons in numerical simulations along these directions. We leave it for future work.

\subsubsection{Resolving axitons}

Axitons are quasi stable oscillons of the Sine-Gordon equation with a quickly increasing mass. They form due to the non-linearities of the axion evolution equation, which includes negative (attractive)
interactions~\cite{Kolb:1993zz,Kolb:1994xc,Kolb:1994fi,Kolb:1993hw,Kolb:1995bu}. We describe them in Sec.~\ref{sectionaxitons} but we can already advance a challenge for the simulations. Their physical radius
is approximately given by the Compton-wavelength of the axion $1/m_A$. In order to resolve them we need $m_A a \ll \pi $, which is exactly the opposite requirement than~\eqref{timeNyqvist}. In other words, we
either resolve the axitons, or evolve the higher momentum modes until they are non-relativistic, but not both.

A witty way out, used already in~\cite{Kolb:1993hw}, is to switch off the growth of the axion mass at some intermediate time where axitons are still well resolved. It can be thought as advancing the saturation of the axion mass from the physically motivated $\ctc$ derived in Eq.~\eqref{ctc} to a suitable moment. Axitons dissipate shortly after the axion mass saturates and it is possible to advance the axion field until all modes are non-relativistic, satisfying Eq.~\eqref{timeNyqvist}.

In this paper we opt for a different strategy. We will argue that the most relevant effects of axitons, at least at intermediate scales, happen at early times and we want to study them as physically as
possible. Once these effects are captured, we will stop our simulations before axiton dynamics can suffer too grave resolution issues, i.e. before the fastest modes are non-relativistic. For $\n=7$, we find
$\ct_{\rm Ny}=5.4$ and $6.4$ for our benchmark simulations $L=6,N=4096$ and $L=6,N=8192$, respectively. Using $k_{\rm Ny}/4$ instead of $k_{\rm Ny}$ the relevant times are $\sim 4.0$ and $4.7$ respectively,
which are our typical ending times. After the simulations end, we evolve the fields analytically with a linearised equation of motion, which does not have self-interactions and thus no pseudo-breathers. This
effectively allows late time axitons to diffuse away, but it preserves the density fluctuations created by the first and more relevant axitons. Typically we apply the linearised evolution until $\ct\sim 6$
when all modes are non-relativistic, although the role of the highest momentum axions is always very small. We further justify this procedure in Sec.~\ref{sectionaxitons}.

\subsubsection{Discretising the Laplacian}
The later point is strongly related with the discretisation of the Laplacian in the equation of motion \eqref{eom2}. Speed and simplicity of the numerical code suggests the use of the nearest neighbour formula 
\be
\ddot{\cf}(\mathbf{x}) = \sum_{j=\pm 1}^{\pm 3} \frac{\cf(\mathbf{x}+a\mathbf{\hat{j}}) - \cf(\mathbf{x})}{4a^2} + m_A^2\,\z^3 + \lambda\cf(\mathbf{x})\left(\left|\cf(\mathbf{x})\right|^2 - z^2\right),
\ee
which is a direct discretisation of Eq.~\eqref{eom2}. Here $\mathbf{x}$ represents the spatial coordinates within our discrete lattice and $\mathbf{\hat{j}}$ is a unit vector in the direction $j$. The nearest
neighbour derivative introduces $\mathcal{O}(a)$ discretisation errors whose size we estimate by employing a spectral propagator for smaller volumes up to $2048^3$,
\be
\ddot{\cf}(\mathbf{x}) = \sum_p p^2 e^{-i\mathbf{p}\mathbf{x}}\sum_{\mathbf{x}'} \cf(x') e^{i\mathbf{p}\mathbf{x}'} + m_A^2\,\z^3 + \lambda\cf(\mathbf{x})\left(\left|\cf(\mathbf{x})\right|^2 - z^2\right),
\ee
where the discrete derivatives have been substituted by the Fourier transforms, $\mathbf{p}$ is the corresponding momentum vector and $p$ its modulus. The all to all derivative in the spectral propagator
removes most of the discretisation errors and reproduces exactly the expected dispersion relation of the scalar field $\cf$, but it still feels the finite lattice spacing through the ultraviolet cutoff given
by $2\pi/a$. High energy axions can feel the discrete nature of our simulation. Hence, even with the spectral propagator a continuum extrapolation ($a\rightarrow 0$) is necessary to recover the right physical
behaviour of the fields.

\subsubsection{Time-stepping}
Time stepping is usually the least of the problems. We have implemented different symplectic algorithms including KDK and DKD leapfrogs, the generalised time-reversible Omelyan of 2nd, 4th and 6th order
\cite{Omelyan2002} and the 4th order McLachlan-Atela (McA) optimised method for quadratic kinetic energy~\cite{McLachlan1992}. Usually we identify our fastest mode frequency $w_{\rm max}=\sqrt{\ms^2+k_*^2}$
and measure $d\ct$ as a fraction of its period. Here $k^2_*=3 (\pi N/L)^2$ for the spectral propagator and $k^2_*=12/a^2$ for the nearest neighbour formula. Convergence for the McA integrator is very good for
$w_{\rm max}\ct<2$, and we use values $w_{\rm max}\ct\lesssim 1.5$. To compare with the time stepping in other works, note that this can be written as
$d\ct = 1.5/w_{\rm max}\sim 1.5 a /\sqrt{1.5^2+12}\sim 0.4a$, much below the Courant condition for McA (which is a four-step integrator).

\subsubsection{Initial conditions} 

We have experimented with different ways of setting initial conditions. The preconception of the patchy axion field that takes random values at causally disconnected regions inspires to start with a random
value of $\theta$ at each position of the grid and then smooth the distribution with some iterative method until the desired correlation length is achieved, as done in ref~\cite{Fleury:2015aca}. However, for
very large grids it is advantageous to directly build the PQ complex field as a sum of modes as in~\cite{Gorghetto:2018myk},
\be
\Phi(\veca x) = \sum_{\veca k} \widetilde \Phi(\veca k) e^{i \veca p\cdot \veca x},
\ee
where $|\widetilde \Phi(\veca k)|$ are picked from an exponential distribution
\be
|\widetilde \Phi(\veca k)|^2 \sim \exp(-|\widetilde \Phi(\veca k)|^2/b),
\ee
and the average value is exponentially suppressed above a certain critical momentum $k_{\rm cr}$
\be
b=\exp(-k^2/k_{\rm cr}^2)\,.
\ee
The phase of the Fourier modes is chosen randomly. Once the modes $\tilde \Phi(\veca k)$ are generated, we perform a FFT to build $\Phi(\veca x)$ and normalise $\Phi(\veca x)=\ct_0$ except around the strings.
In the string cores $\Phi/\ct_0\to 0$ as the gradient energy density of the axion increases, (c.f. Eq.~\eqref{tension}), so we compute and give the saxion field the value that would correspond to the straight string
solution. This decreases saxion breather modes around the string, but it is only efficient when the typical string curvature is small.

To relax the saxion field as much as possible we evolve the system through an initial phase with an extra damping in the saxion direction. Although this helps a bit in reducing breather modes, it seems to have
very little influence on our final results. 

The initial conditions are expected to produce an axion field smooth at comoving length scales $L_{\rm cr}\sim 1/k_{\rm cr}$. Defining $k_1\sim\frac{1}{L_1}$ the typical momentum at $\tau_1$, we have verified
that choosing 
\be
k_{\rm cr} =c_{oe} \frac{L}{L_1}\frac{k_1}{\ct} \,,
\ee
and $c_{oe}\sim 1$, the initial string density parameter starts very close to its scaling value. 

The initial time is constrained by the maximum string tension we can give to our strings. If the axion field is random in disconnected patches, it stores an energy density of order
$f_A^2/\ell_H^2\sim f_A^2H_1^2/\ct^4$, but the PQ potential can only be simulated with a central height $V_{\rm PQ}(0)=\lambda_\phi f_A^4/8 = \lambda f_A^2 H_1^2$. If we want to keep the saxion field in the
$\m=f_A$ minimum between strings, the potential energy has to be much larger than the axion gradients, which for our parameters requires
\be
\ct_0 \ll  \ct_{\rm PQ} \sim \lambda^{-1/4}.
\ee
The conformal time $ \ct_{\rm PQ}$ would correspond conceptually to the time when the PQ symmetry would be spontaneously broken by the fields fluctuations. The simulations of
\cite{Hiramatsu:2010yu,Hiramatsu:2012gg} started before $\ct_{\rm PQ}$ and go through the phase transition smoothly (introducing an extra term in the potential) to ensure a physical string network. The practical
problem is that before $\ct_{\rm PQ}$, the string energy does not dominate the dynamics. Collapsing a string loop requires a time comparable to their length $t\sim \ell$ and can only start after the loop enters
the horizon, but at $\ct<\ct_{\rm PQ}$ loops can be ``destroyed" immediately by the roll-over-the-top dynamics once they enter the horizon thanks to the large gradients. Thus, when the the symmetry is finally
spontaneously broken, it is expected to be under-dense with respect to the scaling solution by a factor of $\sim 2$. This very naive argument seems to support the results of~\cite{Hiramatsu:2012gg}, that give 
$\xi\sim 0.5$ with respect to recent simulations that attempt to build initial conditions in the scaling regime~\cite{Fleury:2015aca,Gorghetto:2018myk}.

Another reason to use $\prs$ strings is that $\ct_{\rm PQ} \sim 1/\sqrt{\lambda_{\prs}}= H_1/\ms$, only dependent on the the square root of $\lambda$, and therefore a much earlier start of the simulation is
allowed.

\section{Simulation results and axion spectrum}\label{spectrum}

The main concern of this publication is to study the density contrast which is created in the axion field from misalignment, the decay of the strings and domain walls, and from axitons. However, we commence with a discussion of the string density and the axion spectrum created in the decay of strings. These results are intended to aid the understanding of some features of the axion distribution presented in the subsequent sections. They also allow us to crosscheck our simulations with previous findings. In particular the spectrum of axions emitted by strings has sparked many discussions in the past. Indeed, we reproduce recent claims \cite{Fleury:2015aca,Gorghetto:2018myk} by which the axion number density is dominated by IR and the axion energy density is dominated by the UV. Naively extrapolated to physical string tensions this result conflicts with dark radiation limits from BBN and CMB. Still, we do not think our simulation's results are in conflict with observations. This is due to a cut-off, expected at that scale which corresponds to the horizon size at the PQ transition,  but unobservable in our simulations due to our choice for setting the initial conditions. Thus we assume that our simulations represent well the axion spectrum at our scales of interest and that this spectrum can be extrapolated until a harmless cut-off.

\subsection{Evolution stages in the density contrast}

\begin{figure}[tbp]
\begin{center}
\includegraphics[width=0.45\textwidth]{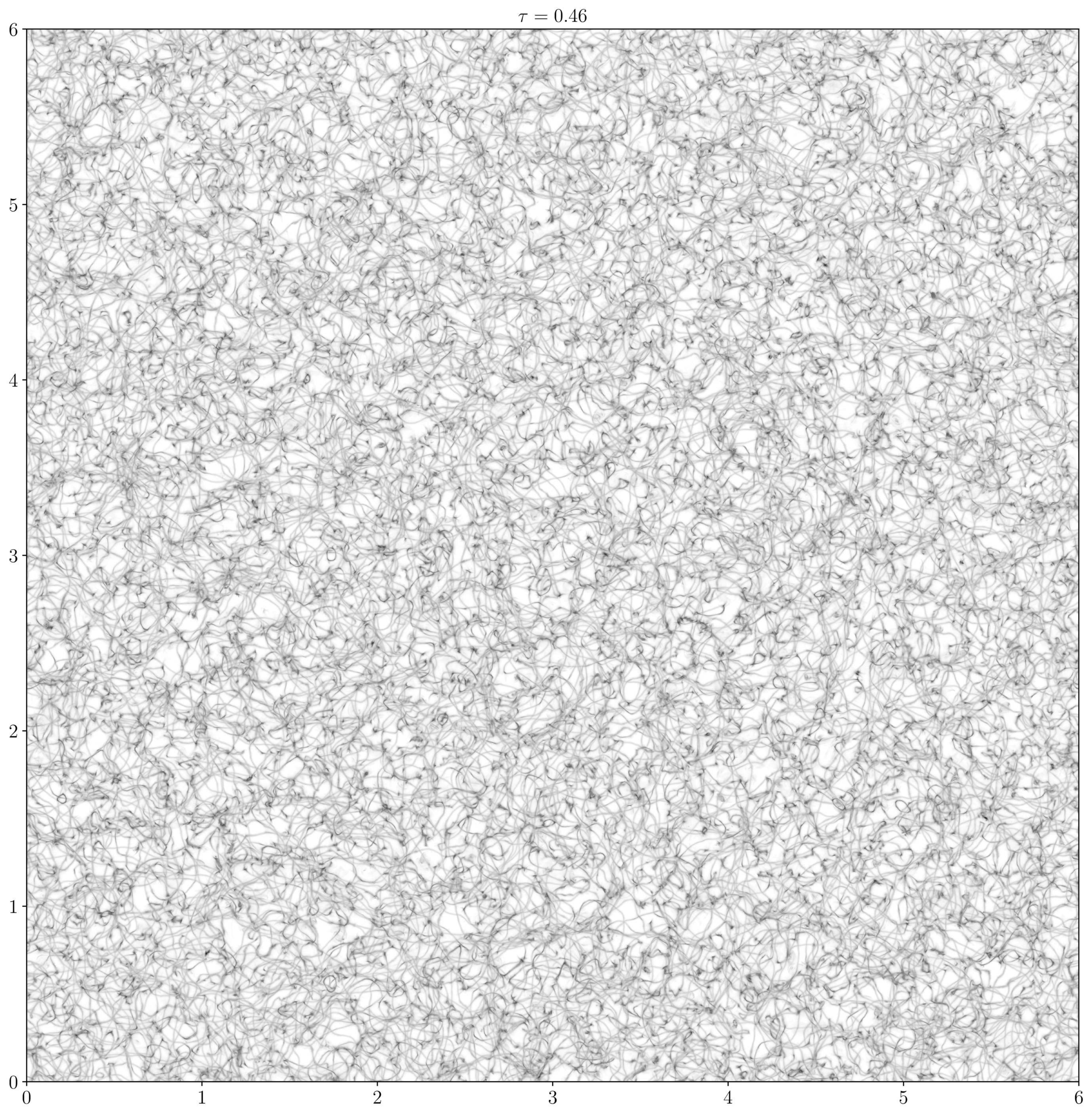}\includegraphics[width=0.45\textwidth]{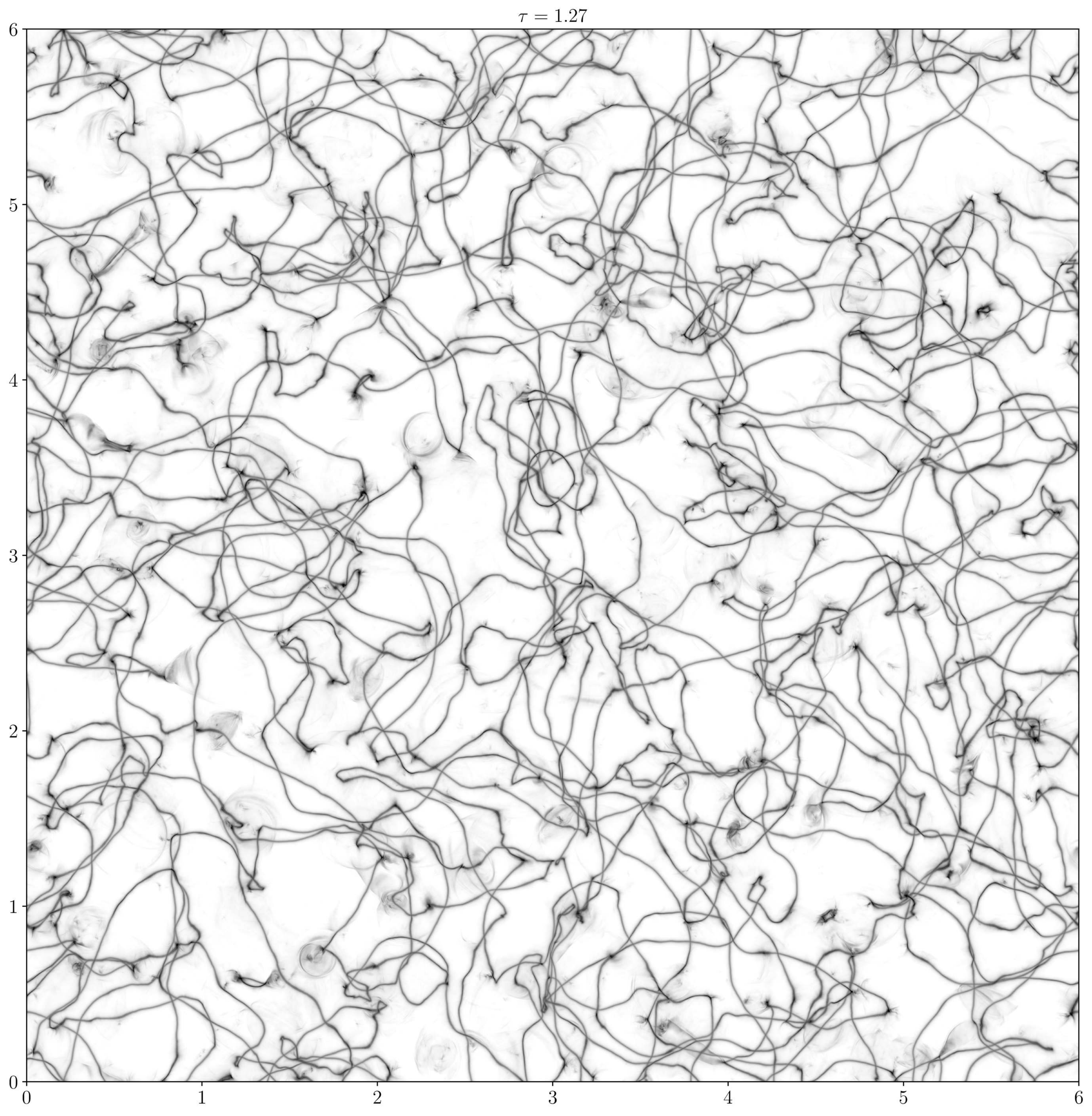}
\includegraphics[width=0.45\textwidth]{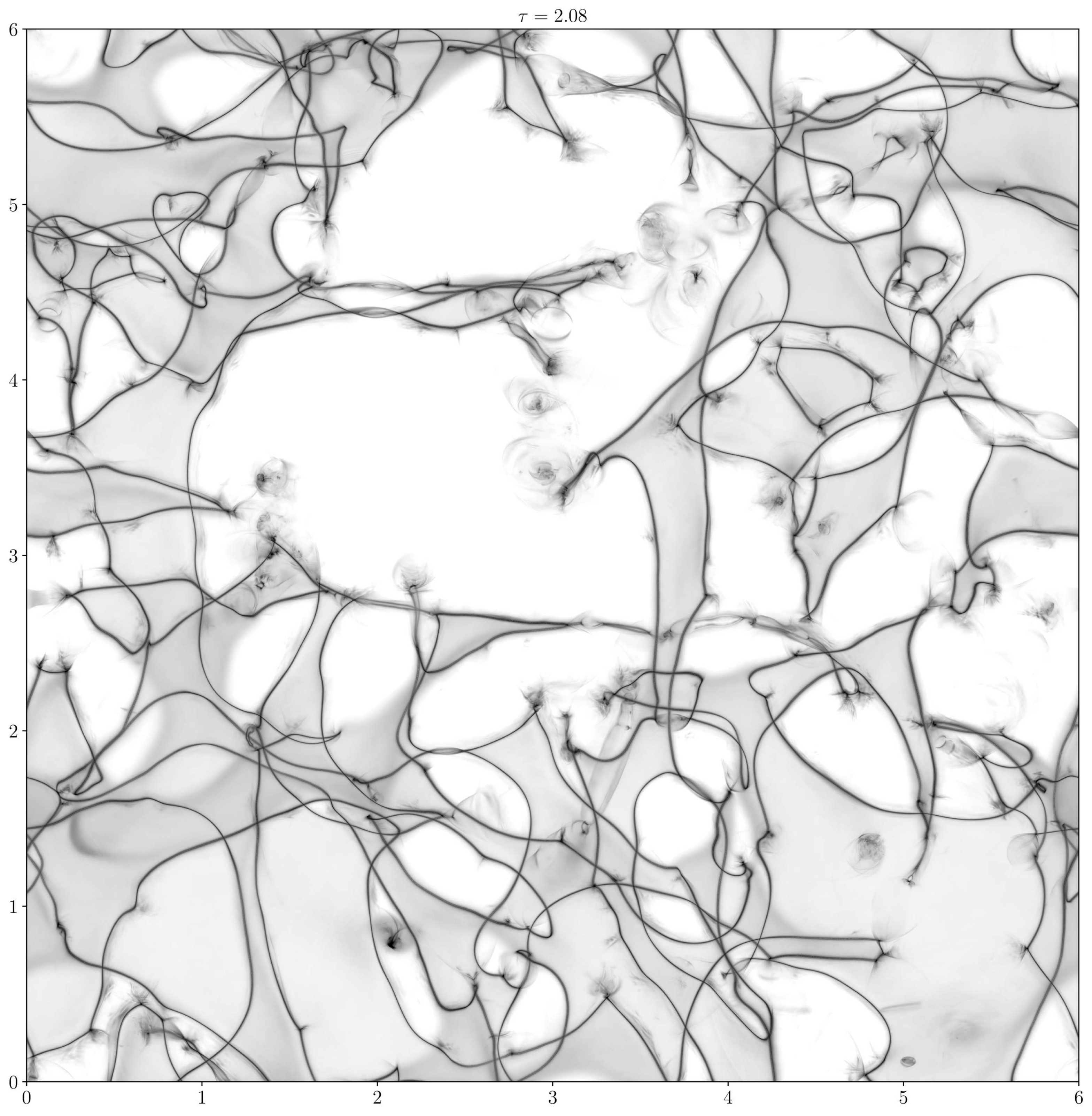}\includegraphics[width=0.45\textwidth]{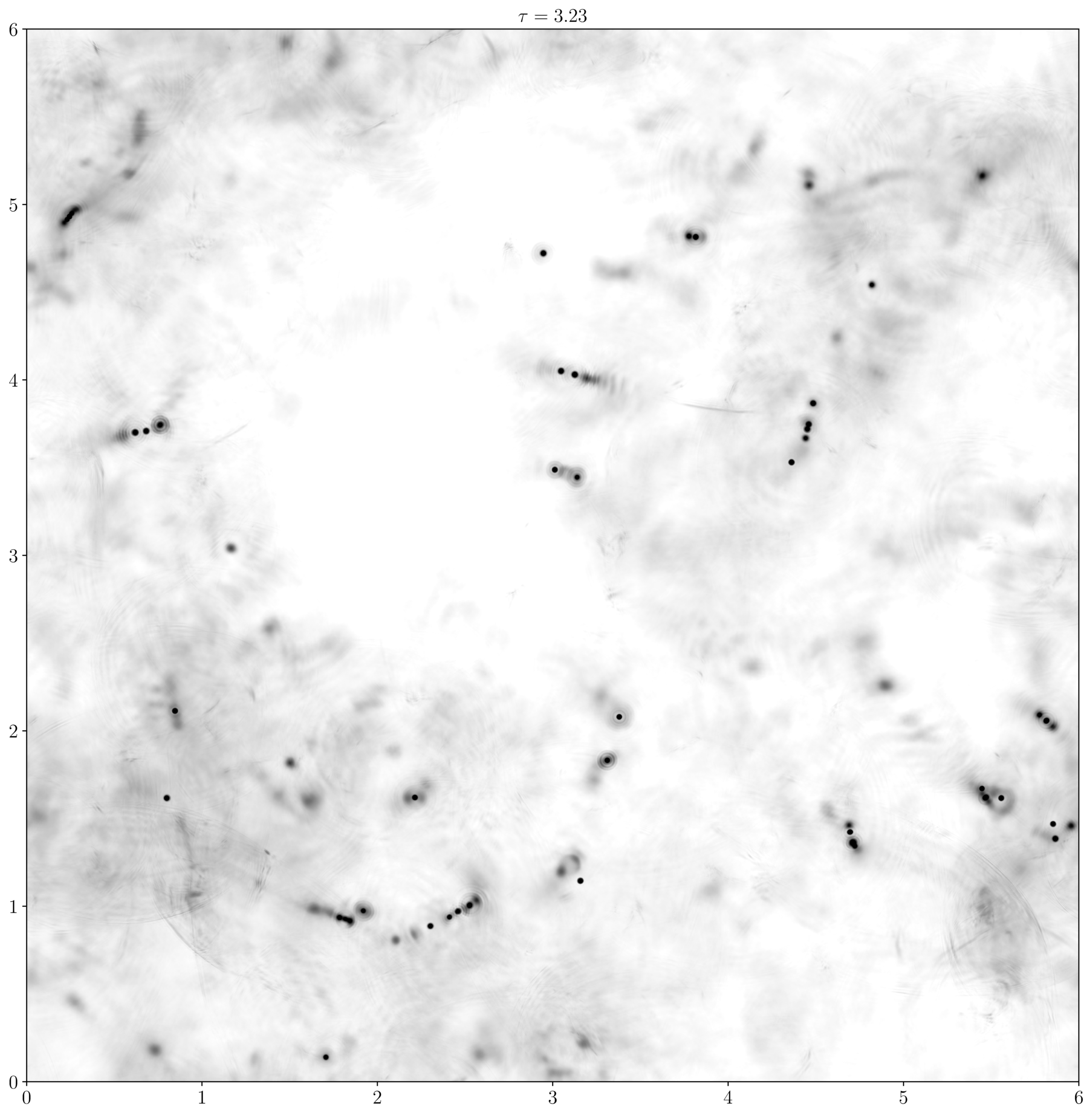}
\caption{3D$\to$2D projection plots of the axion density squared $\int dz (\varrho(\veca x)/\bar\varrho)^2$ for several values of $\ct$. The densest structures distinctly appear in the plots for the 4 stages of
the evolution of axion dark matter simulations: string-network scaling (up-left to up-right), domain walls attached to strings pulling the strings into destruction (down-left) and frozen dark matter field with
axitons (down-right). The simulation parameters are $L=6L_1, \ms a=1.0,\n=7$ and $N=4096$.}
\label{pics}
\end{center}
\end{figure}

The sequence of events in our axion dark matter simulations is illustrated in Fig.~\ref{pics}. The four images depict a 3D$\to$ 2D projection plot of the axion density squared $\varrho^2_A$ summed along the
dimension perpendicular to the image. Only the most dense objects are revealed in this images. The upper left and right images show the early times of the simulation when cosmic strings are in the scaling
regime. The densest points correspond to the strings, and in particular the cusps where strings were cut and collapsing loops that relax very fast by emitting relativistic axions are visible. Relativistic
axions can be seen as shock waves when zooming in the images. At the maximum zoom one can also see the string cores as under-dense regions in the core of the strings. At time $\ct\sim 2$ (lower left) the domain
walls have developed and are clearly visible between the strings. The walls are already pulling the strings at a fast pace. Typically the cusps cannot relax faster than the strings are pulled, so they are
dragged with the strings. Again we see very dense axion radiation from these cusps.
The lower right plot shows a more relaxed environment where strings and walls have already collapsed. Interference patters can be clearly recognised in several regions. Some waves have the very short wavelength
that we see in previous plots originated in the cusps. However we see other characteristic wavelengths too. The most conspicuous objects at late times are ultra dense spots that can be associated with axitons,
ultra-dense lumps of the axion field where $\theta$ reaches $\sim\pi$. They have appeared in places where the density was relatively high and are surrounded by spherical waves (axions) that emanate from them.
Axitons are so dense that we need to over-saturate them in order to be able to see the surrounding density field. 
Overall, the density squared seems quite correlated with the previous plot showing the domain
walls because the latter form in regions of the largest misalignment $\theta\sim \pi$. 
The smooth component of the final density (\emph{squared}) seems to come from these large $\theta \sim \pi$ regions. But it
is filled with interference patterns from waves that have longer wavelengths than those we observed in previous epochs.

\subsection{String density parameter}

\begin{figure}[tbp]
\begin{center}
\includegraphics[width=\textwidth]{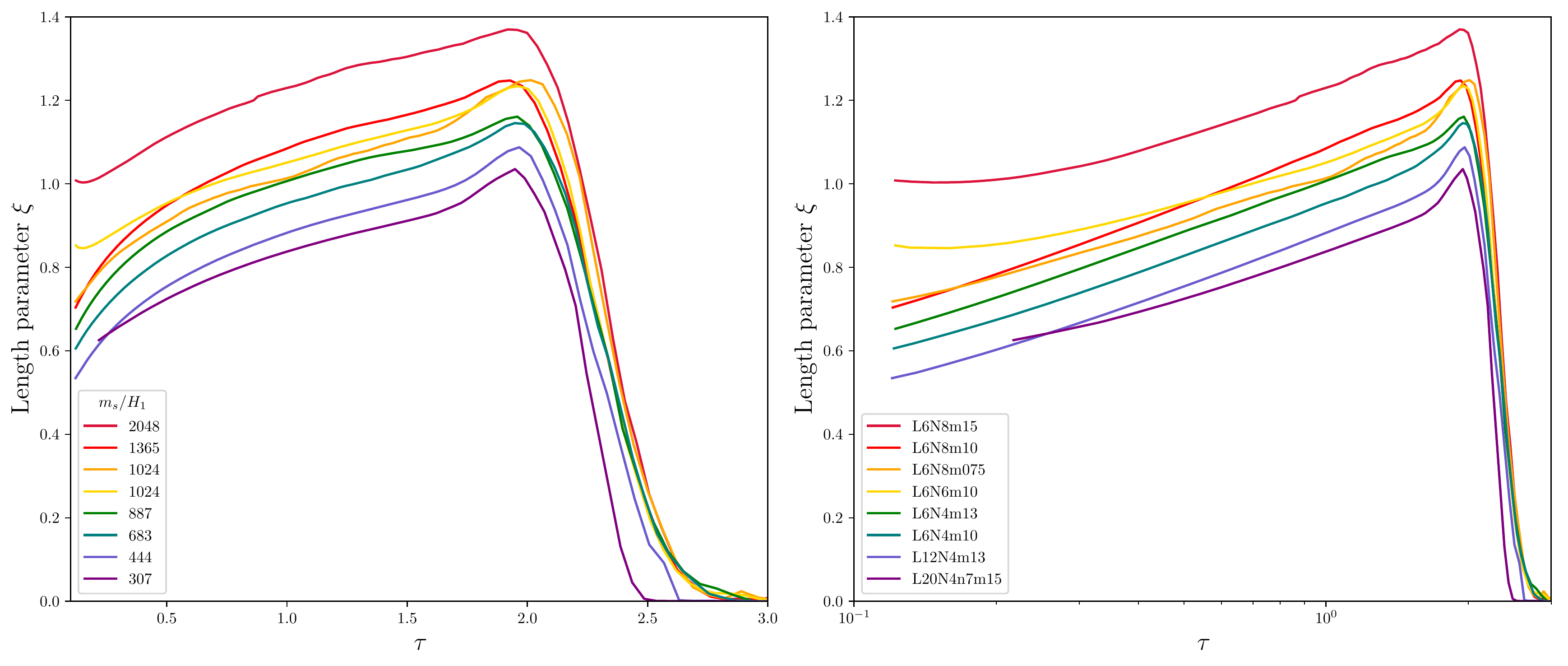}
\caption{String length $\xi$ as a function of conformal time $\ct$ in our simulations for PRS strings with different tensions as represented by the parameter $\sqrt{\lambda_{\textrm{PRS}}} = \ms/H_1$. The
legend in the right plot shows also the length of the box, the number of points of the grid along one dimension and the $\ms a$ parameter. For instance $L6N8m10$ means $L=6L_1, N=8192, \ms a=1.0$. Other values
of $N$ are $6144, 4096$ and $\ms a =0.75,1.0,1.3,1.5$.}
\label{xievol}
\end{center}
\end{figure}

The string density parameter is a nice gauge to understand the scaling regime of our simulations. In Fig.~\ref{xievol} we show the evolution of $\xi$ as a function of conformal time for $\prs$ strings of
different tension, represented by the tension parameter
\be
\sqrt{\lambda_\prs} = \frac{\ms(\ct=1)}{H_1}=\frac{\ms(\ct=1)}{m_A(\ct=1)}.
\ee
The string length in the full simulation is computed as the number of plaquettes in our grid pierced by a string using the method of \cite{Fleury:2015aca}, see also~\cite{Yamaguchi:2002zv,Yamaguchi:2002sh}. Our initial conditions tend to be slightly over-dense,
but they converge fast to a scaling value $\xi\sim 1$  that depends on $\sqrt{\lambda_\prs}$ and grows slightly, as discovered in~\cite{Fleury:2015aca,Gorghetto:2018myk}. The dependence of $\xi$ with
$\log(\ms/H)$ seems linear with a slope of $\alpha\sim 0.22$ in agreement with the recent results of~\cite{Gorghetto:2018myk}.

The scaling regime comes to a halt between $\ct=2$ and $3$ when axion potential is sufficiently high that the domain walls are able to pull the strings binding them together, leading to a violent collapse. The
destruction of the network is slightly delayed for the strongest string tensions, because the domain walls need more tension to drag the string's higher inertia. The sudden increase in string length around
$\ct\sim 2$, before the collapse, is due to the fact that some loops, which were collapsing around a region where $\theta\sim 0$, discover a domain wall that is pushing them in the extending direction to
annihilate with nearby strings. Therefore, some loops actually grow in length before annihilating.
 
\subsection{Axion spectrum}\label{section:axionspectrum}

\begin{figure}[tbp]
\begin{center}
\includegraphics[width=\textwidth]{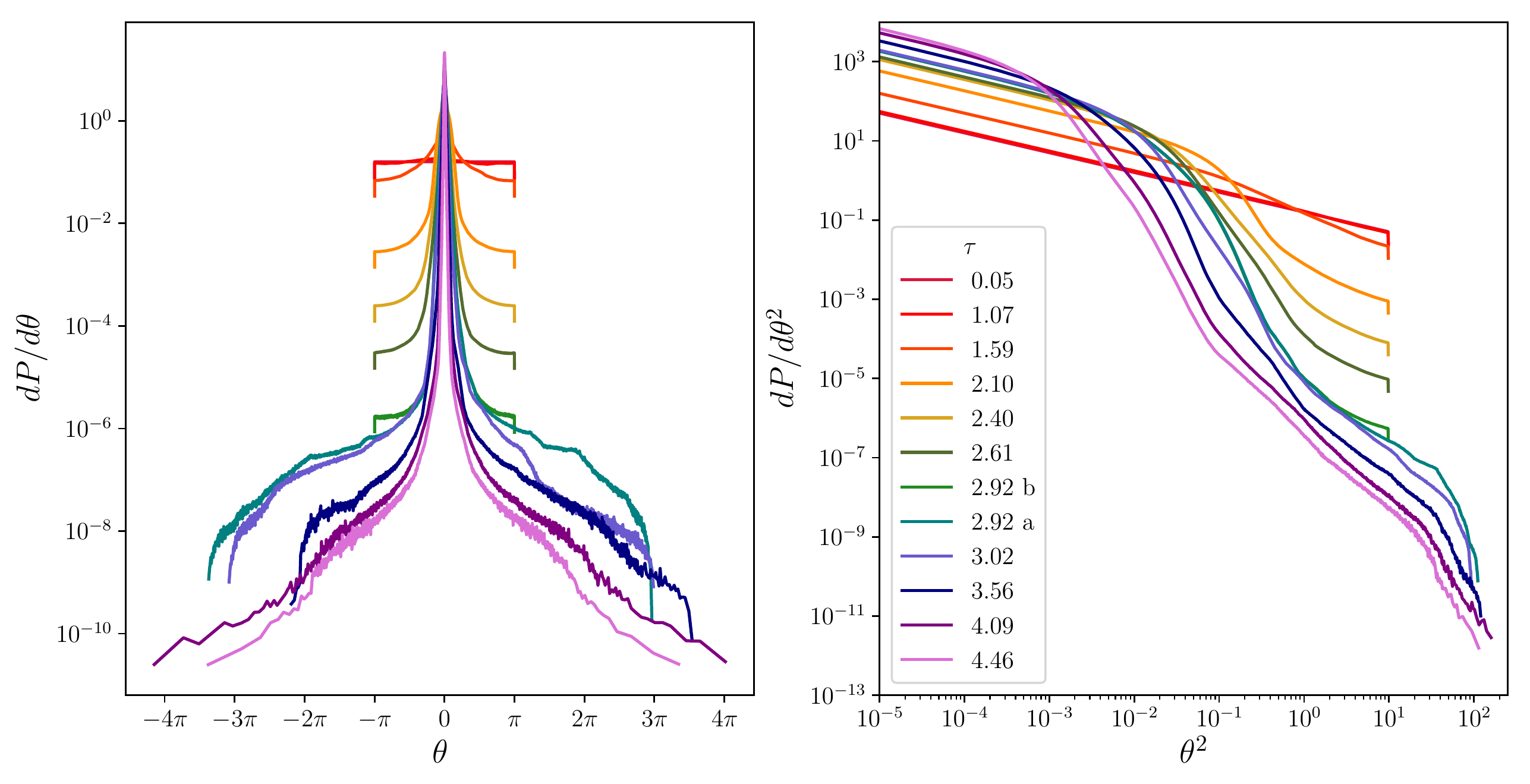}
\caption{Evolution of the distribution of $\theta$ and $\theta^2$ as a function of the conformal time $\ct$ in one of our biggest simulations with $L=6L_1 ,N=8192,\n=7.0$.}
\label{thetaevol}
\end{center}
\end{figure}

After the destruction of the string network, we switch to axion-only simulations evolving the axion field with Eq.~\eqref{eom11} and neglecting $R_{\ct\ct}$,
\be
\label{eomctheta}
\partial_{\ct\ct}\ctheta -\triangle \ctheta + \ct^{\n+3}\sin(\ctheta /\ct)= 0.
\ee
As the Universe expands, the amplitude of the axion field oscillations decreases, and on average $\theta$ becomes much smaller than one. To illustrate the effect we show in Fig.~\ref{thetaevol} the evolution of
the distribution of values of $\theta$ in one of our largest simulations. One can clearly see a flat distribution $\theta \in (-\pi,\pi)$ at early times $\ct\lesssim 1$ that starts to peak at $\theta\sim 0$
afterwards. The peak continues to sharpen, but there are always points of our simulation with $\theta\sim \pi$ due to the presence of domain walls and strings. By $\ct\sim 2.9$ the string-wall network has
disappeared and we switch to $\theta$-only simulations. Our mend-$\theta$ processes builds a continuum field that in this case already contains some values $\theta>\pi$. As time evolves, the central peak
increases and sharpens but there are always a small number of points with $\theta\gtrsim \pi$, even reaching $4\pi$, which we interpret as the values of the axion field in and around the cores of axitons. The
distribution  around $\theta\sim 0$ is more clearly visible in the right plot, where we show the distribution in log-scale as a function of $\theta^2$. At small $\theta$ $dP/d\theta^2\sim 1/\theta$, which
corresponds to a flat distribution in $\theta$. It suffers a strong cut-off at a value of $\theta^2$ that decreases with time and determines the average value $\langle \theta^2 \rangle$, which is
$1.7/(m_A\ct^3)$. Further, one can see a tail $dP/d\theta^2 \sim 1/\theta^4$, which we attribute to the axitons. That tail does not contribute significantly to $\langle \theta^2 \rangle$, but will dominate
$\langle \theta^4 \rangle$, see below.

When $\theta$ is small we can approximate the sine in Eq.~\eqref{eomctheta} as its argument, in which case the equation of motion is linear in $\ctheta$ and can be solved by Fourier modes through the ansatz\footnote{
Translating between discrete and continuum with  $\int d^3{\vec k}/(2\pi)^3\leftrightarrow \sum_{\vec k}/V$. The units of $k^3$ are the ones of $V$. Note that $\cctheta_{\vec k}$ has usually units of comoving
volume, but here we have included a factor $L_1^3$ to make it dimensionless and of natural size.},
\be
\label{modes}
\ctheta(\veca x) = \frac{L_1^3}{V}\sum_{\veca k} \widetilde\ctheta(\veca k) e^{i \veca k\cdot \veca x}\,,
\ee
where $V$ the comoving volume of our simulation. The mode's amplitude $\ctheta(\veca k)$ satisfies the equation of motion
\be
\partial_{\ct\ct}\widetilde\ctheta(\veca k) - w_{k}^2 \widetilde\ctheta(\veca k) = 0\,,
\ee
with a c-angular frequency\footnote{We reserve $\omega$ for frequencies related to the usual time $t$ and use $w$ for those related to conformal time, which differ by an extra factor of $R$,
$\omega dt\leftrightarrow w d\eta$.}
\be
\label{disp}
w_k^2(\ct) = k^2L_1^2 + \ct^{\n+2}.
\ee
Recall that the modes are quantised in a discrete periodic grid as
\be
\veca k = \frac{2\pi}{L} \veca n\,,
\ee
where $\veca n$ is a vector of integers $m = -N/2-1,...,0,...,N/2$ when the number of points $N$ along each dimension is even.

The way in which Eq.~\eqref{eomctheta} remembers the expanding Universe and the ensuing damping is through the increase of the axion mass term in the dispersion relation~\eqref{disp}.
In the adiabatic limit, i.e. when $w_k \gg \ct\partial_\ct w_k$, the WKB solution for one mode can be written as
\be
\label{solumode}
\widetilde \ctheta(\veca k)(\ct) = \sqrt{\frac{w_k(\ct_0)}{w_k(\ct)}}\(c_+ e^{iW_k(\ct)}+c_- e^{-iW_k(\ct)}\)\,, \\
\ee
with $c_\pm$ complex-conjugated coefficients that depend on the initial conditions. The phase is
\be
\label{Wphase}
W_k(\ct) = \int^\ct_{\ct_0} w_k(\ct')d\ct' \to \partial_\ct W_k = w_k.
\ee
Note that the amplitude of each Fourier mode decreases as $w_k$ increases with time.

By virtue of solution~\eqref{solumode}, the quantity
\bea
\frac{1}{2w_k}|\partial_\ct\widetilde\ctheta(\veca k)|^2+\frac{1}{2}w_k|\widetilde\ctheta(\veca k)|^2\,,
\eea
is an adiabatic invariant, i.e. it is conserved and can be interpreted as the number of axions with comoving momentum $\veca{k}$ in a comoving volume.
The justification is as follows. The quadratic part of the axion Hamiltonian ${\cal H}_0$ (energy density) is
\be
{\cal H}_0 =\frac{(f_AH_1)^2}{\tau^4}   \(\frac{1}{2}(\partial_\ct\ctheta)^2+\frac{1}{2}(\nabla \cctheta)^2 + \frac{\ct^{\n+2}}{2}\psi^2\)\,,
\ee
and the energy $E$ in a comoving box of volume $V$ can be expanded in the axion modes of Eq.~\eqref{modes} as
\be
E = \int_V d^3x~ R^3 {\cal H}_0  = R^3 \frac{(f_A H_1)^2}{\tau^4} L_1^6
\int \frac{d^3 \veca k}{(2\pi)^3}\(\frac{1}{2}|\partial_\ct\widetilde\ctheta(\veca k)|^2+\frac{1}{2}w_k^2|\widetilde\ctheta(\veca k)|^2\).
\ee
Taking into account that the energy of an axion ``quantum" is
\be
\omega_{k} = \sqrt{\frac{k^2}{R^2} + m_A^2} = \frac{R_1H_1}{\ct} w_{k} \,,
\ee
we can define the \emph{occupation number}, i.e. the number of axions per phase-space cell as
\bea
\label{nk0}
n(\veca k) &=& \frac{1}{\omega_{k}}\frac{dE}{V d^3 \veca k/(2\pi)^3} =
\frac{f_A^2}{H_1^2}\frac{L_1^3}{V} \(\frac{1}{2 w_k}|\partial_\ct\widetilde\ctheta(\veca k)|^2+\frac{w_k}{2}|\widetilde\ctheta(\veca k)|^2\),
\eea
which, as we know, is an invariant if we can neglect self-interactions of the axion field. Assuming statistical isotropy, the relevant occupation number is the angle averaged version
\bea
\label{nk}
n(k) &=& \int \frac{d\Omega}{4\pi} n(\veca k) =
\frac{f_A^2}{H_1^2}\frac{L_1^3}{V}
\left\langle\frac{1}{2 w_k}|\partial_\ct\widetilde\ctheta(\veca k)|^2+\frac{w_k}{2}|\widetilde\ctheta(\veca k)|^2\right\rangle_{|\veca k|=k},
\eea
that we will be showing in this paper.

The total number of axions per unit volume is the quantity directly related to the dark matter,
\be
\frac{N_A}{V} =\int \frac{k^2d k}{2\pi^2} n(k) =
\frac{1}{L_1^3}\frac{1}{2\pi^2}\int \frac{dk}{k}(k L_1)^3 n(k).
\ee
Note that the natural size for $n(k)$ is $f_A^2/H_1^2$, which is a huge number. The number of axions per comoving volume have the expected units of $n(k)/L^3_1=H_1 f_A^2R_1^3$.

\begin{figure}[htbp]
\begin{center}
\includegraphics[width=\textwidth]{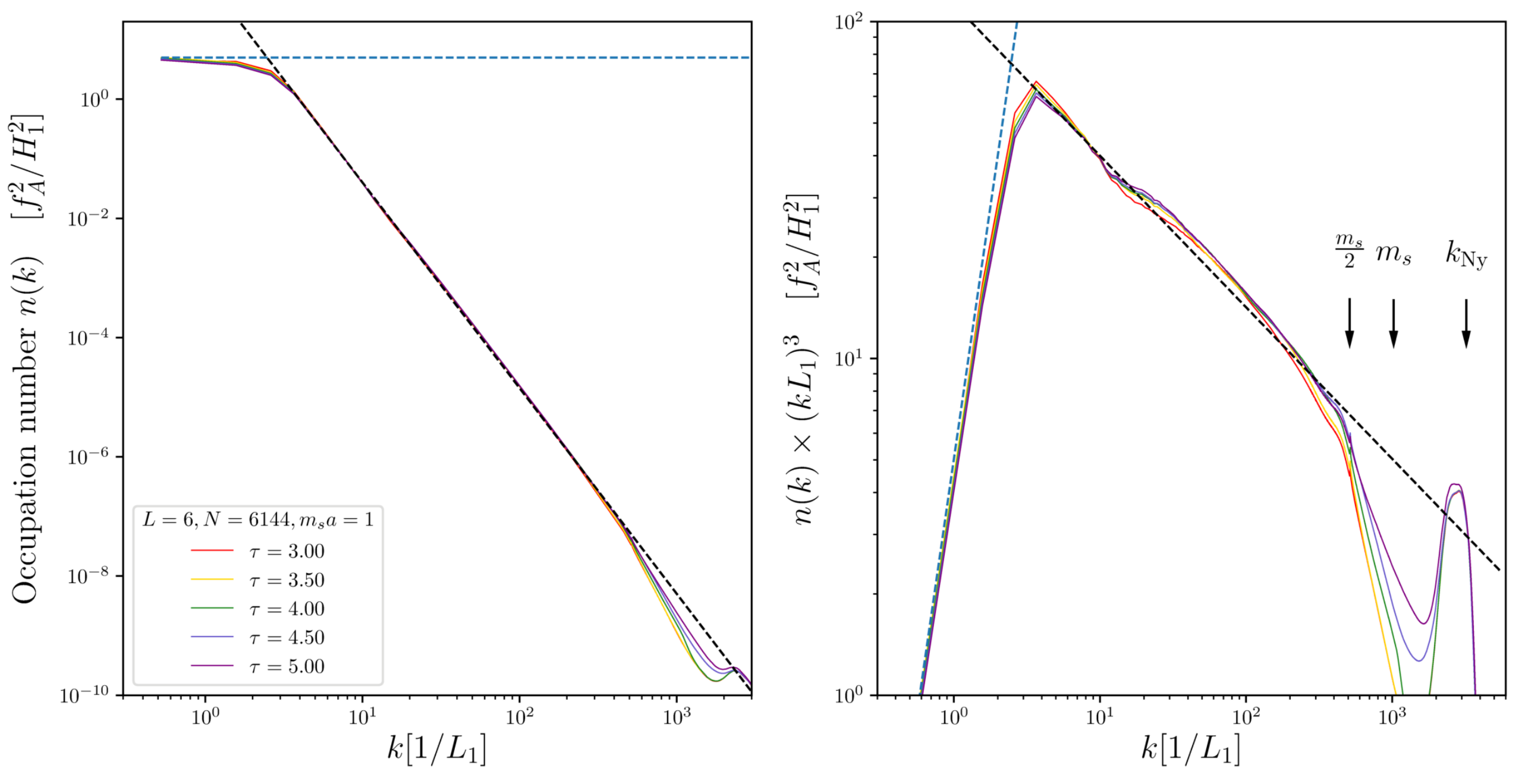}
\caption{On the left, occupation number $n(k)$ at several values of $\ct$ as a function of the comoving momentum. On the right, contribution of each logarithmic interval of comoving momentum to the axion
density $N_A/V$. The arrows indicate important changes of behaviour, namely the UV cut-off given by the string tension ($\ms/2$), and the Nyqvist frequency ($k_{\rm Ny}$). Both plots have been generated after
the string-wall collapse for benchmark simulations in $N=6144$ grids with $\ms=1.0$ and $\n=7$.}
\label{spectrumPlot}
\end{center}
\end{figure}

In Fig.~\ref{spectrumPlot} we show the spectrum of axions obtained from our simulations (left) and the contribution of each logarithmic $k$ interval to $N_a/V$ (right). These we obtained from our benchmark
simulations with $L=6L_1$ and high tension $\ms a = 1.0$ performed in 6144$^3$ boxes. We show averaged spectra for increasing times, $\ct=3,3.5,4,4.5,5$.
The main features of the spectrum are:
\begin{align}
\nonumber
1.-& \text{low-momentum (IR) cut-off} & \quad k_{\rm IR} \simeq (2\sim 3) \, L^{-1}_1, \\ \nonumber
2.-& \text{intermediate power law}    & \quad n(k) \sim 112 \(\frac{1}{k\, L_1}\)^{3+d} \quad, \quad d\sim 0.45 , \\ \nonumber
3.-& \text{UV cut-off at}             & \quad k_{\rm UV} \simeq \ms/2
\end{align}
and a bump around the Nyqvist frequency $k_{\rm Ny} = \pi N/L$. The UV cut-off (3) starts at values slightly below $\ms/2$, where a peak corresponding to saxion decay giving $k=\ms/2$ axions can be guessed.
It continues until $\ms$ where we start suffering the effects of finite resolution at $k_{\rm Ny}$.

We find five remarkable features of this spectrum to comment. The first one is that at $k\sim 3/L_1$ the axion spectrum has a small bump over the power-law before decreasing. These low-momentum axions are the
ones usually associated with the misalignment mechanism/domain walls.

Second, the axion occupation number spectrum is harder than $k^{-3}$. If we were interested in the total number of axions we could extrapolate this result to arbitrary values of $\ms$ safely because the density
of axions is dominated by the low-energy part of the spectrum,
\be
\label{axionnumber}
\frac{N_A}{V} =\int \frac{k^2d k}{2\pi^2} n(k) =
\frac{1}{L_1^3}\frac{1}{2\pi^2}\int \frac{d k}{ k } \frac{k^3}{k_1^3} n(k).
\ee

Third, the above extrapolation, however, reveals a potential UV issue. The reason is related to the findings of~\cite{Gorghetto:2018myk}, claiming that the number density of axions radiated from strings is
dominated by the low-$k$ part of the spectrum but the \emph{energy} is dominated by the UV. Our results clearly corroborate these findings~\cite{Gorghetto:2018myk} against previous claims of
\cite{Hiramatsu:2010yu,Hiramatsu:2012gg}, which find an almost exponentially decreasing spectrum.

The issue concerns the energy density in axions, which is given by integrating $n(k)$ over an extra factor of $\omega_k$. In a comoving volume $V$ we have therefore
\be
\frac{E_A}{V} = \int \frac{k^2d k}{2\pi^2} n(k) \omega_k =
\frac{m_A}{L_1^3}\frac{1}{2\pi}\int \frac{d k}{ k } \frac{k^3}{k_1^3} n(k)
\sqrt{1 + \frac{k^2}{m_A^2(t)R^2}}.
\ee
The hardest axions are those emitted from strings and have physical momentum $k/R=\ms\sim \ms$ when the string network collapses. With $\prs$-strings, one could naively think that the cut-off at high momenta,
$k_h$, should precisely correspond to those axions,  $k_{\rm h}/R_1\sim f_A$, i.e. $k_{\rm h}/k_1\sim f_A/H_1$. If the cut-off is so large that the hardest axions are relativistic, i.e.
$\omega_h\sim k_h/R\gg m_A$, the energy would be dominated by them since $E_A\propto \int k^2 dk k^{-3.45} k/R\sim k^{0.55}/R$. Therefore axions from strings would behave as dark radiation, and what is more
important, the amount of dark radiation would exceed the cold axion dark matter component. Plugging some numbers, one can easily check that this would be the case even at CMB times. 
However, this cannot be the case because the total energy of the string network and radiated axions, $O(H^2 f_A^2)$, is smaller than the background radiation $O(H^2 \mpl^2)$. 
Indeed, we believe that the extrapolation of the spectrum is not justified. In fact, one expects a softening of the spectrum at $k/k_1\sim \sqrt{\ms^2/H_1 \mPl}$ and yet another at $k/k_1 \sim \sqrt{\mPl/H_1}$ before reaching the
maximum $k/k_1\sim \ms/H_1$ (assuming radiation domination without many non-standard degrees of freedom until a Temperature $T\sim \ms$). With $\prs$-strings the two last cut-offs coincide, because $\ms$ is
decreasing in time, and cannot be distinguished. The first cut-off is, however, more relevant and it is related to the change of behaviour at the comoving momentum related to the horizon size at the PQ phase
transition. We have explicitly chosen not to simulate the PQ phase transition, to have a picture as close as possible to the physical one and a maximum separation of scales. However, this choice comes at a
disadvantage.
We think that using $\prs$-strings is still better than physical strings for the increased dynamical range, but the extrapolation cannot be carried in earnest beyond $\sqrt{\ms^2/H_1\mPl}$, where a softening is
expected. Within this much smaller cut-off, axions with a $\sim k^{-3.45}$ spectrum contribute negligibly to dark radiation at BBN times and behave as cold dark matter already at times much earlier than matter
radiation equality. We will extend this discussion in a future publication~\cite{Inpreparation}. For what concerns us in this paper: we assume that the spectrum of axions is well represented by our simulations
around $k\sim k_1$, and that can be extrapolated until a harmless cut-off $\sim k_1 \sqrt{\ms^2/H_1 \mPl}$.

\begin{figure}[tbp]
\begin{center}
\includegraphics[width=\textwidth]{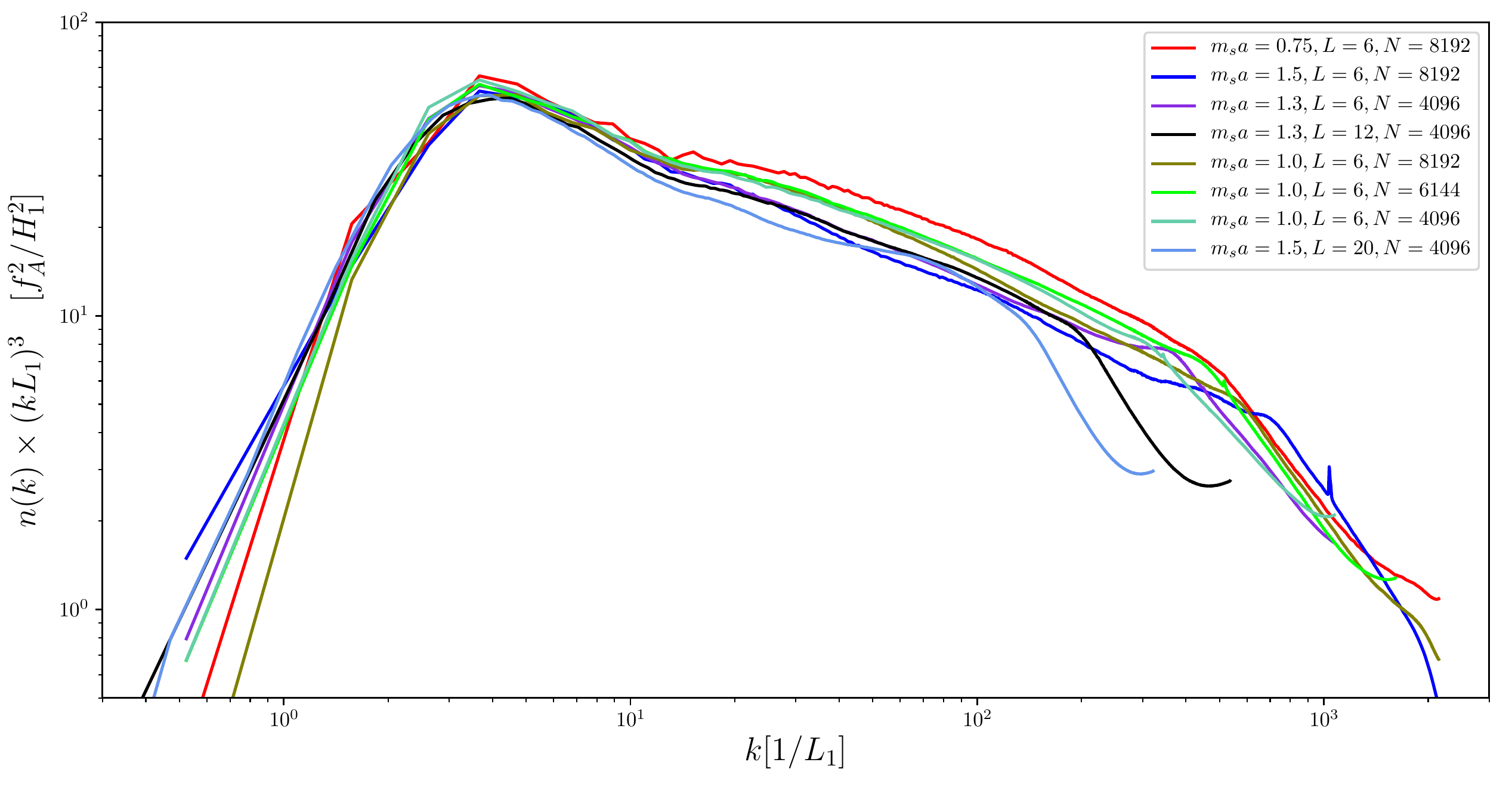}
\caption{Comparison between spectra at $\ct=4.5$ for different string tensions. The large variance at small-$k$ is due to the comparatively few modes, specially for $L=6 L_1$ simulations (those with $L=12,20$
converge much better at low-$k$).}
\label{spectrumComp}
\end{center}
\end{figure}

The fourth remarkable aspect is quantitative. When we compute the total number of axions from Eq.~\eqref{axionnumber}, we find $N_A/V\sim 8 $ in units of $H_1f_A^2R_1^3$, i.e. around half the naive misalignment
contribution, \eqref{misalignment}. This confirms previous findings of reference~\cite{Fleury:2015aca}. A comparison of the spectra obtained with different core resolutions, $\ms a$, shows that the
normalisation of the intermediate power law decreases slightly with increasing $\ms a$, see Fig.~\ref{spectrumComp}. But the low-$k$ region is not very sensitive to the string tension and the overall effect on
the final number density is small. Obtaining a number smaller than the ``naive" misalignment contribution shall surprise some at first glance but it is not completely unexpected. At least part of the effect
comes from the fact that the typical axion momentum is not $0$ but of the order of $H_1$, and the energy of partially relativistic axions redshifts faster, as advanced in~\cite{Ballesteros:2016xej}. The axions
radiated from strings do not seem to contribute much to the total axion number. We understand this as each axion taking a large energy from the string network, in agreement with the results of
\cite{Gorghetto:2018myk}. This work also emphasises the need of studies at increased tension. We believe that our results provide partially that evidence, but a dedicated study the differential production rate
will be required. We are aware of further efforts in this direction, and we will present our own analysis in a future publication.

The fifth aspect that deserves a comment is that the occupation number does change to some extent with time, although not very much, as visible in Fig.~\ref{spectrumPlot} (right). We observe a small decline of
the integrated number of axions of the order of $d(N_A/V)/d\ct\sim -0.04$ in the $\ct\sim3-4.5$ range. It is inversely correlated to $N_A/V$ and increases slightly as the lattice spacing decreases $a\to 0$.
This effect has very little impact on the axion abundance, given the uncertainty implied by simulating strings with small $\kappa$, but certainly is visible. Moreover, the decline seems to level off around
$\ct\sim 4.5$. This changes are likely due to axion-self-interactions. The equations of motion \eqref{eomctheta} have non-linearities, which mix the modes to some extent. Expanding the sine beyond the leading
order $\sin\theta = \theta - \theta^3/6+...$ reveals a cubic term, which would be the dominant interaction when $\theta$ is small. This interaction comes from the first non-trivial term of the QCD potential, a
quartic self-interaction $V_{\rm QCD}\ni -\chi \theta^4/24$.
With the quantum-field-theoretic viewpoint in mind, this interaction term allows axion-scattering processes $2A\to 2A$, which do not change the number of axions but can reshuffle them in energy. Higher order
terms can mediate Axion-number changing processes like $4A\to 2A$, that can covert low momentum axions into more relativistic ones. A close inspection to Fig.~\ref{spectrumPlot} (right) reveals such trend: it
seems that low-$k$ axions ($k/k_1\lesssim 10$) are being up-scattered. The effect is particularly visible above the cut-off, where there is a vast deficiency of axions. These dynamics can be probably understood
in terms of the turbulent thermalisation mechanisms~\cite{Micha:2004bv}. The spectrum tends to a self-similar solution, that allows the transfer of particle number to reach some maximum entropy configuration.
While it is tempting to try to understand the dynamics in terms of~\cite{Micha:2004bv} we will not do so. The reason is that we believe that keeping only the first order interactions is not completely justified
 due to the presence of axitons, regions where $\theta\sim 1$ even at the late times shown in Fig.~\ref{spectrumPlot}. These regions are relatively small and do now show prominently in the number spectrum,
which is proportional to $\theta^2$. However they are quite prominent when we consider the dark matter fluctuations, which are $\propto \theta^4$. In the axiton cores fully non-linear dynamics is at play. The
net effect on our simulations, however, seems to be  compatible with the simple conversion of low-$k$ axions into more relativistic ones.

In conclusion, our axion spectrum seems dominated by low-momentum axions but with a relatively hard spectrum, in agreement with the recent literature~\cite{Fleury:2015aca,Gorghetto:2018myk}. We have pushed the
string tension by increasing the grid spacing by a factor of 4 or more compared to those references and still our results and the general picture converge. It also converges with the picture drawn by the new
technique to simulate string-networks with high string tension~\cite{Klaer:2017ond}. Further studies on the dynamics of the string network evolution will appear elsewhere.

\section{Dark matter distribution}\label{density}

After establishing the layout of our simulations and explaining how our results relate to previous studies, we now turn to our main topic of interest, the study of density fluctuations, commencing with the discussion of their two-point correlations. These would be sufficient to fully characterise the density field if the fluctuations where of purely Gaussian nature. As established in Sec.~\ref{minicluster} this is not entirely the case, and in particular we find that the phases of individual Fourier modes of the density contrast show strong correlations. Nevertheless, the two-point correlation function is an important tool to understand at which scales and times the individual processes in the axion field, like misalignment, string and wall decay and axitons, contribute to perturbations in the axion energy density. 

As stressed previously, we assume that the qualitative features of the density contrast are well represented by our simulations, despite the smallness of the string tension parameter. Further uncertainties to the density contrast are introduced by the existence of axitons. In the study of the density contrast, we presume that only the earliest axitons contribute significantly to perturbations at length scales relevant for miniclusters and use a late-time approximation scheme, which neglects self interactions. However, choosing the exact onset time for this late-time approximation introduces some arbitrariness to our simulations. We observe that the number of axitons created depends on the grid spacing, with coarser grids producing more axitons. Finally, the effect of axitons on the net axion number, though small, is not fully clarified yet. Section \ref{sectionaxitons} discusses these effects and the uncertainties on the density contrast introduced by axitons at length. Summarising, we believe that we can understand and describe axitons sufficiently well for the study of density perturbations and their two point correlations, presented here. We further extend the discussion of density fluctuations in Sec.~\ref{minicluster} to account for non-Gaussianities and give more details on the actual properties of miniclusters.

\subsection{Evolution of the dimensionless variance}

In this paper we are mostly interested in characterising the distribution of axion dark matter at comoving length scales $\sim L_1$ to study the formation of miniclusters. We define the local density contrast
as
\be
\label{deltadef}
\delta(\veca x) = \frac{\varrho(\veca x)-\bar \varrho}{\bar \varrho}\,,
\ee
where the axion energy density is computed as
\bea
\nonumber
\varrho(\veca x) &=& \frac{f_A^2}{2} (\partial_t\theta)^2
+\frac{f_A^2}{2 R^2} (\nabla\theta)^2 + \chi(T)(1-\cos\theta) \\
 &=&
\frac{(f_AH_1)^2}{\tau^4}  \(\frac{1}{2}(\partial_\ct\ctheta+\ctheta/\ct)^2+\frac{1}{2}(\nabla \cctheta)^2 + \ct^{\n+4}(1-\cos(\ctheta/\ct))\).
\eea

\begin{figure}[tbp]
\begin{center}
\includegraphics[width=0.48\textwidth]{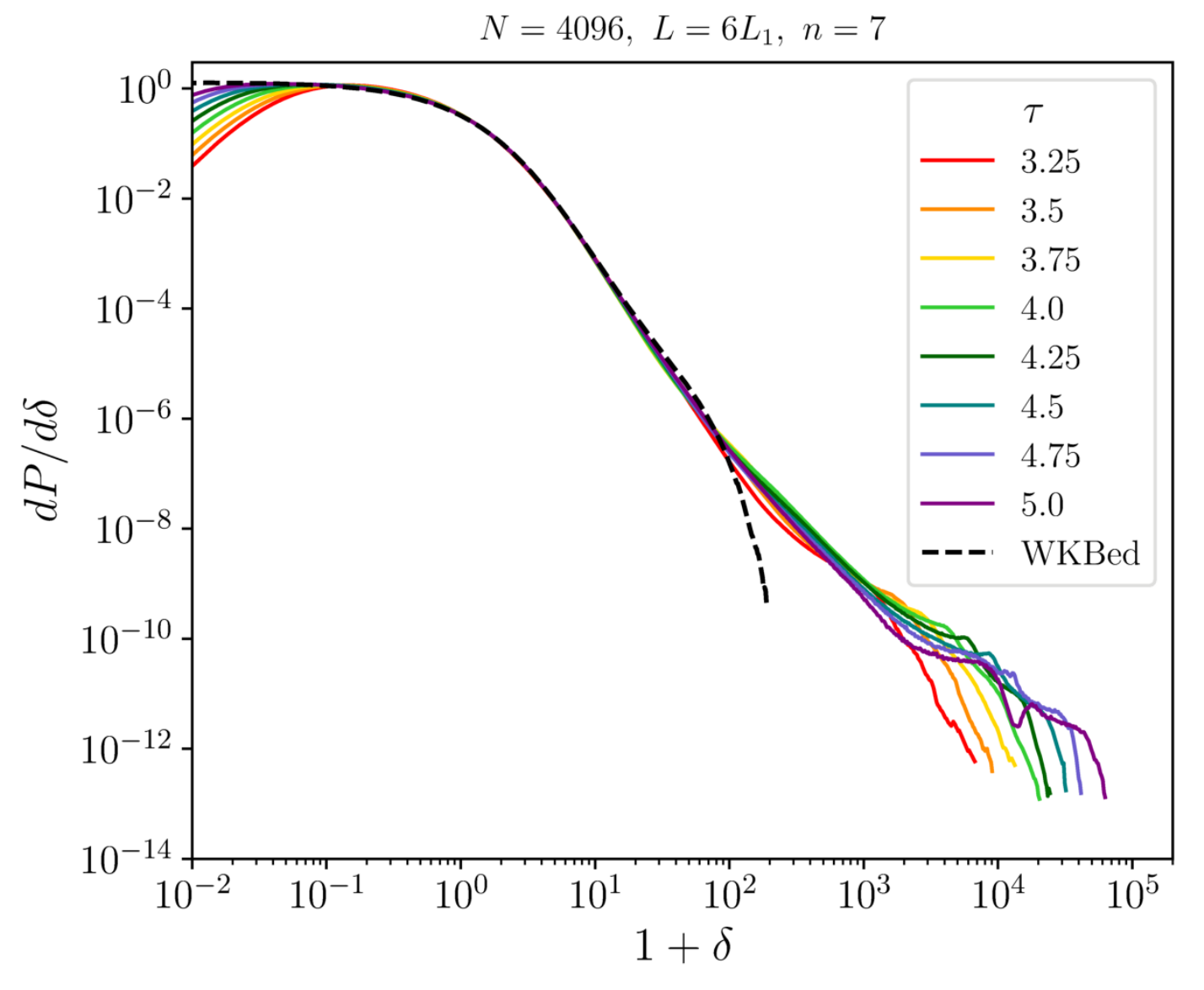}
\includegraphics[width=0.48\textwidth]{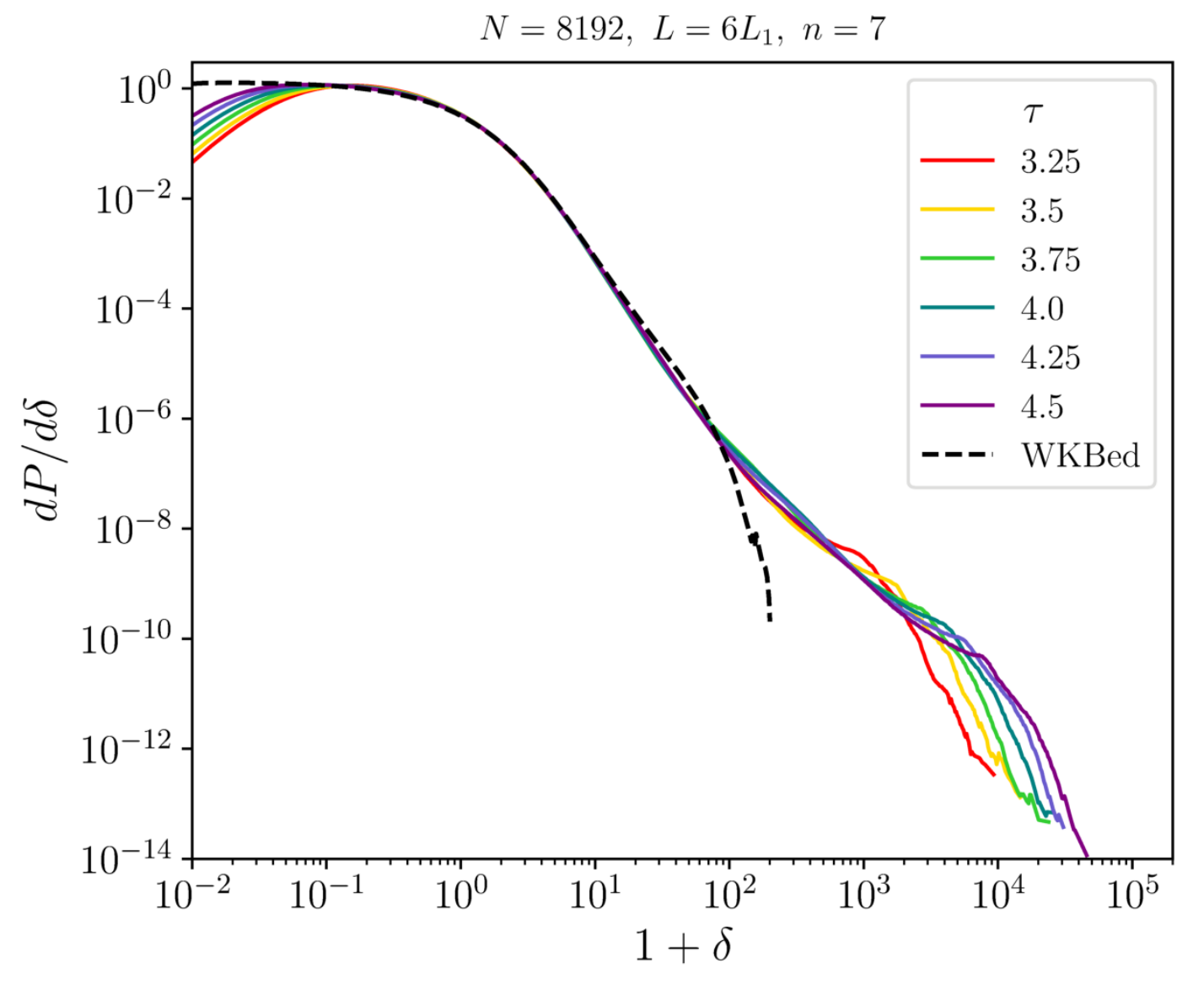}
\caption{The distribution of overdensities in each point of our simulation grid as a function of conformal time. These results where obtained on grids of length $L=6 L_1$ using $\n=7$ and $N=4096$ (left) or
$N=8192$ (right). The dashed black line, labelled `WKBed', is obtaining by free-streaming the axion field at the end of the simulation.}
\label{dpddelta}
\end{center}
\end{figure}

In Fig.~\ref{dpddelta} we show how the distribution of values of $\delta(\veca x_i)$ in each grid position $\veca x_i$ evolves in time. The results of our benchmark simulations with $L=6,N=4096,\ms a = 1.5$ are
presented on the left and $N=8192$ on the right. For better display in the log-log plot, we use $\delta +1$ as abscissa. The distribution peaks around $\delta\sim 1$, decreases very sharply ($\delta^{-3.5}$)
for intermediate values and hardens at $\delta\sim 100$. The slope of this high-density tail is $dP/d\delta\propto 1/\delta^{2}$, which turns out to be very important. Finally, one can identify a distinctive
peak after which the density drops faster again. It corresponds to a density $\varrho\sim V_{\rm QCD}(\theta=\pi)=2m_A^2f_A^2$, which increases in time as the axion mass grows.

\begin{figure}[tbp]
\begin{center}
\includegraphics[width=0.48\textwidth]{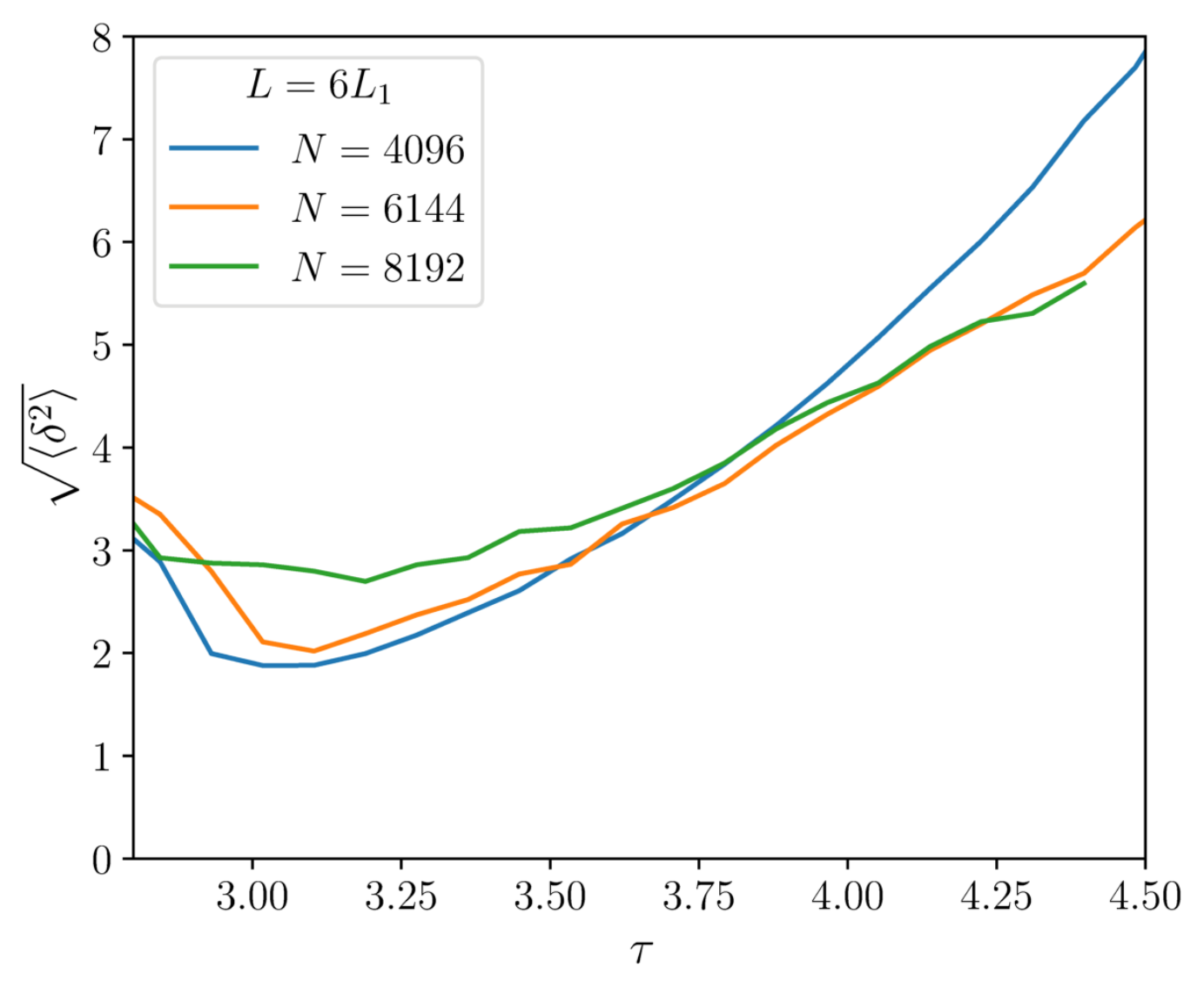}
\caption{Evolution of the density variance as a function of $\ct$ for $N=4096,6144,8192$ simulations with size $L=6L_1$.}
\label{delta2evol}
\end{center}
\end{figure}

Since $dP/d\delta$ is a probability distribution, we can compute its moments and analyse which parts of the distribution contribute the most to them. For instance, the first moment is the average contrast
$\langle \delta \rangle$, which by definition  \eqref{deltadef} is zero. When we compute the  average density $\propto 1+\delta$,
\be
\langle \delta +1 \rangle = \int d\delta \frac{dP(\delta)}{d\delta} (\delta+1)  = 1 \,,
\ee
we see that indeed this integral is dominated by values of $\delta\sim {\cal O}(1)$ and not affected by the hard tail.
However, higher moments starting from $\langle \delta^2\rangle$, are indeed sensitive to the cut-off.
As the cut-off increases with time, so does $\langle \delta^2\rangle$, see Fig.~\ref{delta2evol}.

The explanation for this curious phenomenon is the presence of a few axitons~\cite{Kolb:1993hw} in the grid. Before describing them let us first look at the spectrum of density fluctuations.

The Fourier transform of the density contrast field,
\be
\widetilde \delta(\veca k) = \int d^3{\veca x}\, e^{i\veca k\cdot \veca x}\, \delta(\veca x)  \,,
\ee
is a capital tool to understand the distribution of dark matter. In particular, it allows for a spectral decomposition of the average fluctuation,
\be
\langle \delta^2(\veca x)\rangle = \int \frac{dk}{k} \Delta^2_k,
\ee
where the dimensionless variance in a comoving volume $V$,
\be
\label{Delta2def}
\Delta^2_k = \frac{k^3}{2\pi^2} \frac{1}{V}
\langle |\widetilde\delta(\veca k)|^2 \rangle_{|\veca k| = k}\,,
\ee
is essentially the angle-averaged $|\widetilde \delta(\veca k)|^2$ multiplied by $k^3$ to give the contribution to the density fluctuations per logarithmic interval.

\additionalinfo{The density fluctuations smoothed over a spherical region of comoving size $\sigma$ can be also computed from a similar integral
\be
\label{}
\langle \delta_\sigma^2\rangle = \int \frac{dk}{k} \Delta^2_k e^{-k^2 \sigma^2}
\ee
where we have defined the volume of the region with a Gaussian window function, $W\propto e^{-|\veca x-\veca x_0|^2/(2\sigma^2)}$.}

\begin{figure}[htbp]
\begin{center}
\includegraphics[width=0.95\textwidth]{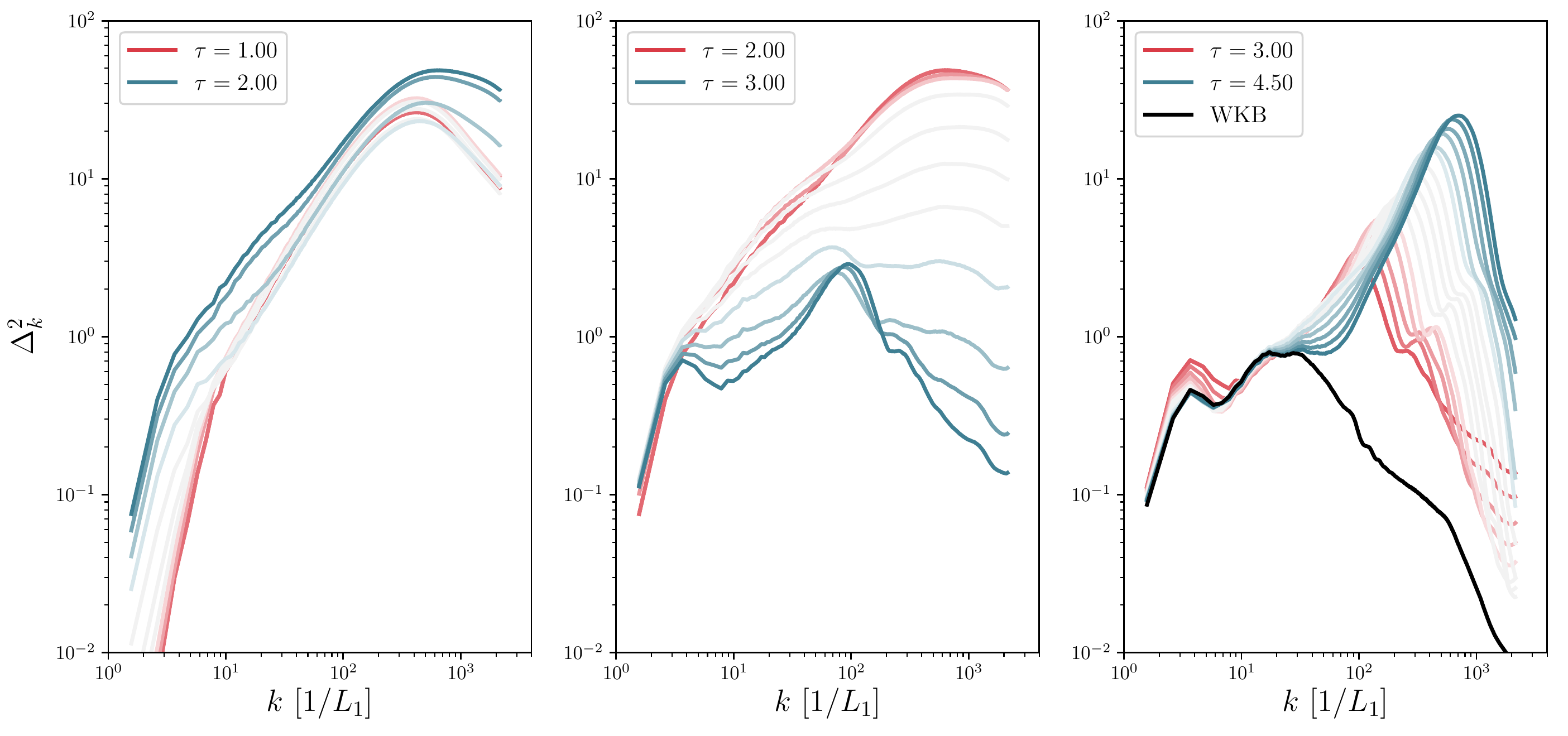}
\caption{Time evolution of the dimensionless variance of axion energy density fluctuations as a function of the momentum. Results are shown for an average over $L=6L_1$, $N=8192$ and $\n=7$ simulations.
Different lines show the time evolution. For simplicity the legends quote only the earliest and latest time shown in each plot, the remaining colours interpolate between these limits. The time difference
between steps is $\Delta\ct=0.1$. The three plots show the evolution through the three periods of our simulation: axions with strings (left), network destruction (center) and non-relativistic period with
axitons (right).}
\label{dimvar}
\end{center}
\end{figure}

The dimensionless-variance resulting from our finest simulations, namely $L=6L_1$ boxes using $N=8192$ grids, is shown in Fig.~\ref{dimvar}. We can identify three stages of the axion field evolution. The first
period of relevance is before $\ct\sim 2$ where the string network is in the scaling regime. The low-$k$ cut-off decreases, signalling the increase of coherence length of the axion patches as the horizon size
increases. The fluctuations grow at all scales presumably due to string and domain wall radiation. In the second plot, we show the collapse of the string-wall network and the axion field becoming
non-relativistic at the same time. The low-$k$ cut-off is almost frozen, and most of the changes happen at higher momenta. There, axions are still relativistic, can free-stream and decrease the fluctuations at
short distances.
The peak, clearly visible at $\ct=3$, seems to be drifting towards large $k$. At $\ct\sim 3$ we switch to $\theta$-only simulations. The  right plot shows that the trend of the peak to shift to higher $k$
continues. Eventually, the peak will reach the resolution of our grid. The peak is related to the same few points of large density contrast that we found studying $dP/d\delta$ and corresponds to the density
fluctuations caused by axitons. Axitons are pseudo-breathers of the Sine-Gordon equation, that are constantly flashing and re-collapsing emitting relativistic axions of momentum $k\sim m_A$. We will see that
they re-collapse becoming pseudo-stable due to the fast increase of the axion mass with decreasing temperature (increasing time) but once the mass acquires its zero-temperature value they are bound to diffuse
away after their last flash around $\ct \sim \ctc\sim 16$ (c.f. Eq.~\ref{ctc}). With the current resolution of our grids we cannot reach these times at which the critical temperature is achieved. We will argue, however, that there
are good reasons to believe that the density fluctuations created at later times due to axions emitted by axitons are largely irrelevant at the typical scales of miniclusters. Therefore, before the axiton-core
size reaches our resolution $\sim a$ we switch-off the axion self-interactions. The axions can then only free-stream, and axitons vanish extremely fast from $\Delta^2_k$. Since all axion modes are in the deep
adiabatic regime by then, the free-streaming evolution is adiabatic and can be done using the WKB approximation \eqref{solumode}. The resulting power spectrum after the WKB evolution is shown as black lines in
Fig.~\ref{dimvar}. The curve gives precisely what we expect from a free-streaming. The distribution at relatively large scales is unaffected by it, but small scale-fluctuations are erased roughly to the level
they had before the axitons formed. Small residual oscillations around $k\sim 100/L_1$ can arise from a number of different effects. To date we have not elucidated their exact origin, but they do not modify
significantly our conclusions.

\begin{figure}[tbp]
\begin{center}
\includegraphics[width=0.85\textwidth]{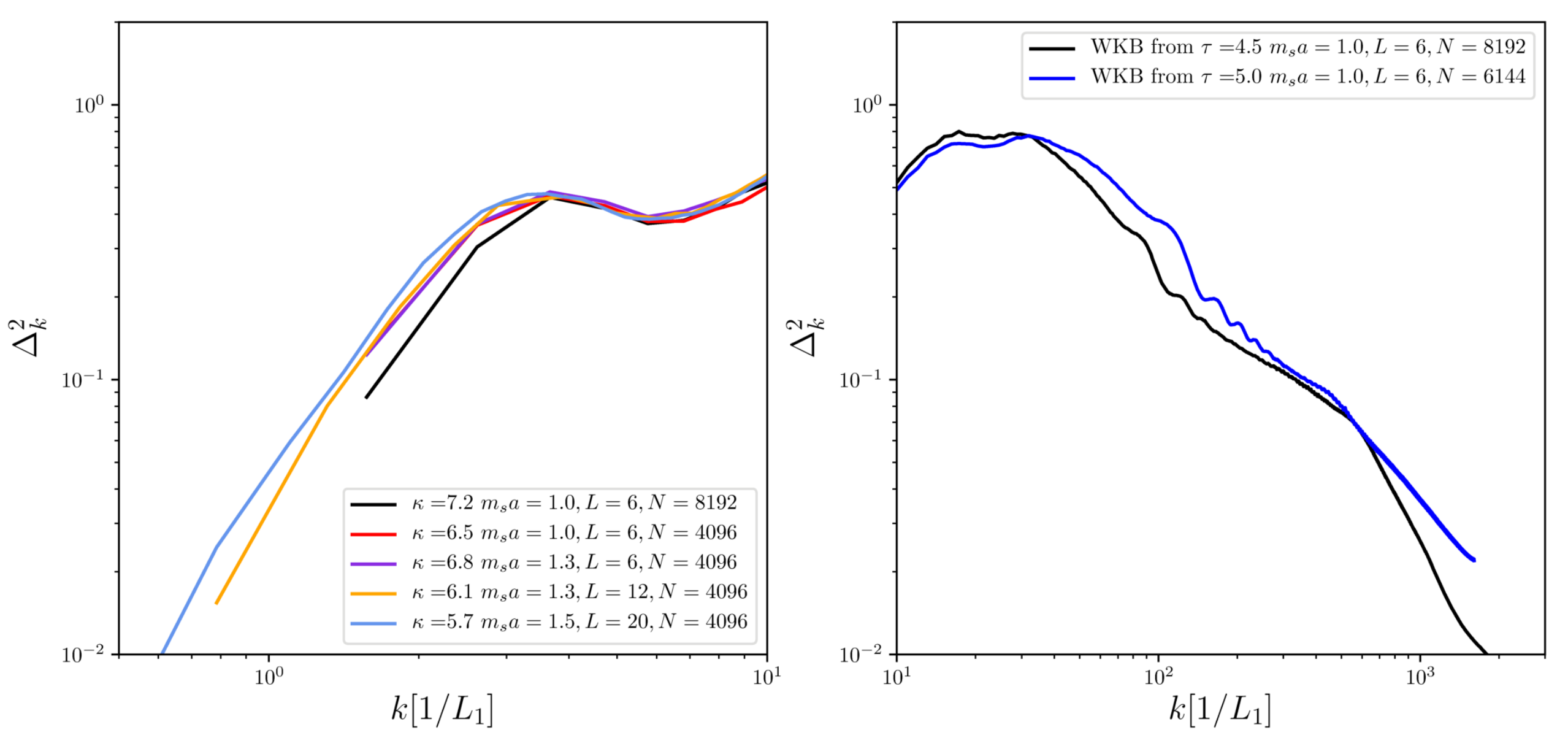}
\caption{Spectrum of axion dark matter density fluctuations after the WKB procedure to disperse the axitons. On the left we show the low-$k$ part of the spectrum for several simulations with different string
tensions. An increasing tension (here $\kappa$ is evaluated at $\ct=1$) or resolution decreases the low-$k$ tail. On the right we show the high end of the spectrum for two different simulations. The time at
which we stop our simulations affects the $k\sim 100/L_1$ oscillations. We evolve both simulation sets to $\ct=6$ with the WKB solution. The blue line has already resolution problems at $\ct=5$.}
\label{dimvar_summary}
\end{center}
\end{figure}

Fig.~\ref{dimvar_summary} summarises some of our findings for varying string tensions, physical volumes and final times to our simulation. In the left plot we observe how increasing the string tension or the
resolution decreases the low-$k$ tail. In the right plot we show that our choice of the time when we stop the simulations and perform the WKB evolution has only a moderate influence around $k\sim 100/L_1$. The
blue line was evolved until $\ct=5$ in $L=6L_1$, $N=6144$ grids, for which we have $m_A a = 1.4$. The simulations corresponding to the black line were stopped at $\ct=4.5$, for which $m_Aa=0.63$.

Overall the results are largely consistent and reveal a very clear picture. We discuss it in the remainder of this section for the low-$k$, intermediate-$k$ and high-$k$ regions separately.

\subsection{Low-$k$ region: Patches}\label{patches}

In the low-$k$ regime, $\Delta_k^2$ seems to converge to the power law,
\be
\label{lowfit}
\Delta^2_k \simeq (3\pm 1)\times 10^{-2} (k L_1)^3,
\ee
which means that $|{\widetilde\delta}(\veca k)|^2$ becomes independent of momentum, i.e. a white noise power spectrum. 
We interpret this low-$k$ trend as the region where the density field is dominated by the misalignment axions. Different patches of comoving size $L_1$ have different initial values of the axion field and thus
different DM densities. The $\Delta^2_k\propto k^3$ behaviour can be easily understood with a model where the total DM mass $M_t$ in a comoving volume $V$ is put into $N_c$ clumps of mass $M_c$, volume $V_c$
(linear dimensions $L_c\sim V_c^{1/3}\sim L_1$) in random positions. The calculation of the Fourier mode $\delta(\veca k)$ with wavelength $\lambda_k =2\pi/k \gg L_c$ is a volume integral over density that can
be translated into a sum over clump masses time a phase,
\be
\widetilde\delta(\veca k) = \sum_{i\in\{c\}} \frac{M_i}{M_t/V}e^{i\veca k\cdot \veca x}\,.
\ee
The angular mean and variance are then given by
\bea
\widetilde\delta(k) = \langle \widetilde\delta(\veca k) \rangle_{|\veca k| = k} &=& 0\,,   \\
\label{delta2gen}
{\widetilde\delta}^2(k) = \langle {\widetilde\delta}^2(\veca k) \rangle_{|\veca k| = k} &=& \frac{N_c M^2_c}{{M_t/V}^2}  \,,
\eea
as long as the clump-common phases can be considered random. This is the case, given that the clusters are randomly distributed and much smaller than the mode's wavelength, i.e. $k L_c \gg1$. Assuming that essentially
all axion DM mass is in these clumps, we have $M_t=M_c N_c$ and find
\be
\label{delta2mc}
 {\widetilde\delta}^2(k)  = \frac{V^2}{N_c},
\ee
which is of course independent of $k$. We can use this interpretation to estimate the number of clump objects in our simulations. From our result on $\Delta_k$ \eqref{lowfit} we can derive the white noise variance as
${\widetilde\delta}{^2}(k) = 2\pi^2V\Delta^2_k/k^3=0.6 V L_1^3$ and use Eq.~\eqref{delta2mc} to get the density of clumps in our simulations,
\be
\label{ncestimate}
\frac{N_c}{V} \sim 1.7\frac{1}{L_1^3},
\ee
very close to expectations.

The lower and the higher bound in Eq.~\eqref{lowfit} correspond to the largest ($\kappa(\ct = 1) = 7.2$) and smallest ($\kappa(\ct = 1) = 5.7$) string tensions tested in our simulations, respectively. Increasing
the string tension increases the string density and therefore reduces the size of the axion patches because the axion field takes values $0,2\pi$ around the string. Although the string density is controlled by
the horizon size, it is also directly affected by the string tension, which, acting as inertia, seems to be delaying the destruction of strings and thus increasing the density parameter $\xi$. In our
simulations $\xi\sim {\cal O}(1)$ and the difference cannot be too large. But the general prediction is that our patches should be smaller and the peak shifted to higher $k$ as $\xi$ increases. This leads to
the unfortunate conclusion that our results have to be taken with a grain of salt. If the trend towards increasing $\xi$ at higher values of the tension parameter shown in
\cite{Fleury:2015aca,Gorghetto:2018myk,Kawasaki:2018aa} and our Fig.~\ref{xievol} is confirmed, which seems to be the case in the simulations of~\cite{Klaer:2017qhr}, the size of clumps will decrease, its
number $N_c/V$ will increase, $\Delta_k^2$ will be further suppressed and the value of $k$ where it levels up will increase. In our simulation range, $\xi$ increases from $\sim0.8$ to $\sim 1.1$, which probably
cannot account for all the variance of \eqref{lowfit}. However, there are some effects only indirectly related to the string tension which can contribute to this variance as well. Our large tension simulations
use finer grids and therefore more relativistic axions. In these simulations we could therefore afford to use $\ms a = 1$, while for larger physical boxes we used $\ms a=1.3,1.5$. In general lower values of
$\ms a$ produce more axions in the intermediate $k$ range. As relativistic axions free-stream more, the prediction is that large scale fluctuations should decrease. Finally, at this low-$k$ the number of modes
is so small that we cannot neglect a small error due to insufficient statistics, our binning of modes, and finite volume effects in the $L=6 L_1$ simulations.

As already stated at the end of section \ref{sec-digression-axion-yied}, for this work we adopt the assumption that the results of our simulations would not suffer too large changes if the string tension was
increased to physical values. However, a deeper understanding on the role of $\kappa$ in the production of axions will be needed to confirm our conclusions. We will compare the numerically obtained results in
the low-$k$ region to the analytical predictions for misalignment-only production in the following section.

\subsubsection{Comparison with only-misalignment}
At a value $k\simeq (3\sim4)/L_1$, which corresponds to a half wavelength $\lambda/2 = \pi/k \sim L_1$, $\Delta^2_k$ reaches a first peak and softens its behaviour. A small valley at $k\sim 6/L_1$ is clearly
visible in Fig.~\ref{dimvar_summary}. This valley corresponds to the axion modes that enter the horizon a bit before $\ct=1$ and perform just one damped oscillation before becoming effectively frozen.
Qualitatively the behaviour of $\Delta_k^2$ observed in our simulations resembles the expectations from pure misalignment, recently presented in \cite{Enander:2017ogx}. Indeed the $k\sim 6/L_1$ valley is
clearly visible in their Fig.~1 and Fig.~2.

There are, however, a number of discrepancies expected between the results of \cite{Enander:2017ogx} and our simulations. In~\cite{Enander:2017ogx} the axion field is evolved in Fourier-space neglecting
self-interactions from the beginning. The initial misalignment angle is taken from a white noise spectrum, and its amplitude is chosen in order to match the expected average $\langle\theta^2\rangle = \pi/3$. This requires to introduce a cut-off at high-$k$, which in turn influences the normalisation of the low-$k$ region. Therefore, we do not expect our results to exactly agree in magnitude with Fig.
\ref{dimvar_summary}. Moreover, the position of the first peak and valley are not expected to coincide with the linearised approach of~\cite{Enander:2017ogx} either. The main reason is the mentioned presence of
axion strings in our simulation, which act as boundaries for the axion field.

\begin{figure}[tbp]
\begin{center}
\includegraphics[width=0.5\textwidth]{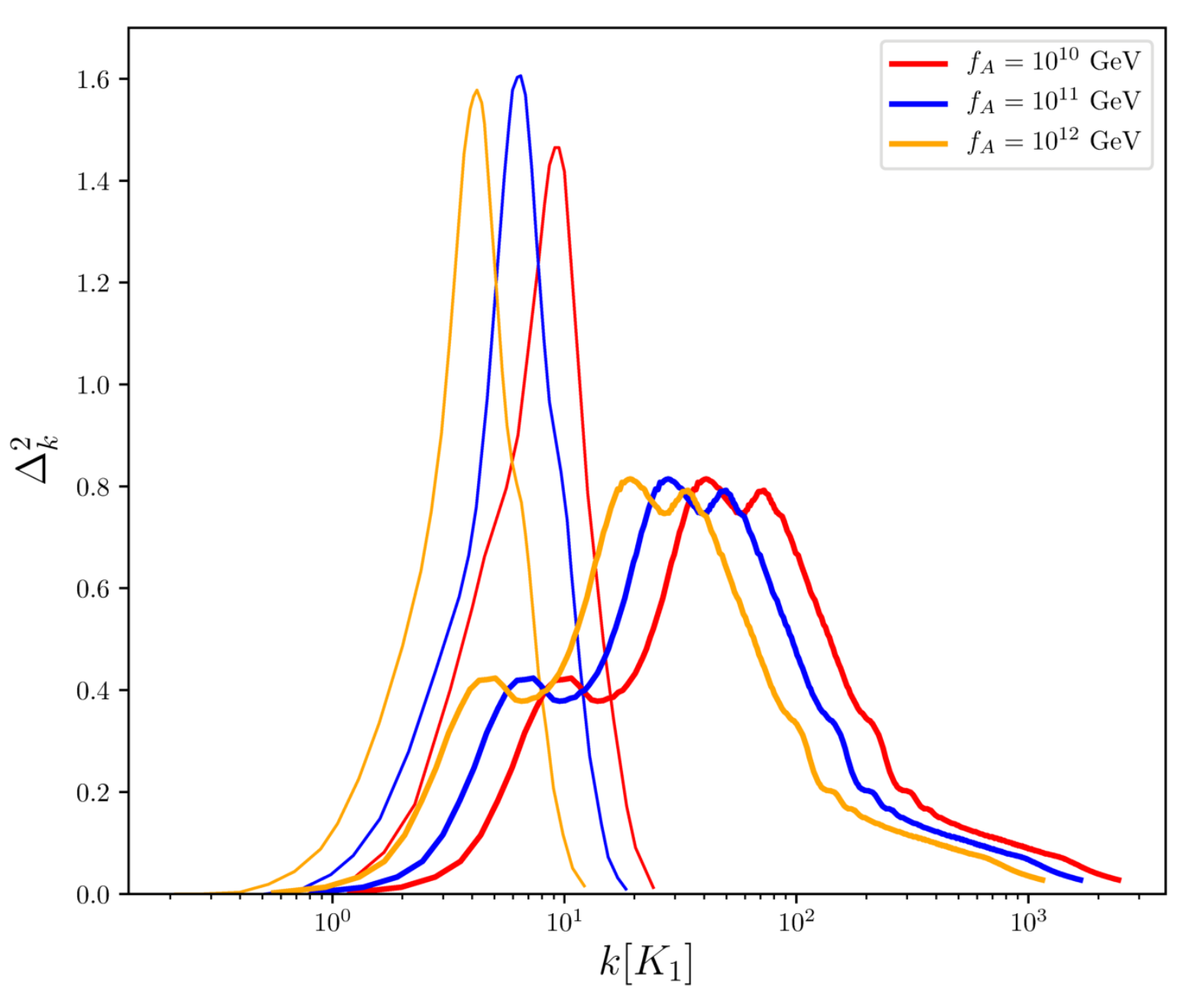}
\caption{Comparison of dimensionless variance from this work (thick lines) to the analytical study of~\cite{Enander:2017ogx} based on linear evolution (thin lines). Here $K_1$ is the comoving wave number that
enters the horizon when $T=1\rm GeV$ as defined in~\cite{Enander:2017ogx}. It corresponds to our $1/L_1$ for $m_A=2\mu$eV.}
\label{enander}
\end{center}
\end{figure}

A comparison of our results with~\cite{Enander:2017ogx} for three values of $f_A$ is shown in Fig.~\ref{enander}, where we {indeed} identify the three effects mentioned: our power spectra (thicker lines) peak
at higher momenta, have generally less power and extend up to much high higher $k$. The main difference is caused by the different initial conditions. The white noise assumed by~\cite{Enander:2017ogx} is only a
good approximation up to $k\sim 3/L_1$ as seen already in our axion spectrum, see Fig.~\ref{spectrumComp}. At higher-$k$ the axions radiated from strings have much less power {than they would if the white-noise
spectrum continued to those scales.} Our approach and the discussion of the previous section  shows that we do not need an artificial cut-off when we use the correct initial conditions.  

Another possible discrepancy is that both calculations differ slightly on the physical parameters used. We have used essentially constant number of degrees of freedom, $g_{*S}$, during the evolution and a
simple power law for the axion mass, $\n=7$, while ~\cite{Enander:2017ogx}  makes full use of the results of~\cite{Borsanyi:2016ksw}. However, {this} cannot explain the large differences observed {for the
final $\Delta_k^2$.}

\subsection{Intermediate length scales}

\begin{figure}[t!]
\begin{center}
\includegraphics[height=0.45\textwidth]{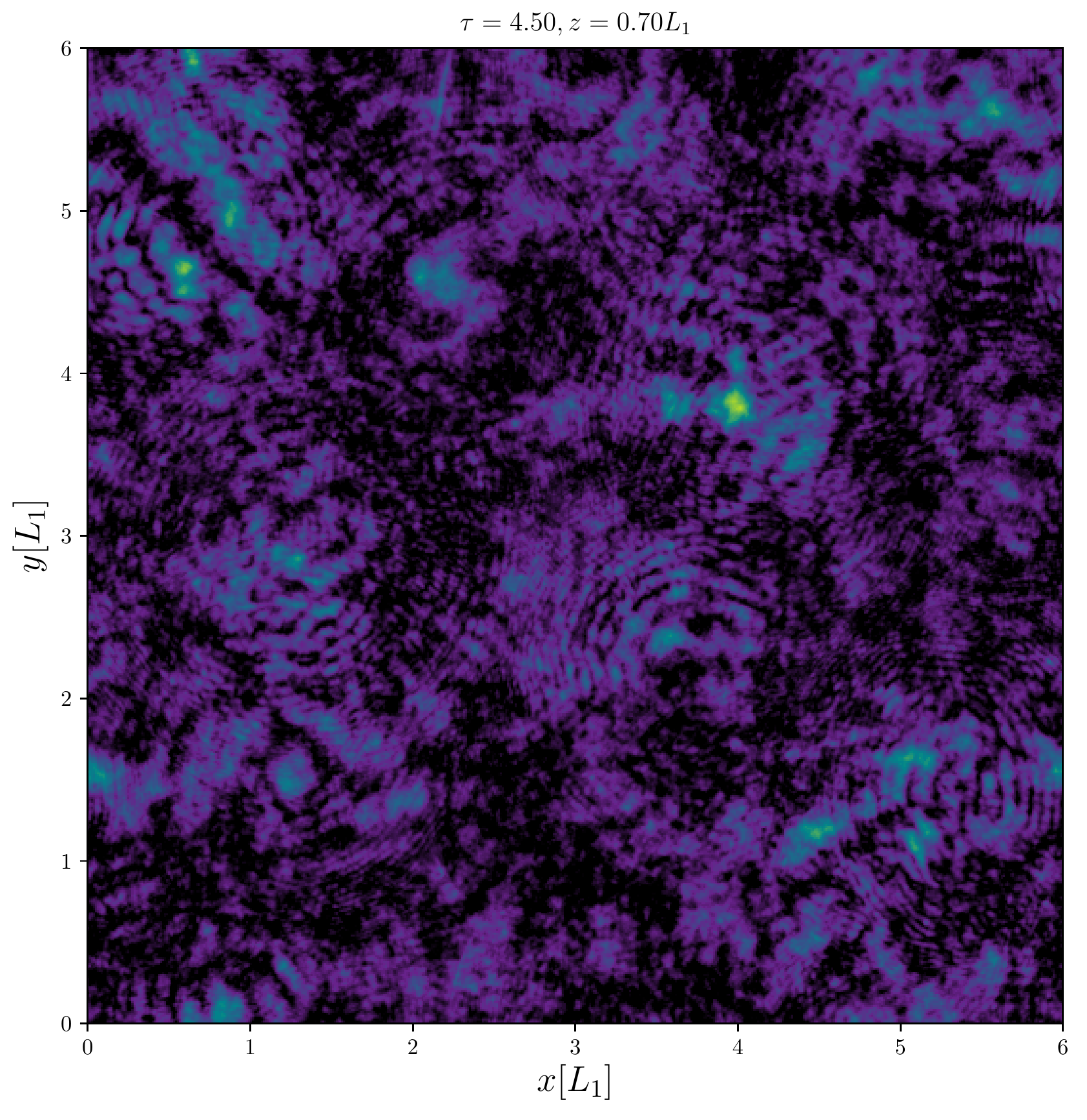} \includegraphics[height=0.45\textwidth]{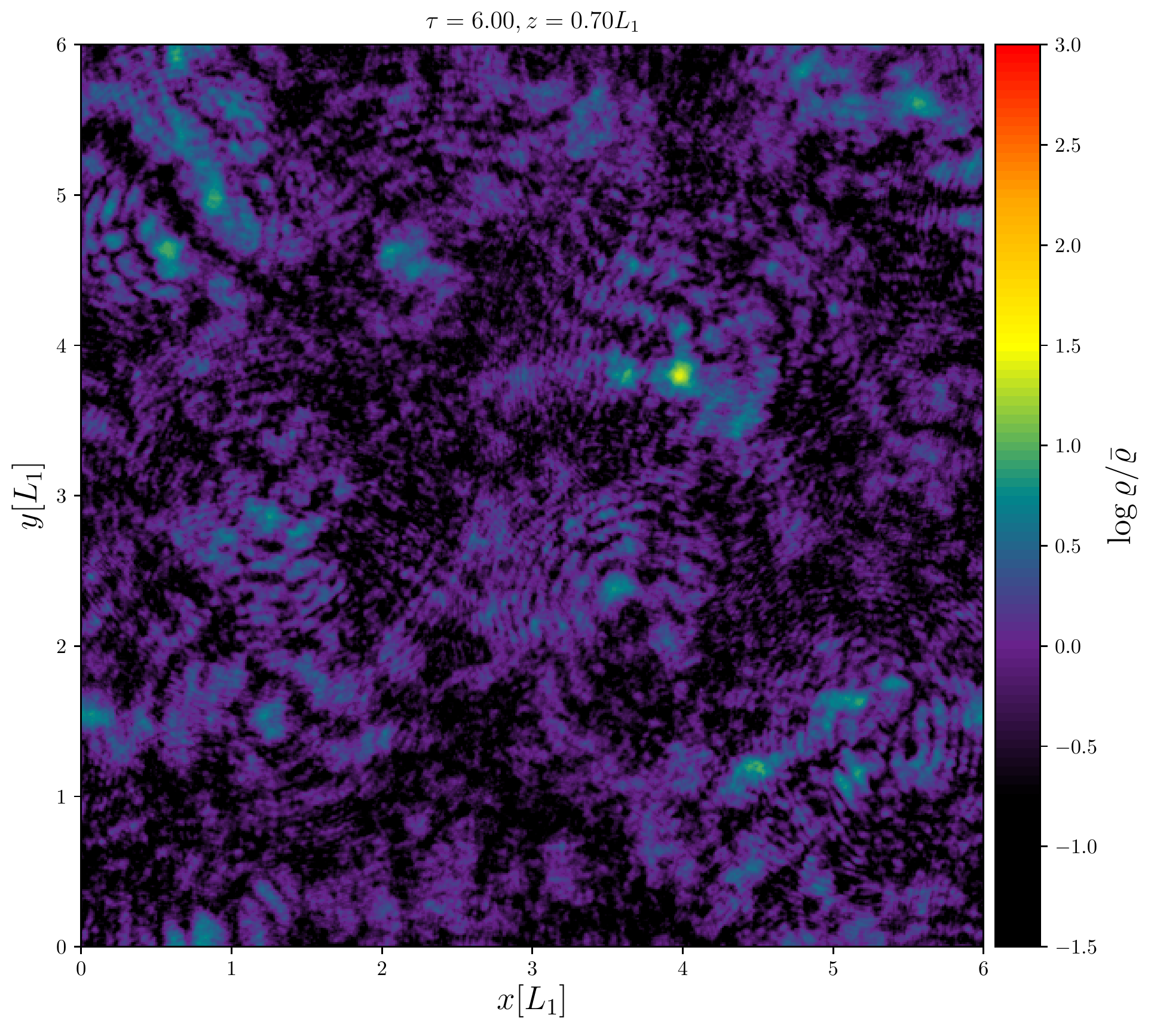}

\includegraphics[height=0.45\textwidth]{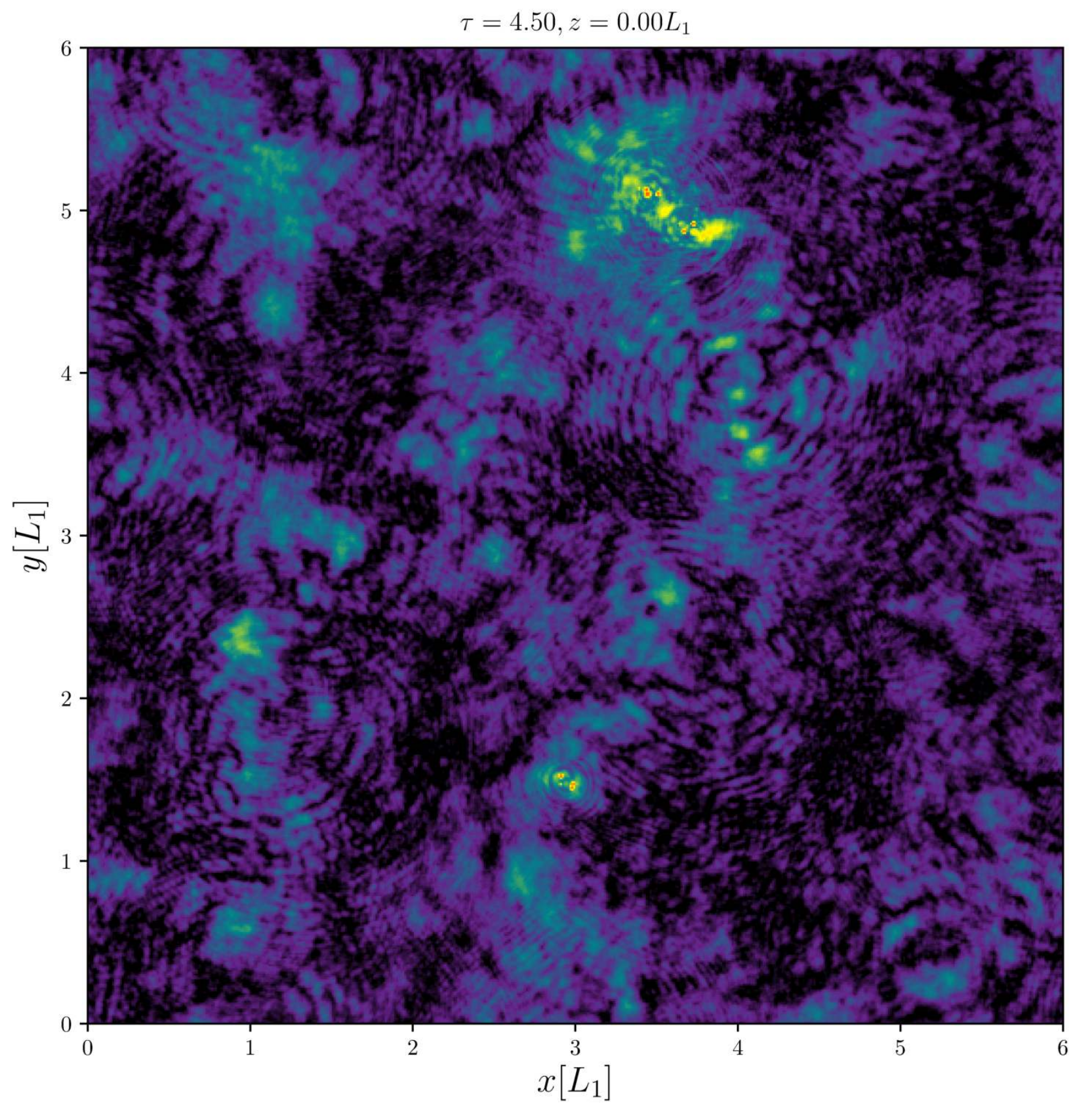} \includegraphics[height=0.45\textwidth]{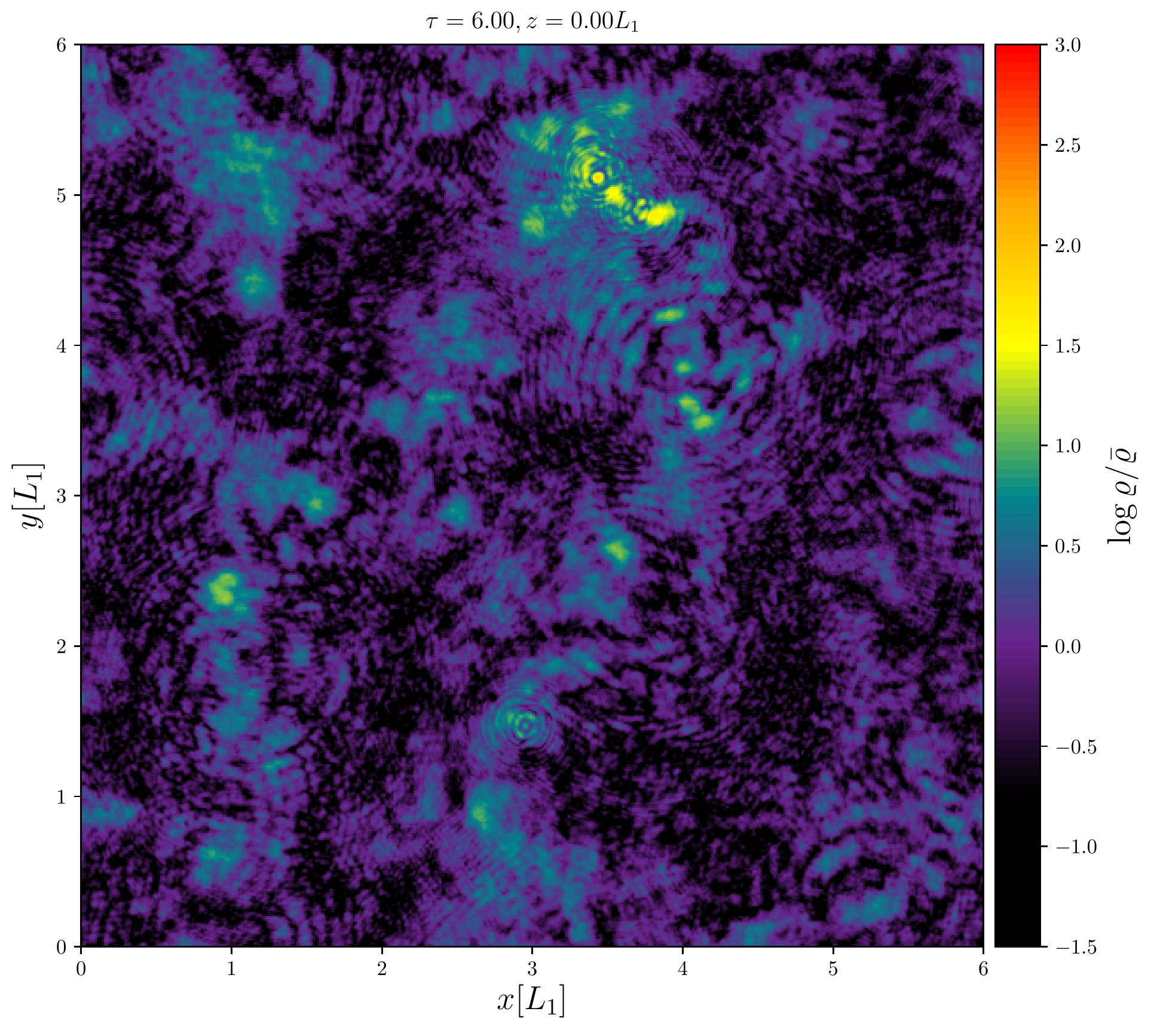}
\caption{2D slices of a benchmark simulation ($L=6L_1,N=8192,\n=7$) showing the density normalised to the average $\varrho/\bar\varrho$. We show two different slices at the ending time of the simulation
$\ct=4.5$ (left) and after a WKB evolution up to $\ct=6.0$ (right). The upper left plot shows a relatively calm slice, while the lower left one shows a slice in which several clustered axitons, with
$\varrho/\bar\varrho$ exceeding $10^3$, can be recognised. After the WKB evolution we see how the calm slice is largely unaffected, whereas the axiton cores in the lower left plot have diffused away. The WKB
evolution does not affect significantly the long and intermediate scales. The axion DM distribution in the WKBed maps is essentially frozen, including the spherical wave fronts and interference patterns at
scales $\lesssim 0.1L_1$.}
\label{densityslices}
\end{center}
\end{figure}

At intermediate length scales, $k\in (3 k_1,100 k_1)$, the variance reaches its maximum and becomes relatively flat. A look at Fig.~\ref{dimvar} (center) suggests that these fluctuations are already present when
the axiton peak develops. However, it is not entirely clear if fluctuations at intermediate length scales can be attributed to the tail of the misalignment mechanism only. The late evolution of Fig.~\ref{dimvar}
(right) shows that late times also see an increase of the power in the $6-20/L_1$ region, although the effect is not large. As this coincides with the slow decrease of large scale fluctuations, it could be
caused by non-linearities of the axion potential compressing intermediate-size over-densities.

Further light can be shed on axion modes in the intermediate $k$-range by the inspection of density contrast maps like the 2D slices shown in Fig.~\ref{densityslices}, which show very interesting characteristic features. The maps are carpeted with circular wave-fronts and their constructive and destructive interferences. Most of them can be traced to the wave-fronts generated by the fast acceleration of the strings by walls and specially their collapse at $\ct\sim 2.5$. The non-harmonic axion potential or, equivalently, axion self-interactions seem to enhance the constructive interference. The interference patterns have a characteristic comoving radius $\sim L_1$ and smaller wavefronts inside. Spherical waves of smaller wavelength seem to emanate from the few axitons. Altogether {the maps} seem to suggest that the fluctuations of intermediate wavelength are related to the string-wall collapse, non-linearities in the axion potential and the first axitons. Possibly there is also a random component from the misalignment tail. The fluctuations appear to be highly non-Gaussian.

\begin{figure}[t!]
\begin{center}
\includegraphics[height=0.5\textwidth]{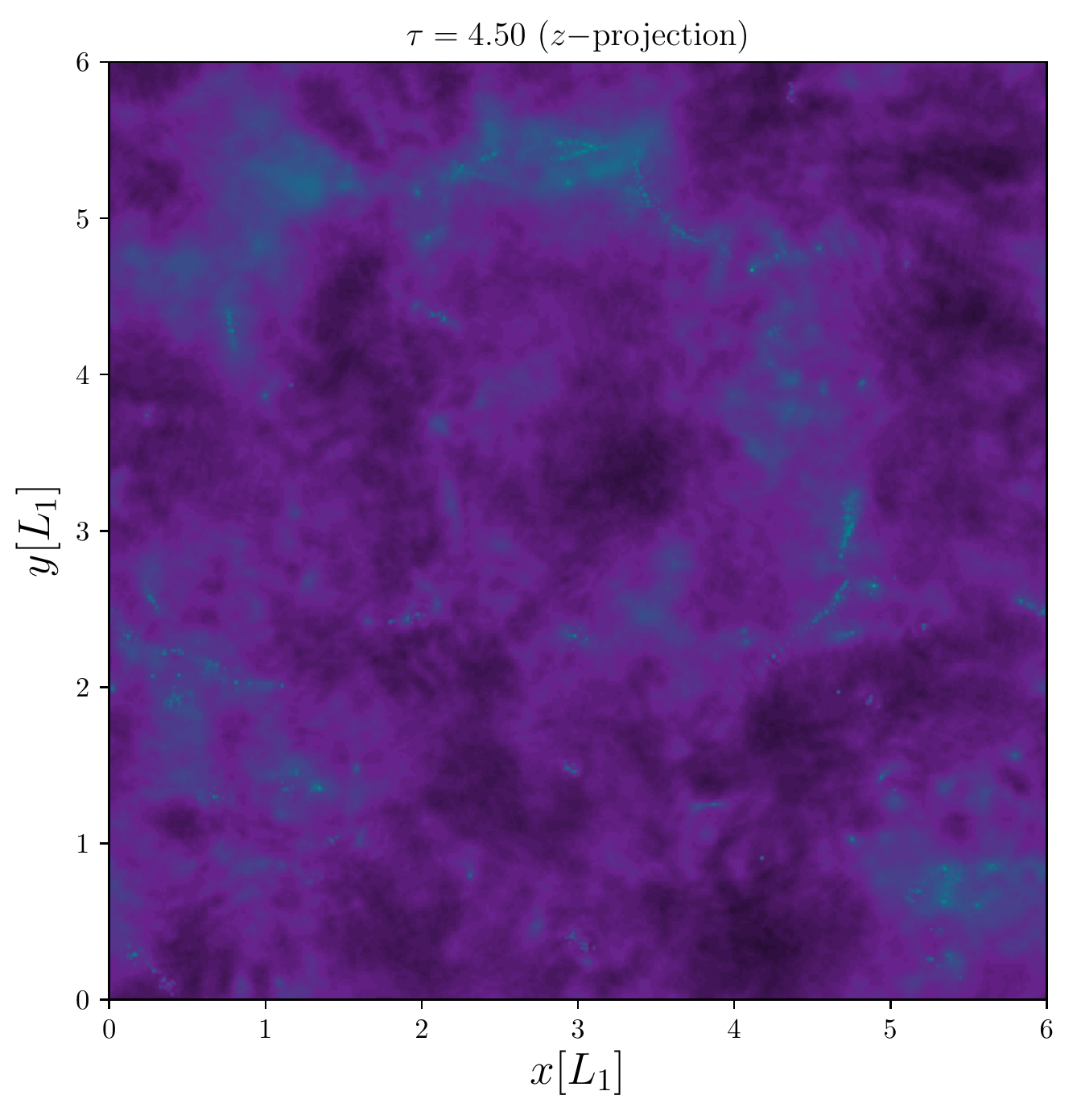} \includegraphics[height=0.5\textwidth]{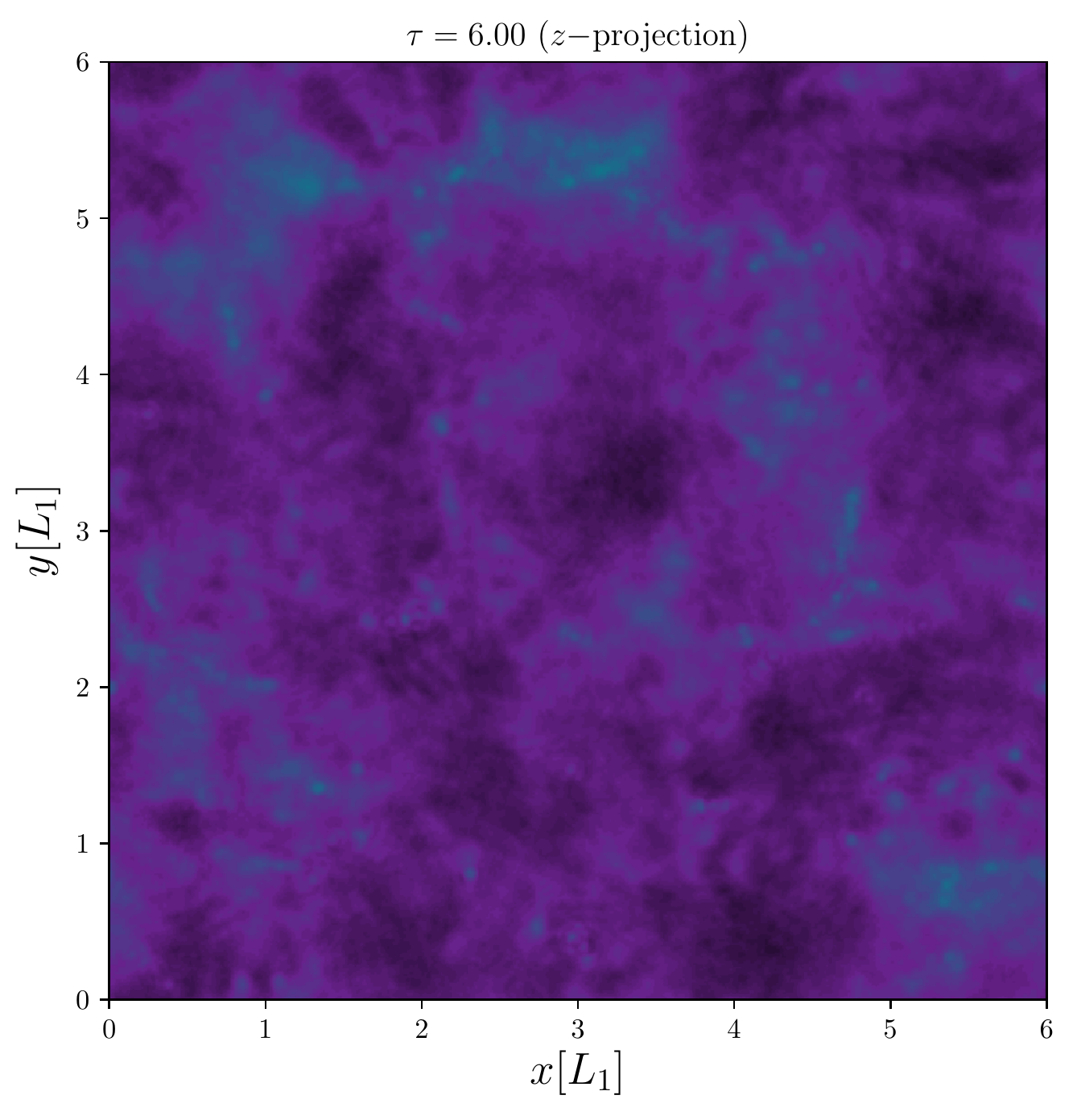}

\caption{3D$\to$2D projection plots of the density normalised to the average $\int_0^L (dz/L)\varrho(\veca x)/\bar\varrho$ at the ending time of our simulation ($\ct=4.5$, left) and after the WKB ($\ct=6$,
right). The colour code and the simulation shown are the same than Fig.~\ref{densityslices}. The axitons show as moderate point-line enhancements that disappears after the WKB. They appear in overdensities
that remain, though.}
\label{densityslices2}
\end{center}
\end{figure}

From the 2D slices one sees quite clearly what the power spectrum tells us in an indirect way. The large scale density field is dominated by scales $\sim L_1$ but ${\cal O}(1)$ fluctuations are very abundant at
smaller scales. The greatest overdensities correspond to axiton cores, which show as red points in Fig.~\ref{densityslices} (low left). Axitons mostly appear clustered in particularly overdense $\sim L_1$
regions. As expected, in projection plots of the density, axitons show very moderately (in contrast to those of density squared shown in Fig.~\ref{pics}), see Fig.~\ref{densityslices2}. However, the
misalignment and the intermediate scale fluctuations are clearly visible.

\subsection{{Small scales:} axiton peak}

\begin{figure}[htbp]
\begin{center}
\includegraphics[width=\textwidth]{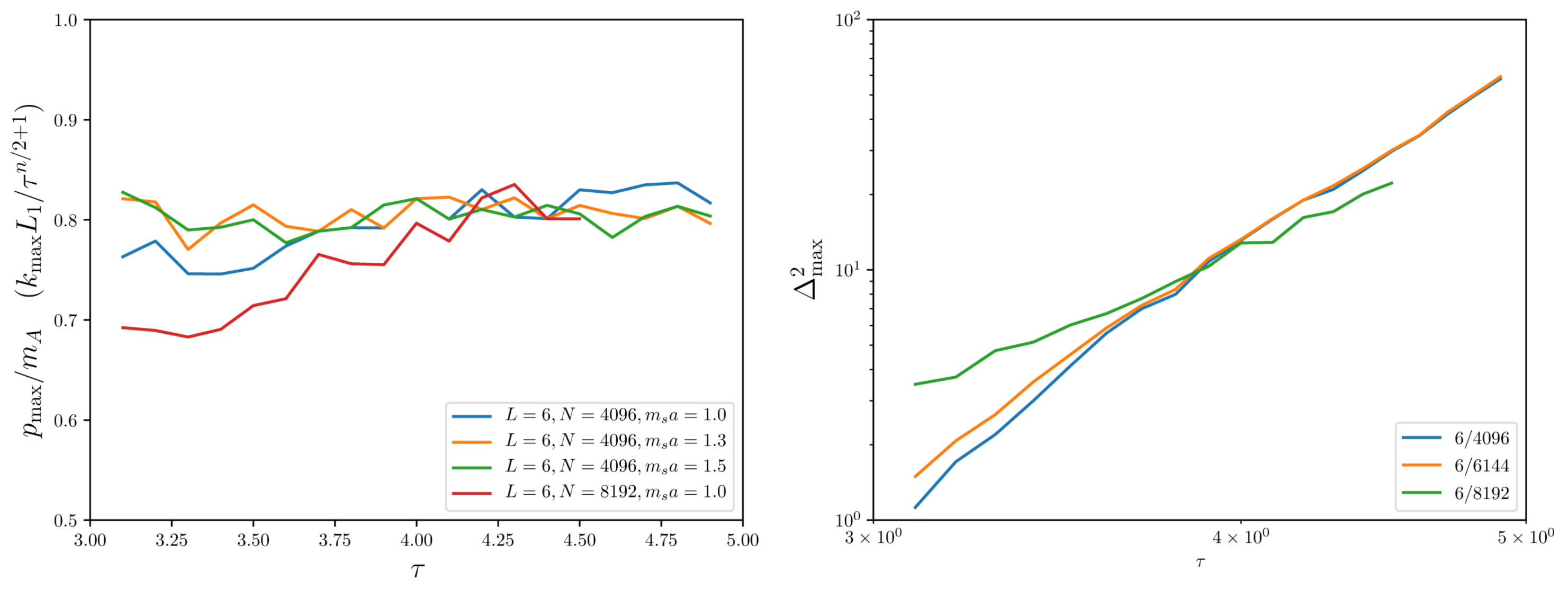}
\caption{Position (left) and height (right) of the axiton peak in the power spectrum for several simulations. The position of the peak coincides with the physical wave number $p_{\rm max} \simeq 0.8m_A$, whereas the peak height depends on the simulation resolution ($a$, expressed in units of $L_1$, shown in the lower right legend).}
\label{spectrumPeak}
\end{center}
\end{figure}

The most notable feature of the spectrum's evolution is the development of the axiton peak. The peak position, denoted by $k_\axit$, matches to good precision the axion mass. Indeed, we find $k_\axit/R=0.8 m_A$
as a function of time for all our simulations. This relation is shown in Fig.~\ref{spectrumPeak}, where we also show how the peak height grows as a function of time. 

\begin{figure}[tbp]
\begin{center}
\includegraphics[width=0.95\textwidth]{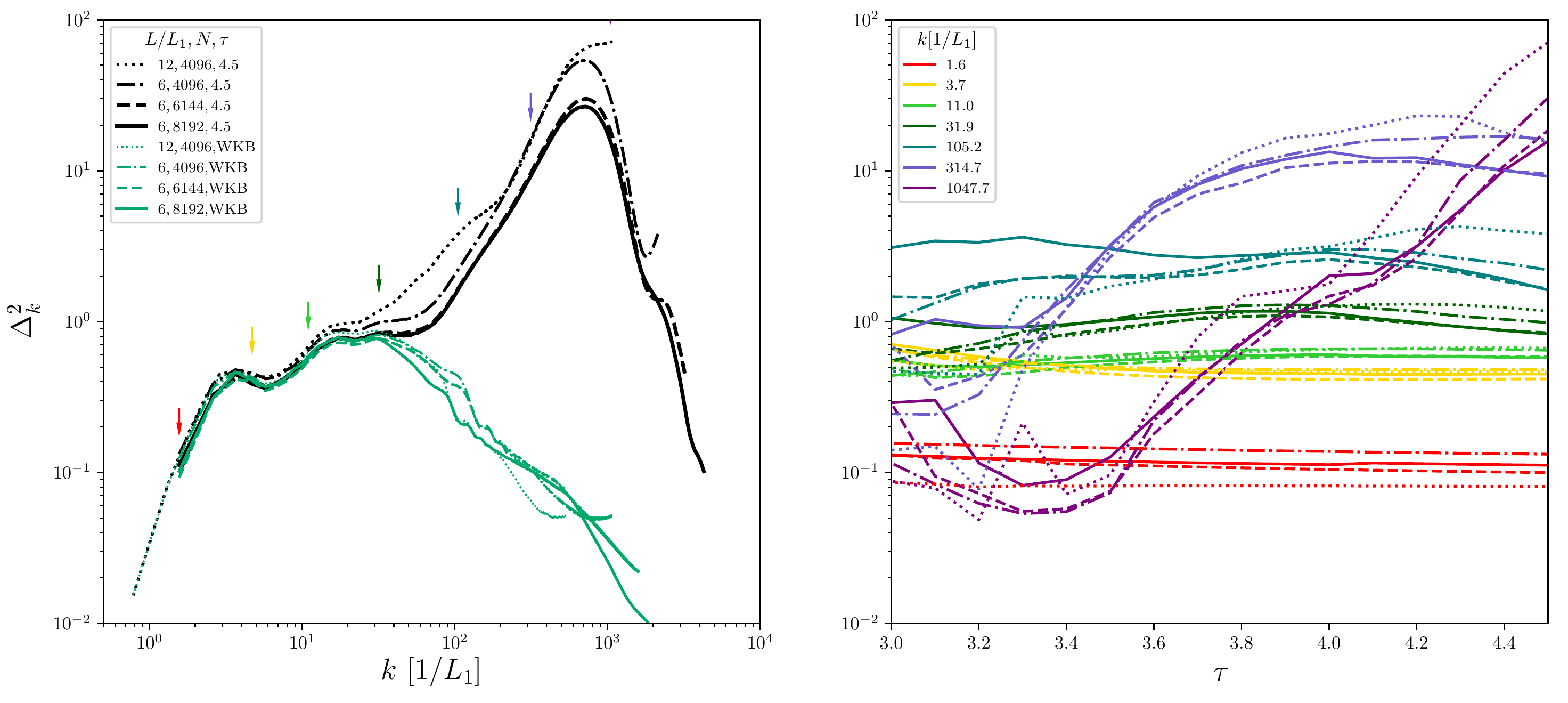}
\caption{On the left, the power spectrum at conformal time $\ct = 4.5$ before (black) and after (green) the WKB smoothing up to $\ct = 6.0$. On the right, the evolution of axion density fluctuations for a few
modes and grid spacings. The modes we show on the right are marked with coloured arrows on the left plot. The different spacings are marked with dots ($a= 12/4096$), dot-dashes ($6/4096$), dashes ($6/6144$)
and solid lines ($6/8192$).}
\label{axievol}
\end{center}
\end{figure}

Finally, to illustrate the dynamics of the individual modes, we show the time evolution of $\Delta_k$ for a few of them in Fig.~\ref{axievol}. Starting by the lowest-$k$ modes, we see once more than the late evolution of long modes is already quite frozen by $\ct \sim 3$. At higher-$k$, starting around $k\sim 30/L_1$, we see a more interesting and peculiar trend. In general, $\Delta_k^2$ seems to decrease at the beginning, then increase and then decrease again. The contrast of mode $k\sim 1000/L_1$ is indeed increasing very fast. A mode reaches its
maximum when its comoving wave number matches the axion mass, that is when $k/R\sim m_A$. The interpretation is that the peak is due to axitons, as already advanced, and their emission of relativistic axions
with momenta $k/R\sim m_A$. As the axion mass increases with time the peak moves towards high-$k$. Once the peak has gone through a given momentum $k$, no more axions of that frequency are emitted (or only a
few) and the already emitted free-stream so their interference patterns and their density fluctuations decrease.

We also note an $\mathcal{O}(1)$ dependence on the lattice spacing, that makes the density fluctuations less severe as the grid spacing approaches the physical limit $a\to 0$. The height of the peak might
reveal also some sensitivity to the initial conditions, which for the $N=6144$ series was slightly more overdense than the $N=8192$ series.

In the next section we discuss axitons~\cite{Kolb:1993hw,Kolb:1994xc} and their presence in our simulations.

\section{Axitons}\label{sectionaxitons}
Besides the smallness of the string tension, axitons are the largest source of uncertainty to the final density contrast of our simulations. Therefore, we devote this section to establishing a better understanding of axitons based on analytical considerations and their appearance in our numerical simulations. Based on these findings, we argue that only the earliest axitons contribute significantly to density perturbations on the typical length scales of miniclusters. This result was already exploited in the previous section alongside with our late-time WKB approximation procedure, which we detail here as well. 

\subsection{Axitons in an expanding FRW geometry}
The Sine-Gordon equation in 3D has an instability that drives the collapse of non-relativistic axion lumps until they reach a physical size of $\sim 1/m_A$, become non-linear and relativistic and decay after a few oscillations by emitting semi-relativistic axions. This is in part due to the periodic feature of the potential, which requires negative self-interaction terms. In the case of the cosine potential \eqref{potential} we have, 
\be
V_{\rm QCD} = \chi(1-\cos\theta)=\chi\(\frac{\theta^2}{2}-\frac{\theta^4}{24}+...\). 
\ee
The negative interactions imply an attractive force, which can overcome the positive pressure due to the axion field gradients and drive the collapse of axion lumps~\cite{Levkov:2016rkk,Eby:2016cnq}. Since the dynamics have been analysed in many several references~\cite{Eby:2016cnq,Schiappacasse:2017ham,Visinelli:2017ooc} in a static, non-expanding background, let us entertain the case of interest, the expanding FRW. 

Consider a lump of non-relativistic axions,
\be
\theta(t,\veca x)= \Theta(t)\cos\(\int^t m_A(t) dt\)  \[\sqrt{2} e^{-\pi\frac{|\veca x|^2}{2\sa^2}}\], \quad \int dV \theta^2 = 2\cos^2\(\int m_A dt\)\sa^3,
\ee
where we allow the axion mass $m_A$, the comoving size $\sa(t)$ and the amplitude of the envelope $\Theta(t)$ to depend on time. The axion field oscillates coherently at a frequency very close to the axion
mass. In the absence of pressure or self-interactions $\sa$ will remain constant and the amplitude will decrease in time.

The energy of the configuration including the first order self-interaction is
\bea
U &=& R^3\int dV \(\frac{f_A^2}{2}(\dot \theta)^2+\frac{f_A^2}{2R^2}(\nabla \theta)^2
+\chi \(\frac{\theta^2}{2}- \frac{\theta^4}{24}+...\) \) \\
&=& M + \frac{3\pi }{4} f_A^2 \sa R \Theta^2  - \frac{1}{32\sqrt{2}}\chi \sa^3 R^3 \Theta^4\,,
\eea
where the kinetic terms and the first potential terms have been combined into the total mass $M$
\be
M= \chi \Theta^2 \sa^3 R^3\,,
\ee
and we have used time averages, i.e. $\cos^2()\to 1/2$ and $\cos^4()\to 3/8$, in the gradient and self-interaction term respectively. Using the conservation of the number of axions, $N$, we have $M=m_A N$, which is accurate up to high-order interaction terms. The energy can then be divided into mass, gradient and self-interactions as
\bea
\label{energylump}
U &=& U_M + U_g  + U_{si} = m_A N +  \frac{3\pi }{4} \frac{N}{m_A \sa^2 R^2}  - \frac{1}{32\sqrt{2}} m_A N \Theta^2.
\eea
Keeping $N$ constant, the gradient energy will be minimised by expanding the lump ($\sa$ increases) while the self-interaction energy is minimised by increasing $\Theta^2$, which requires a contraction. Since $N\propto \sa^3 \Theta^2$, both cannot be achieved at the same time. Indeed, for a given $N$ and $\sa$ there is a critical value of $\Theta$,
\be
\label{critical}
\Theta_c = \sqrt{24 \pi \sqrt{2}} \frac{1}{m_A \sa R}\simeq 10.3 \frac{1}{m_A \sa R}.
\ee
Below this amplitude our lump will \emph{tend} to diffuse away and above it to contract. In an expanding lump the gradient pressure decreases and therefore the lump size will freeze  asymptotically in comoving coordinates. On the other hand, the contracting lump will see its self-interaction term becoming more negative, thus driving a faster compression.

In the expanding Universe $\Theta$ decreases in time to keep $N$ constant as $m_A R^3$ increases. Indeed in the adiabatic approximation we can write
\be
\Theta = \Theta (t_i)\sqrt{\frac{m_A(t_i) R^3(t_i)}{m_A R^3}}\,,
\ee
and therefore both $\Theta$ and the critical amplitude $\Theta_c$ decrease with time. Which one decreases faster is controlled by $\n$, the index of temperature dependence of the axion mass,
\be
\label{thetathetacrit}
\frac{\Theta(t)}{\Theta_c(t)} = \left.\frac{\Theta}{\Theta_c}\right|_{t_i} 
\sqrt{\frac{m_A}{m_A(t_i)}\frac{R(t_i)}{R}} \propto \ct^{\frac{\n-2}{4}} \,.
\ee

When the axion field becomes non-relativistic in the early Universe the axion mass is increasing at a furious pace, $\n\sim 7$ so the critical field decreases slower than the amplitude due to the redshift and a
stable lump can become unstable. For an axion-like particle with constant mass, $\n=0$, this cannot happen. Moreover, when the temperature has dropped below $T_c$, the axion mass saturates and $\n\to 0$ even
for the QCD axion. Under such circumstances, the expansion of the Universe will eventually beat self-interactions for good.

Since our simulations have $\n=7$ and do not reach the saturation of the axion mass, more and more regions could in principle become unstable. However, entering the instability region does not imply the
immediate and complete development of the axitonic instability (collapse) because there is a time associated with its development, and that time-scale is increasing in time (the corresponding term in
\eqref{energylump} decreases as $1/R^3$ when number conservation $N=m_Af_A^2\Theta^2 \sa^3 R^3$ is taken into account).

The time scale of the collapse can be easily estimated in the non-relativistic approximation by using the attractor solution obtained in~\cite{Levkov:2016rkk} in a non-expanding Universe where the axion mass is
constant. To do this we note that in conformal time and conformal coordinates, our axion equation is essentially the same as the one used in~\cite{Levkov:2016rkk} but with a time-dependent mass. Thus, their
solution for the increase of the central amplitude in the limit of a large radius should be valid locally. We estimate the characteristic conformal time of the collapse as
\be
\label{wavecollapseestimate}
\ct_* = \(\frac{\partial_\ct \varrho(0)}{\varrho(0)}\)^{-1}=\frac{|3.99|^2}{|\bar\Theta|^2}\ct^2\,,
\ee 
(in ADM units) where $|\bar\Theta|^2=\Theta^2 \ct^{\n/2} \ct^{3}$ is essentially the number of axions in a comoving volume, which would be conserved in absence of self-interactions, expressed as the amplitude squared that the axion field would have at  $\ct=1$. Recall from Sect.~\ref{section:axionspectrum} that the typical value is $\langle \theta^2\rangle\sim 1.7/m_A\ct^3$, so the collapse time scale is larger than $\ct$
for the typical values of the amplitude. Therefore, only the regions with largest values of $\Theta$ will be driven fast enough to collapse and will preferably do it at early times. Since the gradient pressure
tends to dominate at early times, we expect that the lumps will collapse if $\ct_*<\ct$ at the time when a given lump size becomes over critical, i.e. when  \eqref{thetathetacrit} becomes of the order of 1.

We could build an expectation of the number of collapses from the spectrum of axions by assuming Gaussian fluctuations, for instance, but our first attempts have not been very fruitful due to the necessary $\mathcal{O}(1)$ coefficients involved in the collapse. In any case,  our simulations show clearly that some regions become unstable and collapse. The number of collapses seems to increase in time, which is not
straightforward to understand. In principle, we believe that, given that the time scale for the collapse increases in time, it is most likely that lumps that can collapse do so relatively fast, and then we would expect a decrease of the collapses. However, when a region collapses, the amplitude around its core changes strongly, becoming $\theta\sim 1$. Such a region is the prone to continue re-collapsing and can help nearby lumps, which otherwise might not have collapsed, to collapse in an assisted manner. The previous argument suggests that the number of collapsing lumps should increase very fast, but only around the regions of the first collapses. Indeed, this seems to be the case, most of the new axitons appear very close to the first ones, which implies some complicated, non-linear interactions between them. This is material for a dedicated publication. In the following we discuss only the basic features of axiton dynamics and their relevance to the formation of axion miniclusters.

The instability drives the compression of the lump until it is quenched by the core radius becoming of the order of the Compton-wavelength $\sa R\sim 1/m_A$. In the non-expanding Universe a self-similar solution for the collapse can be found and it is well understood~\cite{Levkov:2016rkk}. The collapse stops because the self-interaction energy cannot be larger than $2 m_A^2 f_A^2$ (including now the full cosine) but the gradient pressure $\sim f_A^2/(\sa R)^2$ can grow indefinitely large by compressing the lump. Equating both estimates shows that the instability halts precisely at a radius of order $(\sa R)^2\sim 1/m_A^2$ where the potential energy saturates the QCD potential. Thus the core of the lump is expected to reach energy densities $\sim \chi$. In fact, taking into account the gradients, even
larger densities can be predicted. We indeed observe these in Fig.~\ref{dpddelta} and in our plots of $\Delta^2_k$ (Figs.~\ref{dimvar} and \ref{dimvar_summary}) by the fact that these spectra have power even above the axiton peak.

Because of the large amplitude of the field, the axion field dynamics in the core becomes completely non-linear. The nonlinear configuration at the core of the collapsing lump was called ``axiton" by Kolb and Tkachev in~\cite{Kolb:1993hw,Kolb:1994xc}. Axitons are strongly related to \emph{pseudo-breathers} or \emph{oscillons} of the Sine-Gordon equation, if not the same thing. To give such a good name a concrete meaning, we underline a clear and crucial difference. While the field in pseudo-breathers oscillates with ${\cal O}(1)$ amplitude only a few times, our non-linear cores are actually quite resilient. Indeed, it seems that oscillations can last as long as the axion mass continues growing as a sufficiently large power of time. Thus we define axitons to be the longer-lived oscillons of the Sine-Gordon equation (or similar equations) when the field mass increases with an index larger than $\n=2$. Since at $T\simeq T_c=150$  MeV, the axion mass becomes constant, axitons will become pseudo-breathers and quickly decay shortly after that temperature. Moreover, the axion self-interactions below $T_c$ turn out to be slightly smaller than those implied by the cosine~\cite{Bonati:2013tt,Bonati:2015vqz}.

Axitons and pseudo-breathers suffer violent oscillations of the axion field in their cores, pulled by self-interactions and pushed out by the gradient force. This produces mildly relativistic spherical axionic waves (axions) that escape the core is small bunches\footnote{The emission of axions can be understood in terms of Feynman diagrams as interactions of order $\theta^6$ and higher are fusing non-relativistic axions into more relativistic axions~\cite{Braaten:2016dlp}. {The emitted spectrum has visible peaks over clear continuum~\cite{Levkov:2016rkk}.}}. This energy loss is what makes pseudo-breathers unstable, but it is not enough to entirely blow up the axitons. Qualitatively, it is quite clear that axitons are so resilient due to the fast increase of the axion mass.

Once a region surpasses the amplitude of the instability, $\Theta>\Theta_c$, its core can collapse but the whole region cannot go back to the stability region due to the usual expansion of the Universe because $\Theta_c$ decreases slower than $\Theta$  (and $\Theta$ is even increasing in the core and surroundings). The energy loss in axions is not very efficient. Similarly, for pseudo-breather it takes $\sim 100-1000$ oscillations to relax, see~\cite{Salmi:2012ta} and references therein. Therefore the region around the core remains largely overdense and prone to re-collapse. The fact that the radiated axions have their mass quickly increased in time also reduces their free-streaming length. This effect points in the same direction, not allowing the lump to get ride of its excess amplitude, although it is probably sub-dominant (at least at late times, see below).

In our simulations, discretisation effects might be playing an important role in the formation and collapse of axitons, as one can deduce from the right pane of Fig.~\ref{spectrumPeak}. Our lattice spacing limits the maximum momentum which axions radiated by the axitons can have to $\sim \pi/a$. Coarser lattices result in lower and less efficient energy radiation and therefore more resilient axitons. Moreover, close to the Nyqvist frequency, the phase velocity of relativistic waves in a Cartesian grid vanishes, which means that axion waves travel very slowly and are slowed down even more by the attractive self-interactions. If we assume that the discretisation effects show as soon as $m_A\sim k_{\rm Ny}/4$, the maximum reasonable value of $\ct$ the simulations should reach becomes $\ct_{max}\sim (k_{\rm Ny}/4)^{\frac{2}{n+2}}$. For the typical values of our simulations $L_1 = 6$ and $N = 4096$, this results in $\ct_{max} \sim 4$, whereas the $N = 6144$ and $N=8192$ cast $\ct_{max} \sim 4.4$ and $4.7$ respectively. Indeed, we see resolution effects near the axitons in the $N=6144$ simulations of Fig.~\ref{picsWKB} of the next section. Resolution effects alone, however, can not totally explain why the height of the axiton peak in Fig.~\ref{spectrumPeak}  does not converge better between the $N=6144$ and $8192$ simulations. We will come back to this issue in the next section.

In reference~\cite{Kolb:1993hw}, spherically symmetric axitons where simulated and shown to be resilient until the axion mass saturates, and even a bit further. In a more recent work~\cite{Levkov:2016rkk}, the collapse was studied in the constant-mass case starting from a gravitationally bound lump over the critical density~\eqref{critical}. The behaviour of the central amplitude in both these simulations is quite similar, despite the fact that in the former case the potential used was $\sim 1-\cos\theta$ (relevant at high $T$) and the latter reference uses the $\chi$PT potential relevant at temperatures below $T_c$ \cite{DiVecchia:1980yfw}. It is interesting to note, thought, that most of the ``axiton" evolution showed in reference~\cite{Kolb:1993hw}, for instance in their Fig.~8, took place after the mass was artificially saturated at $\ct=3.5$. Thus, the axitons of~\cite{Kolb:1993hw} were more precisely pseudo-breathers or oscillons. We will keep the name ``axiton" for the more stable solutions when the axion mass increases very fast.

Our analysis shows that there is a critical exponent of the axion mass, namely $\n=2$, above which an over-dense axion field lump cannot escape the instability region. If the region collapses into an axiton, it
will be resilient. Below $\n=2$ we expect the regions to slowly exit the instability and only transitory pseudo-breathers appear, having their fun for a little period before their inevitable last flash.

On a different note, let us remark that axitons can also appear in the pre-inflationary scenario. In such case the initial conditions of the axion field are homogeneous at $\theta_I$ except for small model-dependent fluctuations (that could have quantum or thermal origins). If $\theta_I\simeq \pi$, these fluctuations are unstable and will grow around $\ct\sim1$ when the axion field starts rolling its potential. This growth can result in O(1) inhomogeneities producing a large number of axitons. We will study this scenario in a different publication.

\subsection{Axitons in the power spectrum}

The existence of axitons is related to the gradient pressure and to the competition between kinetic $\sim p^2f_A^2$ ($p$, physical momentum) and potential $\sim \chi=m_A^2f_A^2$ energy, i.e. between $p^2$ and $m_A^2$. At early times, axions are relativistic, and axitons can not form. This explains why we do not see axitons at $\ct\sim 1.5$ when the low momentum axions become non-relativistic but high momentum axions still dominate the energy density. The first recognisable objects appear around $\ct \sim 2.5$, although the shade of the axiton peak can be traced back to $\ct\sim 2$ in Fig.~\ref{dimvar} (centre). Moreover, there is a trend, quite clear in Fig.~\ref{axievol}, to develop more axitons in simulations with larger lattice spacing, which happen to have less radiation too. This implies that the number of axitons can be sensitive to the axion spectrum and the ``initial conditions" of our simulation.

\begin{figure}[tbp]
\begin{center}
\includegraphics[width=0.45\textwidth]{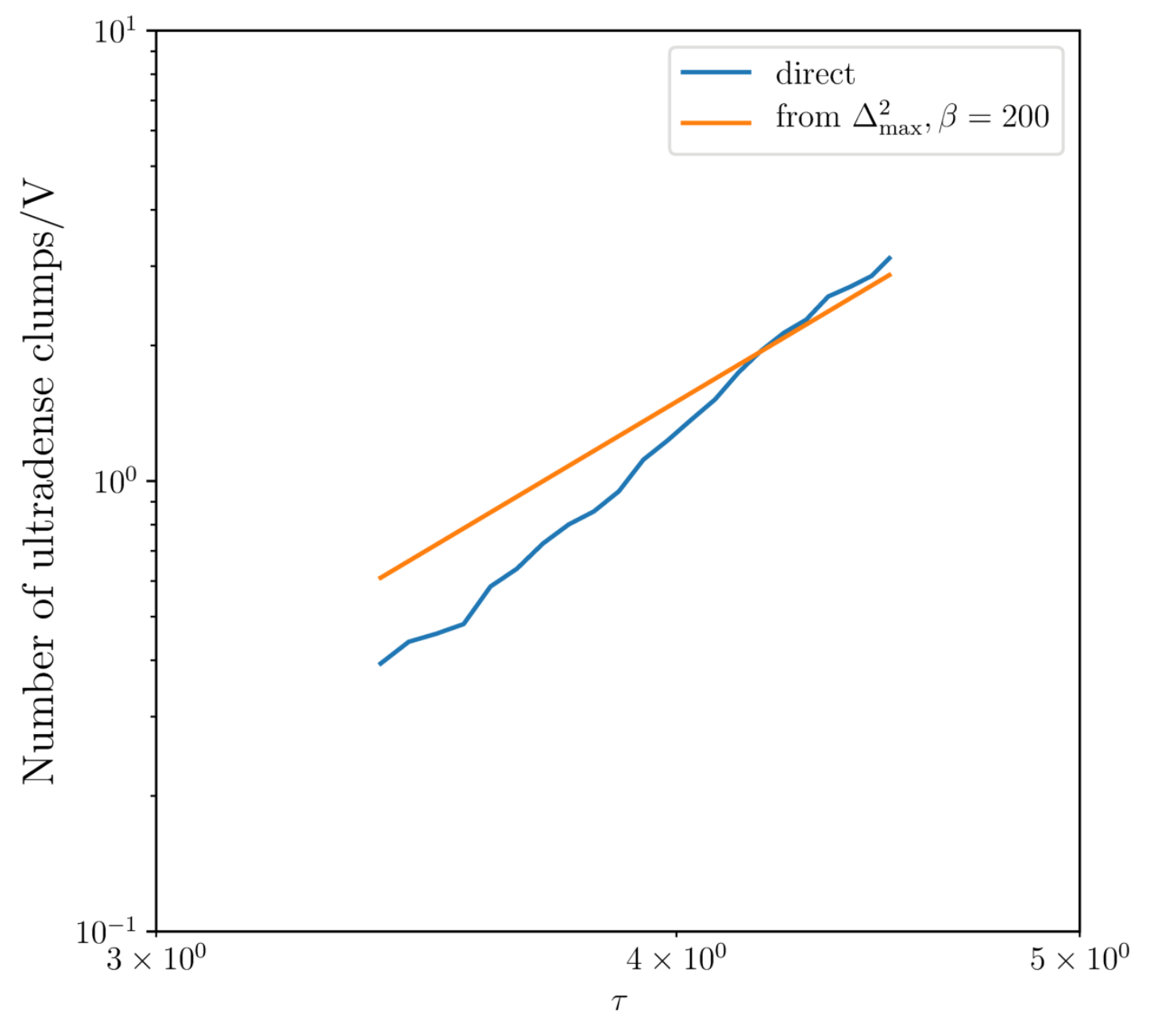}
\caption{Number of axitons per unit volume in a simulation $L=6L_1$, $N=8192$, $\n=7$ and $\ms a=1.0$ (blue) compared to the estimate given by~\eqref{axitonnumber} with $\beta=200$ (orange), as function of $\ct$.}
\label{figAnumber}
\end{center}
\end{figure}

The peak in the axion density fluctuations at high overdensities, c.f. Fig.~\ref{dimvar} and \ref{dimvar_summary}, can be attributed to axitons and the spherical axion waves radiated by them. We can try to describe quantitatively the contribution of the axitons to the variance of density fluctuations as the contribution of a few isolated clumps of ultra-high density, in a similar way that we did for the misalignment patches of size $\sim L_1$ in Sec.~\ref{patches}. In analogy to that case, we expect a white noise contribution to $\langle|\delta(k)|^2\rangle$ until a cut-of given by $p\sim m_A$. In the plot of $\Delta_k^2$ this translates into a $k^3$ power law until $k\sim m_AR$, which is not very far from what  we observe. The axionic waves radiated by axitons keep their comoving momentum as the Universe expands,
and thus contribute mostly to the value of $k$ corresponding to the axiton size when they were radiated. As time evolves, the axions free-stream and they become increasingly more non-relativistic. This diffusion process damps the amplitude of the waves and spreads them in comoving space so their contribution to the power spectrum decreases. This reproduces qualitatively the trend we see in our study of the time-evolution of $\Delta^2_k$ in Figs.~\ref{dimvar} and~\ref{axievol}.

In this simple picture, the number of axitons could be be estimated by the position $k_\axit$ and height $\Delta^2_\axit$ of the peak, because they seem to dominate the variance there. The obvious differences with the misalignment patches case are: 1) the number of cores is increasing in time as new axitons form, 2) their core shrinks (the cut-off increases) and 3) the bulk of the simulation mass is \emph{not} in the axiton cores. Another not so trivial difference is that axitons tend to cluster in regions of already large overdensity and that their radiated waves can interfere constructively and create even more axitons. These effects are spectacularly obvious in the deep zoom shown in Fig.~\ref{picsWKB} (left), which displays a $\delta^2$ projection plot at the end of our simulation, deliberately pushed to the resolution limit.  

For the moment, we neglect correlations between the individual core's positions and between cores and the average density surrounding them. Using the clump estimate for $\delta(k)$ of \eqref{delta2gen} with
$N_c,M_c\to N_\axit,M_\axit$ and Eq.~\eqref{Delta2def} we find,
\be
\label{numbi}
\frac{N_\axit}{V} \sim  2\pi^2 \Delta^2_\axit \frac{(M_t/V)^2}{k_\axit^3 M^2_\axit}.
\ee
We can estimate the axiton core density as $\rho_\axit$, its radius as $1/m_A$ and thus its mass as $M_\axit\sim 2\rho_\axit/m_A^3 \simeq 2f_A^2/m_A$. Reference~\cite{Levkov:2016rkk} shows that the axiton core density can exceed this naive estimate by a factor of 50, see Fig.~2 there. Also our Fig.~\ref{dpddelta} shows densities much larger than the top of the QCD potential. We parametrise our uncertainty in the core energy as $M_\axit= 2\beta f_A^2/m_A$ with $\beta$ a parameter larger than 1. The total DM mass in $V$ is $M_t \sim m_A \times 8 H_1f_A^2 V/L_1^3$, see Sec.~\ref{spectrum}, and our results of Fig.
\ref{spectrumPeak} give the height of the peak $\Delta^2_\axit$. Using that $k_\axit/R  \sim 0.8 m_A$ (recall $R=\ct/H_1L_1$), Eq.~\eqref{numbi} becomes
\be
\label{axitonnumber}
\frac{N_\axit}{V}
\sim 2\pi^2  \Delta^2_\axit \frac{125 m_A^4}{k_\axit^3}\frac{1}{H_1L_1^3}\sim
2.5\times 10^3 \Delta^2_\axit \ct^{\n/2-3} \, \frac{1}{L_1^3}\,.
\ee
Thus, besides a correction factor of $\ct^{\n/2-3}$ --which accidentally is not very steep for $\n=7$-- the peak height would be proportional to the number of axitons per $L_1^3$ volume. Fig.~\ref{spectrumPeak} reveals that the number should be growing at  a furious rate, $\Delta_\axit^2 \sim 12(\ct/4)^5$ for $N=8192$ simulations ($\propto \ct^{8}$ for $N=4096,6144$).

We have analysed one $N=8192$ simulation to study the evolution of the number of axitons as a function of time. As a first approach to the problem, we have saved projection plots of the density squared at
different times. To extract the number of axitons we searched for clusters in which the density squared exceeds the maximum value observed on the grid, divided by $1000$. We note that the number of clumps found
is not sensitive to the exact threshold. Our results are shown in Fig.~\ref{figAnumber}, where we compare them with the value estimated from the variance~\eqref{axitonnumber}. The time dependence is not exact
and the required $\beta\sim 200$ seems relatively high. There are a number of effects which could explain the discrepancy but are difficult to account for: perhaps the core mass estimate needs to include
partially the surrounding axion cloud, perhaps the correlations correlations among the axiton positions are more relevant at late times, or perhaps the interplay with the relativistic axions radiated by the
cores has to be taken into account. Further work along these lines might be needed to understand the evolution of the axion field in the high dense regions where axitons tend to cluster.

A word of warning to top the discussion up. We have shown that the peak height in our simulations depends on the lattice spacing. Therefore we are not sure that we are measuring the correct axiton creation rate in our grids. However, there is a trend for the $\Delta^2_\axit$ peak height to grow slower as the lattice spacing decreases visible in Fig.~\ref{spectrumPeak}. So it might well be that our axiton creation rates have not yet converged. Understanding quantitatively the axiton peak and will require further dedicated studies. In this work we aim to study the final DM distribution, for which axitons are largely irrelevant, at least at intermediate scales, as we will argue in the following.

\subsection{Axiton rings and chains}
At late times, we observe that some axitons are surrounded by a ring (spherical shell in 3D) of non-relativistic axions. We interpret {these axion fossil rings} in the next section.
  
\begin{figure}[tbp]
\begin{center}
\includegraphics[width=0.9\textwidth]{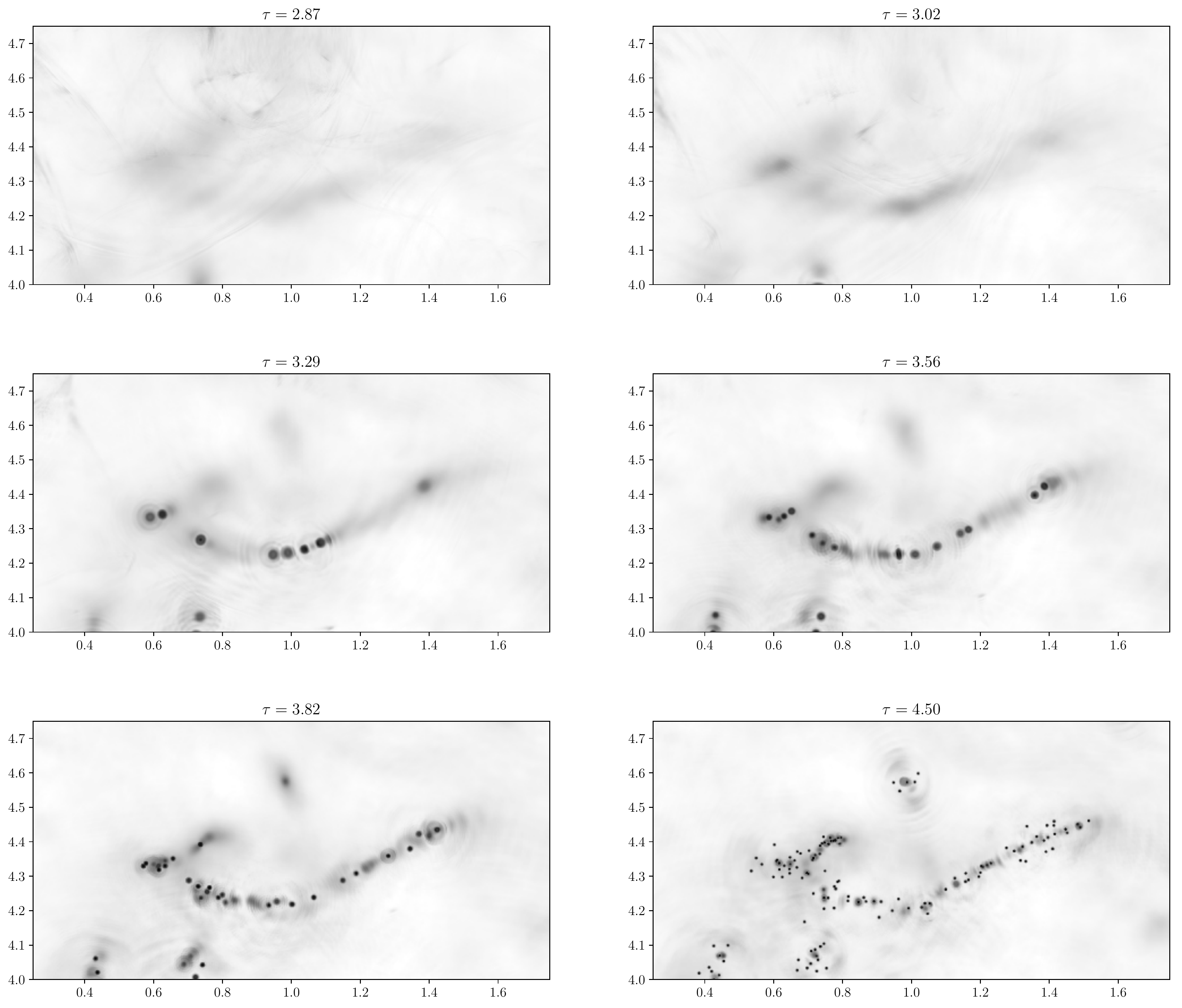}
\caption{Evolution of an axiton chain in a $L=6L_1$, $N=8192$ simulation, from the first overdensity ($\ct\sim 2.8$) up to the end of the simulation ($\ct = 4.5$). The conformal time increases from left to right and from top to bottom. The succession of events shows an increasing number of members in the chain as $\ct$ increases.}
\label{axichain}
\end{center}
\end{figure}

We also find that axitons tend to strongly cluster around each other or, more precisely, they tend to align in sequences.  Fig.~\ref{axichain} shows the evolution of such a structure, taken from a
$L=6,N=8192$ simulation. One can trace the initial overdensity back to $\ct\sim 2.8$, the first two axitons form at the positions $x\sim 0.6$ and $1.0$ around $\ct\sim 3.1$. At $\ct\sim 3.3$ more axitons have arisen from the initial overdensity and one can see how relativistic axions emitted from different members of the chain interfere. The situation at the latest time of the figure is not that clear. Many small axitons have appeared, some of them in what shows as a core at $\ct\sim 3.3$. Notice that the late, small axitons do not seem to radiate sizeably.

\subsection{A simple model for the axiton fossil rings}

A simple model can be put up to discuss the final density of axions surrounding an axiton. The idea is to consider the core as a region where mildly relativistic axions are emitted as a function of time and
follow the free-streaming of these axions from the moment of production until the axion mass saturates and the density is frozen. The physical volume of the emitting region is $V_\axit \sim 1/m^3_A$, the energy
density $\varrho_\axit \sim \chi$ and we assume that axions are emitted at a rate $\Gamma_\axit \sim m_A \beta$ with $\beta\sim \mathcal{O}(1)$ and with a mildly relativistic spectrum, with typical momentum
$p_e=\gamma m_A, \gamma\sim \mathcal{O}(1)$. The average number of axions emitted per unit time can be estimated as
\bea
\frac{dN_e}{dt_e} &\sim& \frac{V_\axit \varrho_\axit \Gamma_\axit}{\omega_{k}} \sim
\frac{m_A^{-3} (m_A^2 f_A^2) m_A \beta}{m_A} ,\\
\label{susy}
\frac{dN_e}{d \ct_e} &\sim& \frac{f_A^2 \beta}{H_1^2 }\frac{\ct_e}{\ct_e^{\n/2}}\propto \ct_e^{1-\n/2},
\eea
where we used $dt_e = \ct_e d\ct_e/H_1$ and $m_A=H_1\ct_e^{\n/2}$. The axion mass is evaluated at the time of the emission. Thus, unless $\beta$ or $\gamma$ change in time, the number of axions emitted by an
axiton is dominated by the \emph{first bursts} with small $\ct_e$,
\be
N_\axit \sim \frac{f_A^2}{H^1_1} \frac{2\beta}{(\n-4) }\frac{1}{\ct_e^{\n/2-2}}.
\ee

\begin{figure}[tbp]
\begin{center}
\includegraphics[width=0.4\textwidth]{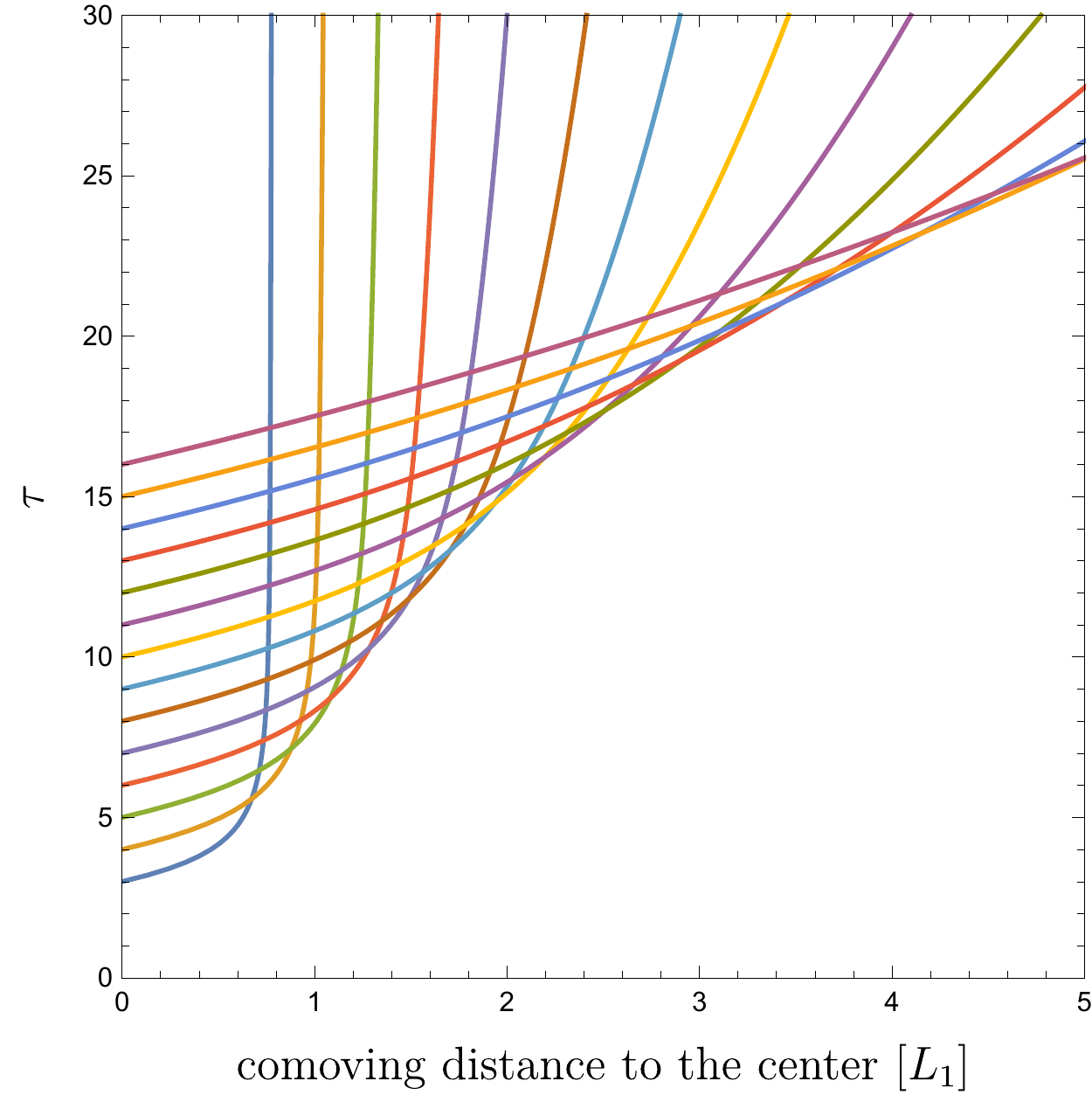}
\caption{\small Comoving space-time diagram showing how axion bursts (wave-fronts represented by coloured lines travelling away from distance=$0$) emitted at different times diffuse away of an axiton core. We
have assumed $\gamma=0.2$.}
\label{axitodel}
\end{center}
\end{figure}

The distribution around the axiton can be traced by assuming that the emitted mildly relativistic axions free-stream a comoving distance,
\be
r(\ct,\ct_e) = \int_{\ct_e}^\ct d\ct \frac{p}{\sqrt{p^2+m_A^2}} = \int_{\ct_e}^\ct d\ct \frac{\gamma}{\sqrt{\gamma^2+(\ct/\ct_e)^{\n+2}}}
\ee
where the physical momentum is $p=p_e R_e/R\simeq p_e \ct_e/\ct$ and we parametrise the emitted momentum by $\gamma$. At emission we have $p_e=\gamma m_A(\ct_e)=\gamma H_1\ct^{\n/2}_e$, which corresponds to a
comoving momentum $k_e = \gamma \ct_e^{\n/2+1}/L_1$. A scheme of the axion trajectories is shown in Fig.~\ref{axitodel}, for which we have switched the axion mass growth from $\n=7$ to $\n=0$ at $\ctc=16$. The
trajectories show a number of interesting features, which we discuss in the following.

First, axions emitted at early times are softer and suffer the most from the increase of the axion mass, so they travel a much shorter distance. These are the axion waves we see in the simulations, and they can contribute to the power spectrum at intermediate-$k$. As time evolves, the axiton core shrinks and the emitted axions are much harder. Those late and hard axions are not only less numerous according to \eqref{susy} but they are also more widespread. Second, the early slow axions seem to accumulate over a characteristic region, where the trajectories form a sort of caustic. This is a dynamical effect due to the adding up of early emitted slow axions and the late time fast burst. When the axiton stops flashing the caustic disappears. After that time the axion distribution at a distance $r$ around the axiton can be
build as
\be
\left.\frac{dN}{dr}\right|_\ct = \frac{dN_e}{d \ct_e} \(\frac{d r}{d\ct_e}\)^{-1},
\ee
which can be also easily extended to include the distribution of radiated momenta.

However, the distribution of radiated momenta already becomes very clear from Fig.~\ref{axitodel}. At times beyond the axion-mass saturation ($\ct>16$) the early axion bursts are closer to the axiton and very numerous. This is both due to their more abundant emission and their slower initial velocity, but the effect is enhanced in 3D when we account for the $1/r^2$ dilution. The distribution of axions around the axiton shortly after its formation almost resembles the final one. The harder axions emitted span a much larger region of space. Note also that when axitons flash for the last time, the core momentum is still of order $1/m_A$, so they resemble very much the last radiated relativistic axions, and thus they are expected to diffuse away very efficiently.

The outcome  of this simple model is two-fold. First, it inspires to build the final distribution of axions just by simply free-streaming the axions after the first relevant axitons have already flashed their first axions. This is precisely what we do, using a WKB approximation discussed in the next section. Second, it predicts a collection of spherical overdensities around the first axitons. We think that these overdensities, clearly seen in our final density maps, contribute to the relatively large density fluctuations at intermediate scales $k\sim (6-100)L_1^{-1}$. The extreme simplicity of the model disregards the very important effect of anharmonicities/self-interactions close to the core. Therefore the qualitative picture has to be considered quite uncertain close to the core, and most likely
cannot be refined to discuss the axion fossil distribution around the core without a proper understanding of anharmonicities/self-interactions.

Note that some of the quantitative aspects of axion emission from axitons are not yet fully understood. In principle we do not know whether the emission corresponds to a net increase of the number of axions in our simulation box or to a decrease. In a non-expanding Universe, axitons clearly convert low-$k$ axions into relativistic ones. In an expanding Universe the situation is more complicated as the equation of state of the core is not quantitatively understood. Attractive self-interactions correspond to a negative pressure that makes the axiton core energy redshift slower than just decoupled matter ($\varrho\sim 1/R^3$). An extreme example would be a region where $\theta=\pi$, anharmonicities cancel the potential entirely, and the axion field in that region would not suffer redshift of its energy at all (it would behave like a slow-roll inflaton). In reality, axiton cores are quite dense and a sizeable part of the energy is in field gradients, which redshifts as radiation ($1/R^4$) so overall it is perhaps more reasonable to think that axitons will decrease the total axion number. Indeed, this is what we observed and discussed in Sec.~\ref{spectrum} when we studied the axion spectrum. Finally, let us mention that although the axion emission of one axiton is dominated by early times, the number of axitons seems to  grow so large than it could overcome this tendency globally. That is to say, late axitons tend to be irrelevant per se in fusing axions, but there could be so many as to shadow the effect of the early axitons. This is certainly not the case during the periods that we have simulated where the dark matter distribution at the largest scales in our simulation is essentially frozen and the decline of the axion number is very moderate and even seems to level-off. Moreover, extrapolating the decrease of the number of axions even to $\ct\sim 16$ appears to be a small effect. Only when we approach the resolution limit, we observe a rise of the axion number in some simulations, but at the moment we cannot be sure that this effect is physical or an artefact of the discretisation. All in all, we think that further work shall be devoted to understand quantitatively the violation of axion number due to axitons but the effects
observed in our simulations seem to be reasonably small.

\subsection{WKBing the axitons away} \label{axitonsWBK}

Our simulations must in principle end before the axiton cores become too small to be resolved by our grid. We have argued that the physics that follows is relatively simple. The axion DM density field at large and intermediate scales is largely non-relativistic and frozen by $\ct\sim 4.5$ in our simulations with $\n=7$. The low-energy axions produced by the first axitons have already been radiated and their axiton rings are mostly frozen too. The only continuing dynamics is the formation of new axitons and their radiation of very relativistic axions. We expect that this continues until $\ctc$ and a bit beyond. We have argued that the the radiated axions are so hard that they free-stream much longer than $L_1$ and do not contribute to new structures at long, intermediate and certainly not at small scales. They can be understood as a diffuse background. Our suspicion is that, as axitons die out, the remnant axions that make their profile will also diffuse away a length scale comparable to the last relativistic axions. In this sense they are not likely to remain as ultra high-density, ultra-small dark matter spots. However, their diffusion is somewhat slowed down by the axion attractive self-interactions, so it is quite likely that some interesting high density remnants are left over. The recent work of Bushmann, Safdi and Foster seems to point in this direction~\cite{buschmann}. At the moment, we can not simulate such small scales and we prefer to leave further speculation for future work.

In this circumstances, we think that the last stages of our simulations are very close to the final distribution of dark matter at large and intermediate scales. As axitons are very social beings, appearing mostly in already dense environments, the small scales in low and average density regions are probably well represented as well. By the time our simulations reach the resolution threshold some semi-relativistic axions still have to diffuse for a little while, and axitons have to eventually disappear. In order to artificially achieve this in the most ``physical" way, we continue the evolution of the axion field \emph{without self-interactions}. This can be done almost analytically and captures perfectly well the free-streaming of the axion field. In order to do this, we compute the evolution of each axion Fourier mode \eqref{modes} separately using the WKB approximation \eqref{solumode}, perform the inverse Fourier transformation and calculate the DM density field at a later, suitable time.

\begin{figure}[tbp]
\begin{center}
\includegraphics[width=0.45\textwidth]{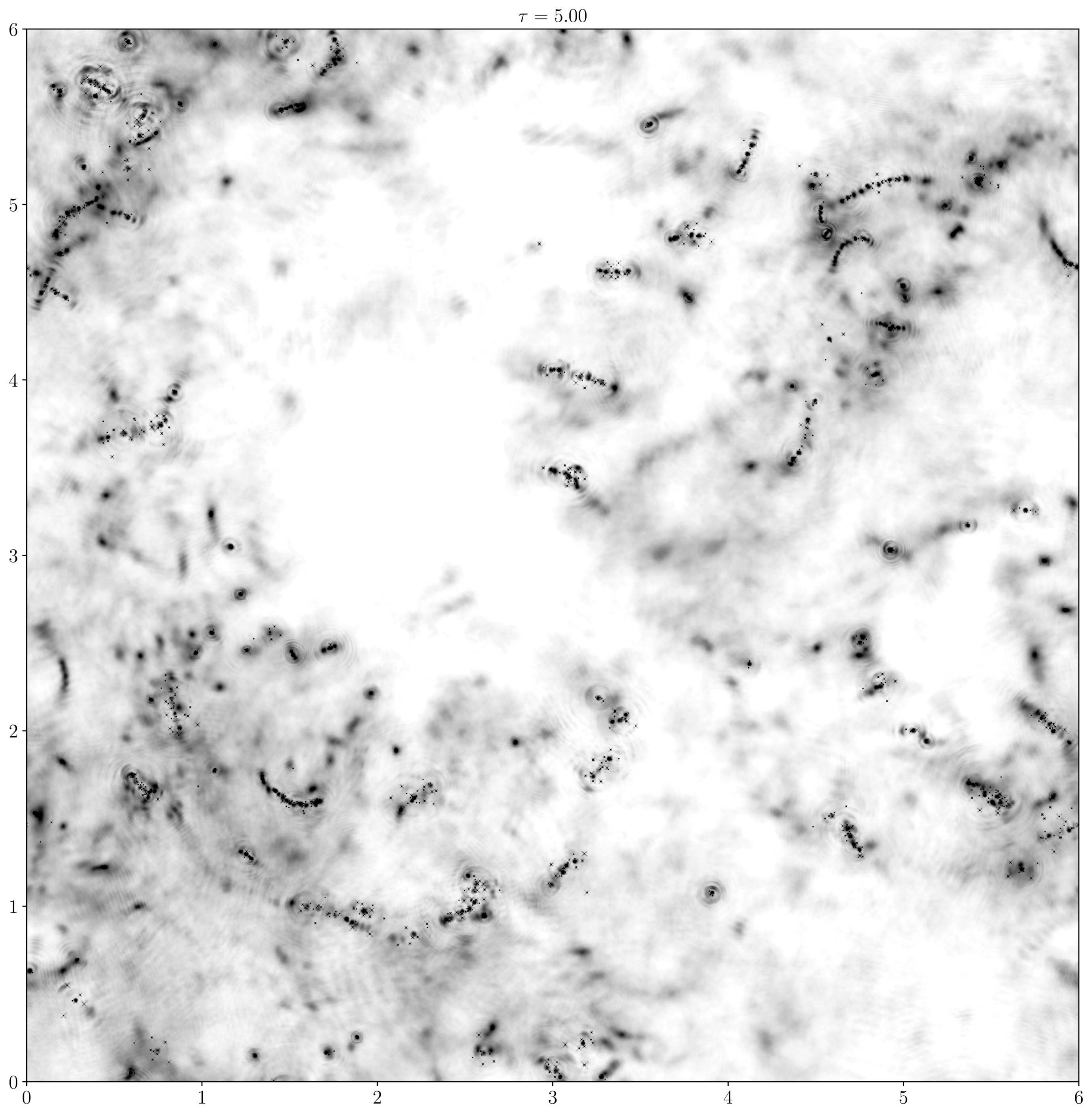}\includegraphics[width=0.45\textwidth]{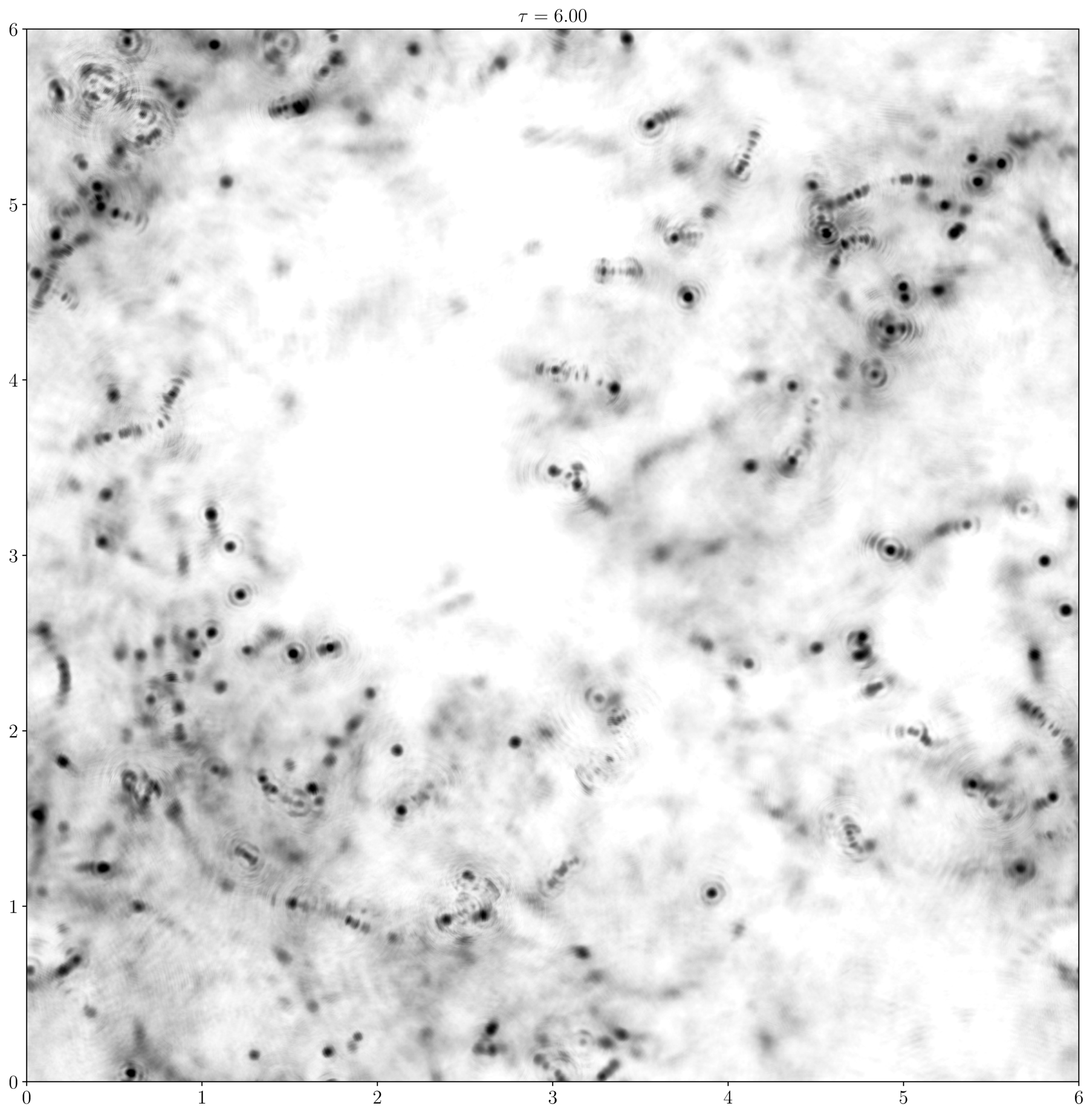}
\caption{3D$\to$2D projection plots of density squared $\int dz (\varrho(\veca x)/\bar\varrho)^2$ before ($\ct=5$, left) and after evolving the axion field with the WKB approximation ($\ct=6$, right).
At those late times only the small scale structure (smallest axitons) is diffused away.}
\label{picsWKB}
\end{center}
\end{figure}

Figure~\ref{picsWKB} (right) shows an example of the application of this WKB filter in the density-squared projection plot. Here we evolved the field until $\ct=5$, where first discretisation effects appear. The WKB is performed only until $\ct=6$, which allows free-streaming of the highest momentum axions by a length $\sim 0.7L_1$. At first sight, the differences are minute. The large and intermediate scales are essentially unchanged. The WKBed map retains the first axiton rings, their cores, and the remnants of axiton chains that were effectively frozen at the end of our simulation. However, when zooming in we discover that the small axitons are in effect dispersed away. The axion field that made the axitons is treated by the WKB evolution like a lump of axions, because self-interactions are neglected. It diffuses
away very fast but the mode amplitude is not lost. Since the WKB neglects self-interactions, a few ``axions" are artificially created by this procedure. Of course, the same error is made when we compute the spectrum of axions with formula \eqref{nk}. Since we have seen that the number of axions is conserved to a very good degree in Sec.~\ref{spectrum}, the effect of this last diffusion can be neglected too. Note that WKB evolution also cures discretisation distortions around the axitons as they contain very few axions that travel very far.

The effect on the power spectrum is well visible in our previous Figs.~\ref{dimvar} and \ref{axievol}. The large scales are left untouched, but the axiton peak is gone, as expected and desired.

The WKB procedure reveals a regime of conspicuous oscillations around $k\sim 100/L_1$ and a power-law trend beyond it. In Fig.~\ref{dimvar_summary} (right) we showed the WKBed results from two types of simulations differing mostly in the time at which the WKB smoothing is performed. We immediately recognise the $\ms/2$ cut-off of the \emph{axion spectrum}, so visible in Fig.~\ref{spectrumComp}, in the spectrum of density fluctuations, Fig.~\ref{dimvar_summary}.  Thus we can connect the density fluctuations at wave numbers larger than $k\sim 100L_1^{-1}$, with those we expect to arise from non-interacting axions with the number spectrum shown in Fig.~\ref{spectrumComp}. To check if this connection is justified we compute the density spectrum, $|\widetilde \delta|^2$, from the axion spectrum, $n(k)$. As nicely described in~\cite{Enander:2017ogx}, if the density fluctuations are uncorrelated, i.e. Gaussian with random phases, we can write the density spectrum as a convolution of the axion spectrum as
\be
\label{deltala}
\frac{1}{V}|\widetilde \delta(k)|^2  = 2(2\pi)^3 \left\langle\right.\frac{\int d^3 {\veca q}\, n(\veca q)\, n(|\veca q-\veca k|)}{\[\int d^3 \veca q\, n(\veca q)\]^2}\left.\right\rangle_{|\veca k| = k},
\ee
where we have adapted equation (3.22) from \cite{Enander:2017ogx} to our notation in the deep non-relativistic limit.

\additionalinfo{They define 
\be
\theta_{\veca k}(t)=\theta_{\veca k} f_k(t) \quad F(\veca k,\veca k') = \dot f_k  \dot f_{k'}+\(\frac{\veca k\cdot \veca k'}{R^2}+m_A^2\)f_k f_{k'}
\ee
The energy density is
\be
\varrho(\veca x)= \frac{f_A^2}{2}
\int \frac{d^3\veca k}{(2\pi)^3}
\int \frac{d^3\veca k'}{(2\pi)^3} \theta_{\veca k} \theta^*_{\veca k'}F(\veca k, \veca k') e^{-i \veca x\cdot (\veca k-\veca k'))}
\ee
its Fourier transform 
\be
\tilde \varrho(\veca q)= \frac{f_A^2}{2}\int \frac{d^3\veca k}{(2\pi)^3}\theta_{\veca k} \theta^*_{\veca k-\veca q}F(\veca k, \veca k-\veca p)
\ee
and the power spectrum, has an expectation value
\bea
\langle|\tilde \varrho(\veca q)|^2\rangle
&=& \frac{1}{V^2} \frac{f_A^4}{4}
\sum_{\veca k}\sum_{\veca k'}
\theta_{\veca k} \theta^*_{\veca k-\veca q}F(\veca k, \veca k-\veca q)
\theta_{\veca k'} \theta^*_{\veca k'-\veca q}F(\veca k', \veca k'-\veca q) \\
&\simeq & \frac{1}{V^2} \frac{f_A^4}{4}
\sum_{\veca k}
|\theta_{\veca k}|^2 |\theta_{\veca k-\veca q}|^2F^2(\veca k, \veca k-\veca q) \times 2
\eea
where the factor of two comes from the two non cancelling averages $\veca k = \veca k'-\veca q$ and $\veca k = - \veca k'$ both contributing equally. In the adiabatic non-relativistic limit, $f$ does not depend
on $k$.

We implement \eqref{deltala} with sums as
\be
\frac{1}{V}\langle |\widetilde \delta(\veca k)|^2\rangle = \frac{2 V}{\[ \sum_{\veca k}  n(\veca k) \]^2} \sum_{\veca k} n(k)\, n(|\veca k-\veca q|)
\rightarrow
\frac{2 V}{\[ \sum_{\veca k}  n(\veca k) \]^2}  \sum_i 2\pi i^2 n(i) \sum_{j=|\mathpzc{q}-i|}^{|\mathpzc{q}+i|} \frac{j}{i \mathpzc{q}}\, n(j)
\ee
where the integers represent moduli of momenta as $k = 2\pi i/L, |\veca k- \veca p| = 2\pi j/L, q = 2\pi \mathpzc{q}/L$, $n(i)$ is the occupation number for $k = 2\pi i/L$. The second sum comes from the
$d\cos$ integral of $\veca k$ phase space.}

\begin{figure}[tbp]
\begin{center}
\includegraphics[width=0.65\textwidth]{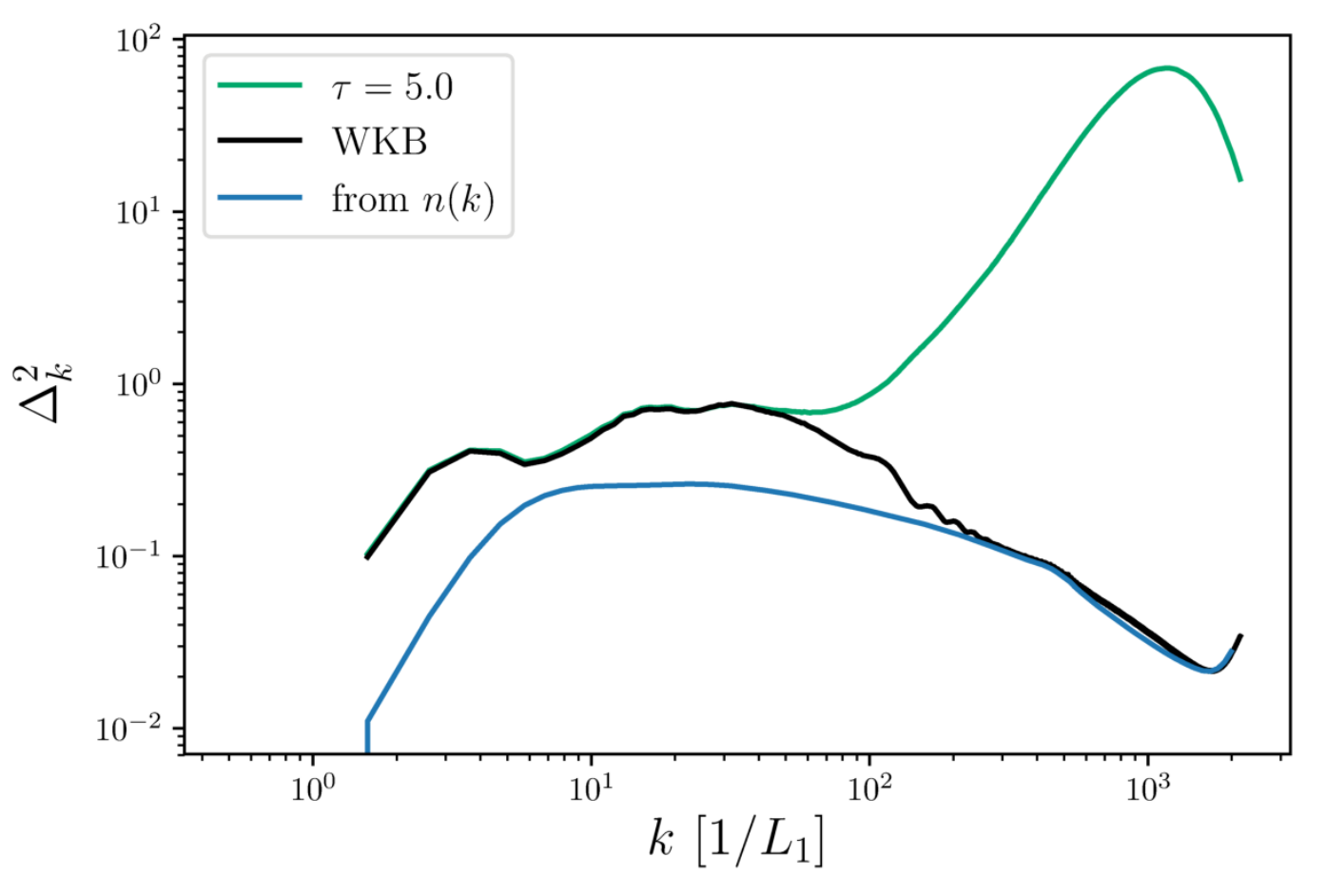}
\caption{Comparison of density fluctuations between the result at the end of a simulation at $\ct=5.0$ (green), after evolving the same data with the WKB approximation until $\ct=6.0$ (black) and the estimate given by the axion number spectrum~\eqref{deltala} with the WKBed data (blue). The results where averaged over a set of $N=6144$, $L=6.0$, $\n=7$ simulations.}
\label{pspnsp}
\end{center}
\end{figure}

In Fig.~\ref{pspnsp} we compare the density fluctuations computed directly after the WKB evolution with the estimate computed from Eq.~\eqref{deltala}. For that we use the average of a set of $L=6$, $N=6144$,
$\n=7$ simulations. The results agree very nicely \emph{in the high energy tail} beyond the $\ms/2$ cut-off ($\sim 500 L_1^{-1}$). In order to make them match so well we had to multiply the latter with a
factor of $2/3$, whose origin is unclear to us at the moment, but might be related to our neglect of the correlations among the density fluctuations. From Fig.~\ref{pspnsp} it is also clear that the
intermediate bump at $k\sim (10-100) L_1^{-1}$ and the region below are not well represented by the uncorrelated hypothesis. But in this regime the WKB evolution did not affect the spectrum at all. The WKBed
spectrum thus still retains the characteristic features of axion dark matter in the post-inflation scenario: the first axiton fossils rings, axiton chains, and interference patters from axion waves radiated
from domain wall and string collapses.

Unfortunately, our WKB procedure has a certain degree of arbitrariness due to our choice for the time when we stop our simulations/start the WKB diffusion and the final time of the WKB. Fig.
\ref{dimvar_summary} (right) reflects the uncertainty on the stopping time. The oscillations at $k\sim 100/L_1$  are clearly dependent on this choice. In this case, however we have a clear bias to chose the
ending time early enough to ensure no large discretisation effects. Probably, part of the excess power of the blue curve is due to fluctuations created in the last stages of the simulation when axitons are not
properly resolved. We have seen that a poor discretisation makes the axiton peak larger. The second part comes from the smaller time left to free-stream. These uncertainties will stay until we can simulate the
non-linear evolution of the axion field until the final freeze out of axitons in over-dense regions. However, we clearly see that the decreasing tail of the spectrum and their impact on large and intermediate
scales can be neglected.

\newpage
\section{Minicluster seeds}\label{minicluster}

\begin{figure}[tbp]
\begin{center}
\includegraphics[width=0.65\textwidth]{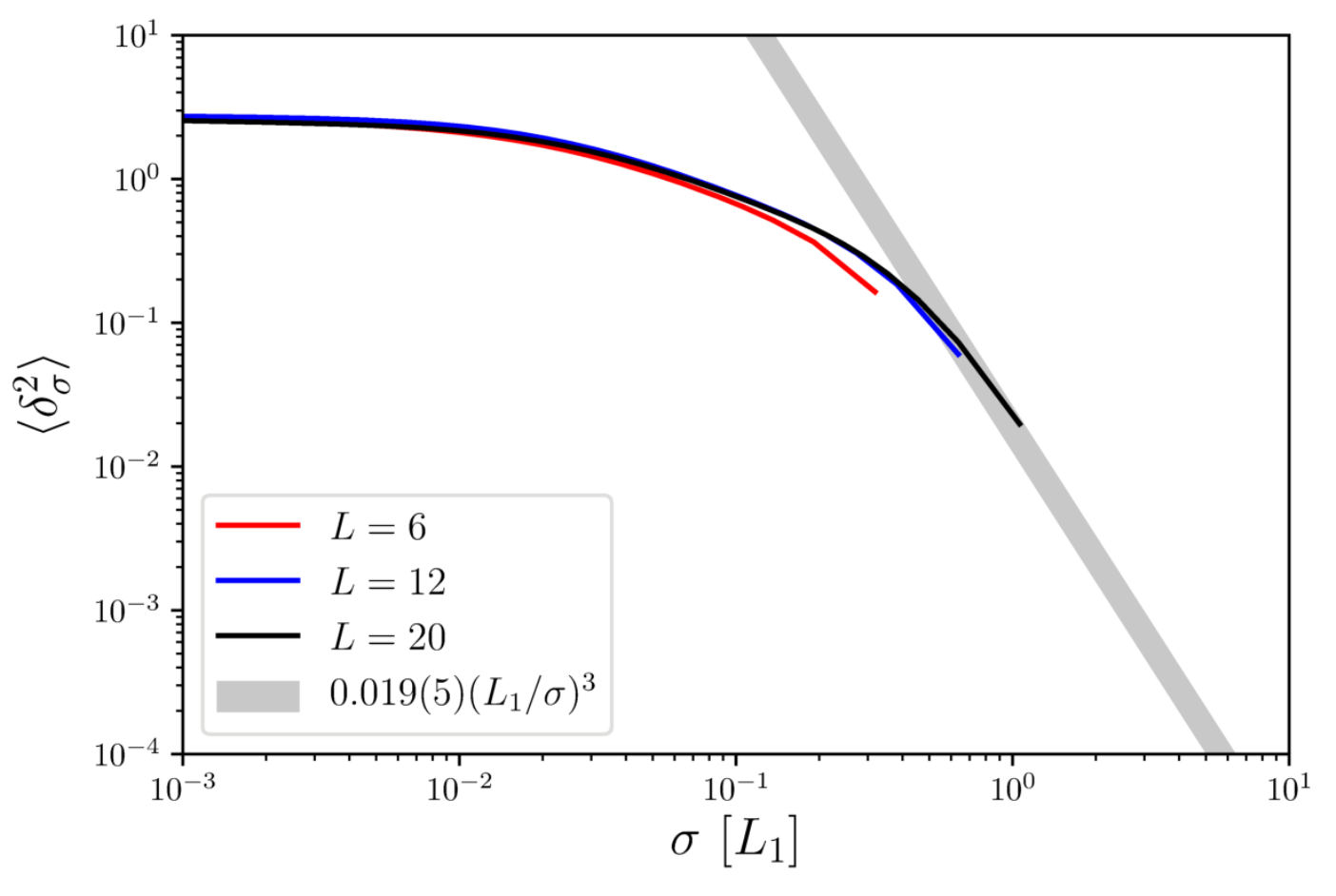}
\caption{Variance of the density fluctuations in regions of width $\sigma$, defined by a Gaussian window function. Only the biggest physical boxes $L=12,20$ enter into the $1/\sigma^3$ region where one averages
over many patches. Our extrapolation at large $\sigma$ is marked as a gray band.}
\label{varianceseed}
\end{center}
\end{figure}

To predict the properties of axion miniclusters, we want to study their spatial clustering beyond the power spectrum because we observe non-Gaussian correlations in the density. To this end we directly study the density maps obtained from our simulations after the WKB procedure, explained in Sec.~\ref{axitonsWBK}, was applied. Commonly, the average length scale of fluctuations is estimated as $L_1$, and the corresponding mass in such a fluctuation would be $\sim L_1^3\, \bar \varrho_A$. Interpreted in a model of spherical collapse \cite{Kolb:1994fi}, one of such regions featuring an overdensity $\delta$, would suffer gravitational collapse at a redshift $\sim z_\mathrm{eq}/\delta$ and relax to a density $\varrho\sim 140\,\rho_\mathrm{eq}\,\delta^3(\delta+1)$ after virialisation. Early simulations broadly confirmed that picture \cite{Kolb:1994xc,Zurek:2006sy}. However, these works presented only the distribution of overdensities on \emph{a point-by-point basis} --similar
what we have done in our Fig.~\ref{dpddelta}, while their clustering was not quantified. More importantly, in all these works the axion density field was obtained from initial conditions which prohibit the formation of cosmic strings and domain walls. These objects are highly non-spherical by themselves, and the typical scales associated with them, the thickness of a string ($\propto f_A^{-1}$) and a domain wall ($\propto m_A^{-1}$), are much smaller than $L_1$.  Certainly, our final density maps show very characteristic interference patters, axiton fossil rings and axiton chains, which pertain to scales smaller than $L_1$. These density patters become frozen in the axion DM distribution due to the extremely fast increase of the axion mass with time. Hence it is interesting to investigate how our more physical initial
conditions, including the dynamics of topological defects, affect the density contrast and to which extent the conclusions drawn from the spherical-collapse model remain valid.

\subsection{Analysis on basis of the power spectrum}

First insight into the mass of fluctuations in clumps of a given radius can be obtained from the power spectrum. In doing so, we implicitly assume Gaussian fluctuations. As emphasised at the end of this subsection, this assumption is not fully justified. However, the estimate servers as a first benchmark in our analysis, which we extend beyond Gaussian fluctuations in the proceeding sections. We define $\delta_\sigma(\veca x_0)$ as the density contrast averaged over a spherical radius $\sigma$ centred at point $\veca x_0$ and use a Gaussian window function $W\propto e^{-|\veca x-\veca x_0|^2/(2\sigma^2)}$. The variance of $\delta_\sigma$ can be obtained from the power
spectrum 
\be
\label{variancedeltasigma}
\langle \delta_\sigma^2\rangle = \int \frac{dk}{k} \Delta^2_k e^{-k^2 \sigma^2}\,,
\ee
and $|\delta(k)|^2$ is distributed as a $\chi^2$ distribution with two degrees of freedom, 
\be
\label{Gaussiandistribution}
\frac{dP}{d|\widetilde\delta(\veca k)|^2} \propto \exp\(-\frac{|\widetilde\delta(\veca k)|^2}{\langle |\widetilde\delta(k)|^2\rangle}\)\,.
\ee
i.e. an exponential and the $\delta(\veca k)$ have uncorrelated phases. Accordingly, also $\delta_\sigma^2$ follows a Gaussian distribution with the variance given by $\langle \delta_\sigma^2\rangle$ in Eq.~\eqref{variancedeltasigma}.

Thus, if the density fluctuations \emph{are Gaussian} we could predict the distribution of minicluster seeds directly from the power spectrum. The variance obtained from our simulations is shown in Fig.
\ref{varianceseed}. As $\Delta^2_k$ is sizeable only in the range $k  \sim (3-100) L_1^{-1}$, the variance saturates below $\sigma\sim L_1/100$, decreases non-trivially up to $k\sim 1/3 L_1$ and then
enters into the $1/\sigma^3$ regime, predicted by the statistics of random patches. 
Of the three box sizes shown in Fig.~\ref{varianceseed} only the largest seems to have entered deep into the white-noise region. These simulations are quite big and do not allow to capture the string-collapse and axiton dynamics with the required detail, which is only achieved in smaller ones. Indeed, comparing the parameters of
these largest simulations with Fig.~\ref{simugraph} we suspect that some of the late domain walls might have ``rolled-over-the-top". However, our largest simulations seem reasonably consistent with finer-grid simulations\footnote{The fact that in Fig.~\ref{varianceseed} the last points in $L=6$ and $L=12$ simulations seem to be lower than the $L=20$ is not to be taken too seriously. We perform the integral \eqref{variancedeltasigma} as a sum, and on the largest scales our $\Delta_k^2$ function is biased due to the small number of modes binned.} and offer much longer modes and better statistics at the low-$k$ end required for $\sigma\sim {\cal O}(L_1)$. From Fig.~\ref{varianceseed} and Eq.~\eqref{Gaussiandistribution} one can read the typical density for a given minicluster seed radius\footnote{Note however that, the Gaussian window volume for $\sigma$ is $(2\pi)^{3/2}\sigma^3$, so our window is equivalent to a hard sphere of comoving radius $\sim 1.555\sigma$.} and estimate how unlikely a given fluctuation would be. For instance, for $\sigma=L_1$, $\langle \delta^2_\sigma \rangle\sim 0.02$ and the probability of finding a $\delta=1$ minicluster-seed $\sim \exp(-1/0.02)\sim 10^{-22}$ is already ridiculously small. Indeed, $\langle \delta^2_\sigma \rangle$ becomes unity for $\sigma\sim 0.1 L_1$, which corresponds to a hard radius $\sim 0.15 L_1$, or diameter $\sim 0.3L_1$, smaller than the naively estimated $L_1$. Already this first look at our results challenges the assumption made in~\cite{Fairbairn:2017dmf,Fairbairn:2017sil} for the typical minicluster radius\footnote{These references use a typical minicluster mass $=(4\pi/3)(\pi/R_1H_1)^3$ where $H_1$ is defined with $3H_1=m_A(t_1)$ so $L_1$ there is even larger than in our work, by about $20\%$}, $\pi L_1$, as too large by a factor of 10.

\emph{In summary, the power spectrum study reveals that the typical minicluster sizes are smaller than previously thought. Also, the typical overdensities are moderate and large overdensities can only be found
in the smallest structures, not in the typical $\sigma\sim L_1$ minicluster seed.} This picture is further confirmed by our analysis of the individual clumps in the following section.

\begin{figure}[htbp]
\begin{center}
\includegraphics[width=\textwidth]{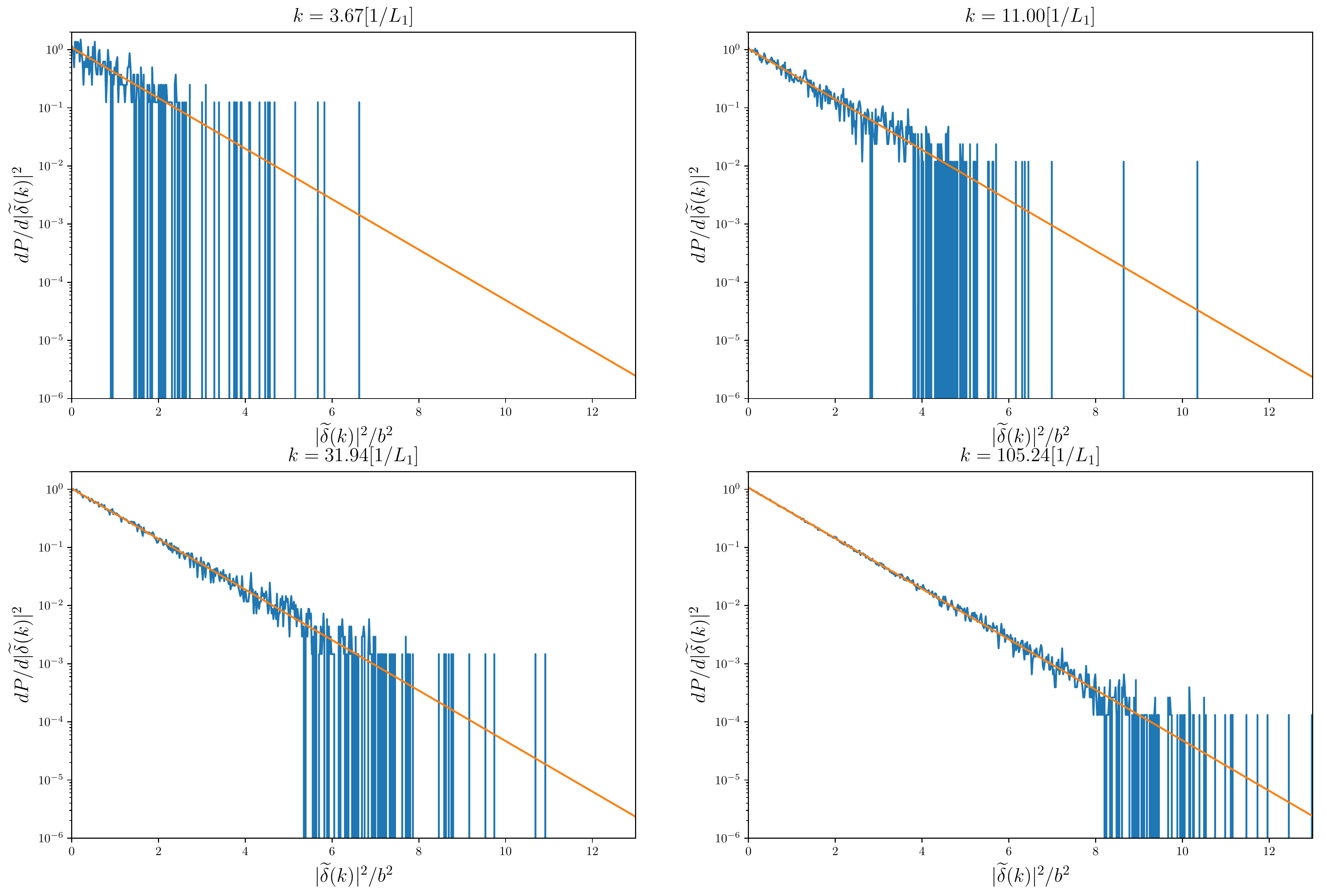}
\caption{Binned distribution of Fourier modes of the axion DM density, $\widetilde \delta(k)$ for a set of 6 $N=6144$, $L=6L_1$ simulations. The distribution is shown for four different momenta. The orange lines represent the Gaussian distribution of~\eqref{Gaussiandistribution} calculated exclusively from the mean $|\widetilde\delta(k)|^2$.}
\label{statis_distri}
\end{center}
\end{figure}

\begin{figure}[tbp]
\begin{center}
\includegraphics[width=\textwidth/2]{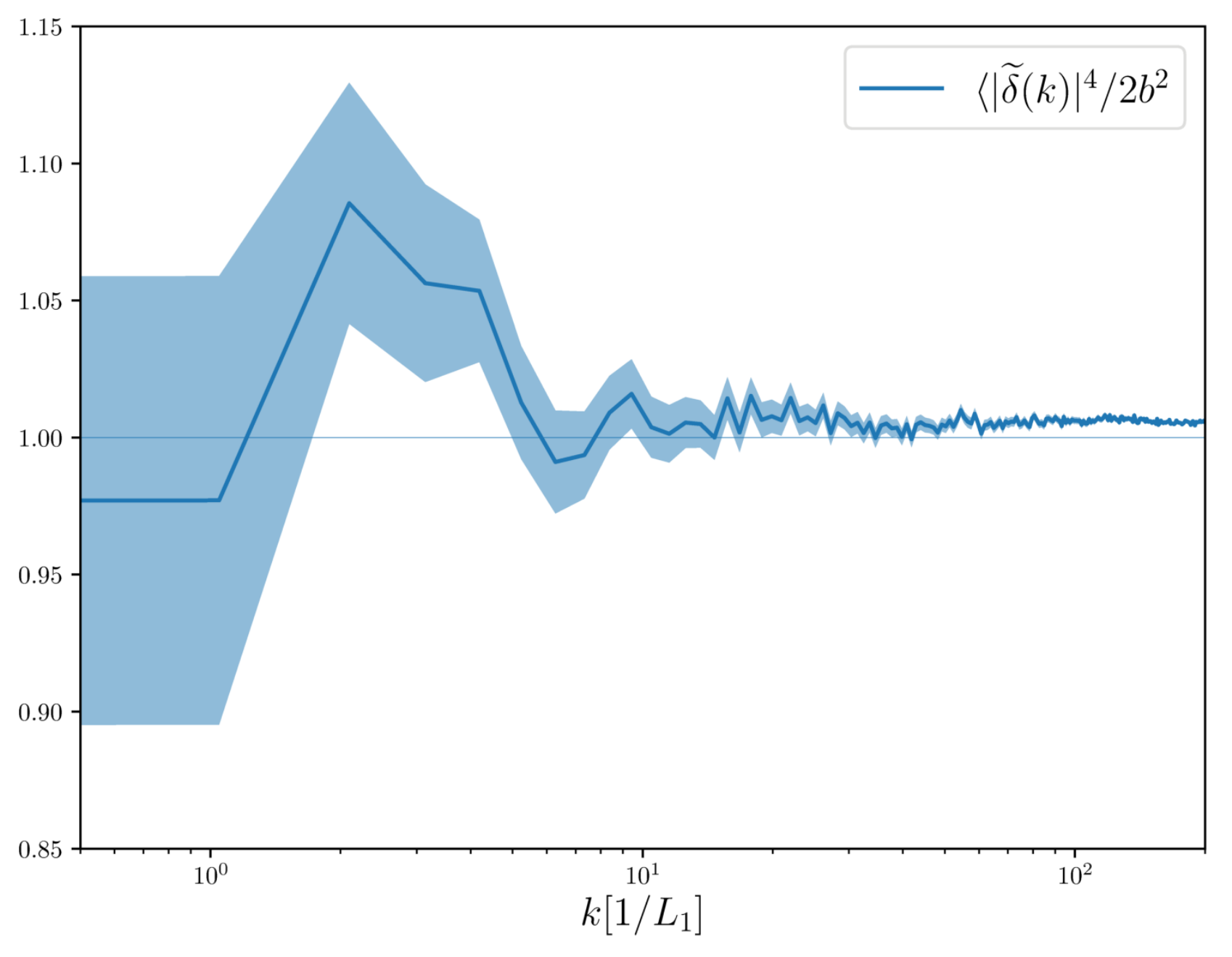}
\caption{Deviations from the Gaussian relations~\eqref{gaussianity} (represented by the line at $y=1.0$ of the momenta of the $|\widetilde \delta(k)|^2$ distribution.}
\label{statis_moments}
\end{center}
\end{figure}

Before going further we want to check to which extent the Fourier modes of the density are Gaussian distributed. To this end, we have collected the final 3D density maps for a few simulations and studied their statistics. In Fig.~\ref{statis_distri} we show the distribution of modes as a function of $|\delta(k)|^2$ for a few representative modes in $L=6,N=6144$ simulations. The orange lines represent the Gaussian distribution of~\eqref{Gaussiandistribution} computed exclusively from the mean $\langle |\widetilde\delta(k)|^2\rangle$. The agreement is excellent, especially at high-$k$, where we have plenty of statistics. There is a small tendency to a little excess at the high-$|\widetilde\delta(k)|^2$ tails. Were the distributions perfectly Gaussian, they would satisfy the following relations for the higher-moments,
\be
\label{gaussianity}
\langle |\widetilde\delta(k)|^4 \rangle =  2 b_k^2\,, \quad
\langle |\widetilde\delta(k)|^6 \rangle =  6 b_k^3\,, \quad
\langle |\widetilde\delta(k)|^8 \rangle = 24 b_k^4\,... \,,
\ee
where we have defined {$b_k \equiv \langle |\widetilde\delta_k|^2 \rangle$} for compactness of notation. In Fig.~\ref{statis_moments} we show the first moment for an ensemble of 26 simulations. The ensemble includes different volumes and grid spacings $L=6,12$, $N=4096,6172$. We see that higher moments do indeed present a small but clear systematic excess over the purely Gaussian prediction. This excess is visible in different sub-ensembles. An oscillatory pattern, similar to that observed in the power-spectrum, is also visible at low-momenta, although is to be taken with a grain of salt due to the low statistics. The error showed is compatible with a purely statistical origin.

The fact that the distribution of modes turns out to be quite Gaussian seems at odds with our expectations. It was already clear from our density maps and projection plots that axion DM is not a random Gaussian field. Moreover, we have attempted to build the power spectrum from the axion number spectrum in Fig.~\ref{pspnsp} and the differences are of the order of 1, not the ${\cal O}(10\%)$ differences visible in Fig.~\ref{statis_moments}. Therefore we are forced to conclude that the phases of the density modes are strongly correlated, i.e. the random hypothesis is not valid. We therefore extend our analysis by a detailed study of the density maps themselves in the following section.

\subsection{Analysis of the final density distribution}

To gain insight into the properties of minicluster seeds beyond the assumption of Gaussian fluctuations, we analyse nine final realisations of our simulations with box size $L=6L_1$,  which were evolved until $\ct=4.5$ and WKBed until $\ct=6.0$. The WKB map gives our most educated guess for the axion dark matter density after it freezes and until the time when gravitational collapse commences. To obtain the overdensity parameter we compute the average energy density of all simulations jointly. Each simulations contains $\mathcal{O}(6^3)$ causally disconnected regions at $t_1$ --slightly more if we are to believe Eq.~\eqref{ncestimate}--  hence statistical fluctuations in the average due to the random nature of the initial conditions are reasonably small. We then normalise each box by the average energy density, extract all grid points exceeding a certain threshold $\delta(\veca x)>\delta_t$ and use a DBSCAN algorithm\cite{scikit-learn} to identify connected regions. 

In what follows we will call these regions ``minicluster seeds''. This analysis of the density contrast does not take into account the distances between neighbouring, over-dense regions. Instead, we consider each of them as an isolated object. Minicluster seeds, identified with a large threshold, might even be connected by overdense regions of some lower threshold. Hence, it is not given that all minicluster seeds will eventually evolve into individual miniclusters. Calling these overdense regions ``minicluster seeds'' establishes a distinction to the term ``miniclusters'', which we reserve for the fully virialised objects.

\begin{figure}[tbp]
\begin{center}
\includegraphics[width=\textwidth/2]{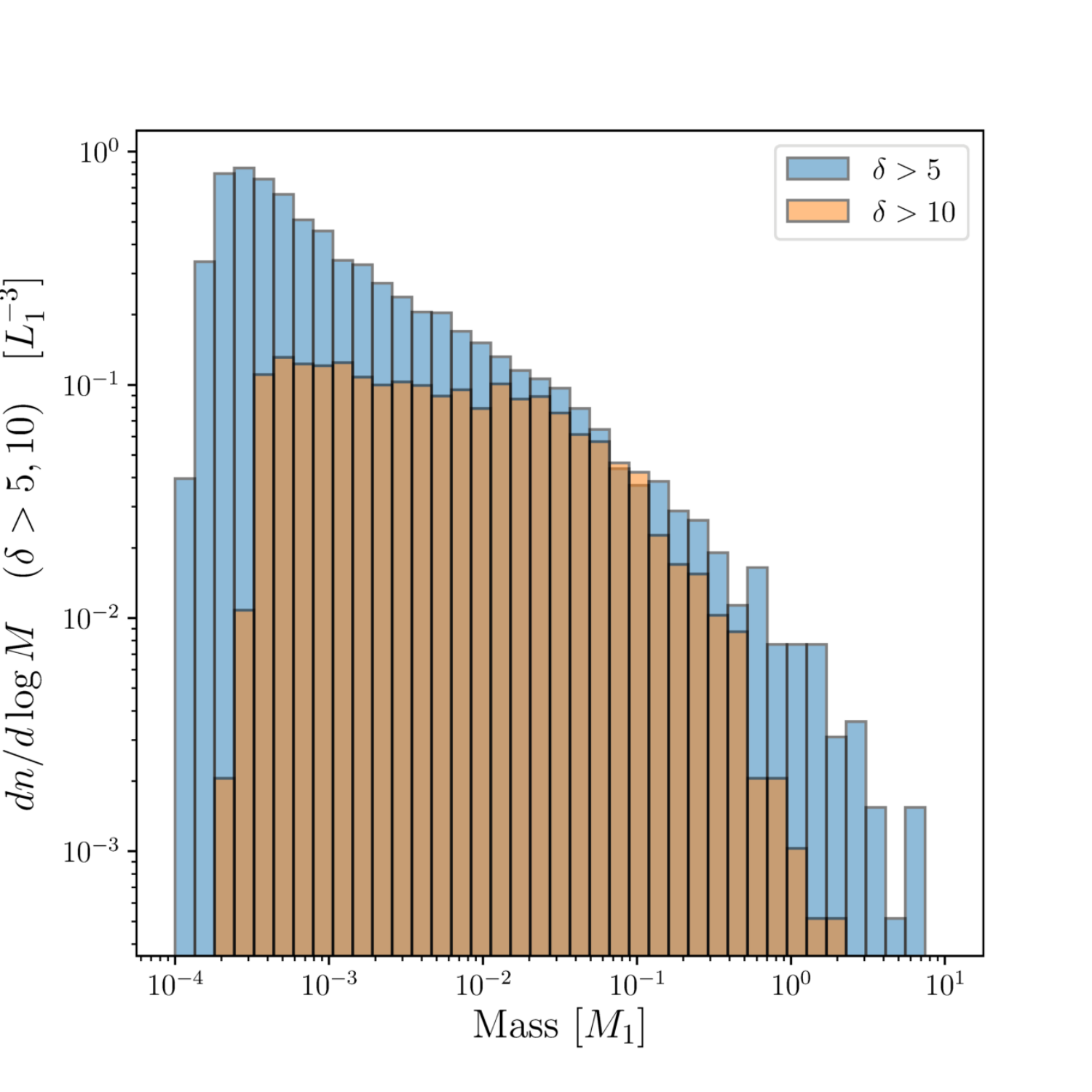}
\caption{Number of minicluster seeds with overdensity $\delta+1>5$ and $\delta+1\ge10$ as function of their total mass. The distribution peaks at extremely low masses, which imply also small radii.}
\label{fig: mc-mass-hist}
\end{center}
\end{figure}

We normalise the mass distribution of minicluster seeds by the amount of DM in a simulation box of length $L_1$
\be
M_1 = 2.1 \times 10^{-12} M_{\odot}\ \frac{\Omega_{Ac}h^2}{0.12}\(\frac{50\,\mu\rm eV}{m_a}\)^{0.49}\,,
\ee
where we have used Eq.~\eqref{Hm1} and allowed for axions to be a sub-dominant component of DM, contributing some fraction $\Omega_{Ac}$ to the critical energy density. The simulations presented here have $L=6L_1$ and hence contain only a total DM mass of $\sim 4.6\times 10^{-10} M_\odot$. Vanilla miniclusters would have a mass $M_\mathrm{mc}\sim M_1$.

The mass spectrum for minicluster seeds with threshold $\delta+1\ge5$ and $\delta+1\ge10$ is shown in Fig.~\ref{fig: mc-mass-hist}. The vast majority of fluctuations has masses much smaller than $M_1$, however, large-mass fluctuations do exist. Their abundance increases as the threshold of the analysis is lowered, indicating that most of the mass in heavy minicluster seeds resides in large regions of moderate overdensity.

\begin{figure}[hbp]
\begin{center}
\includegraphics[width=\textwidth]{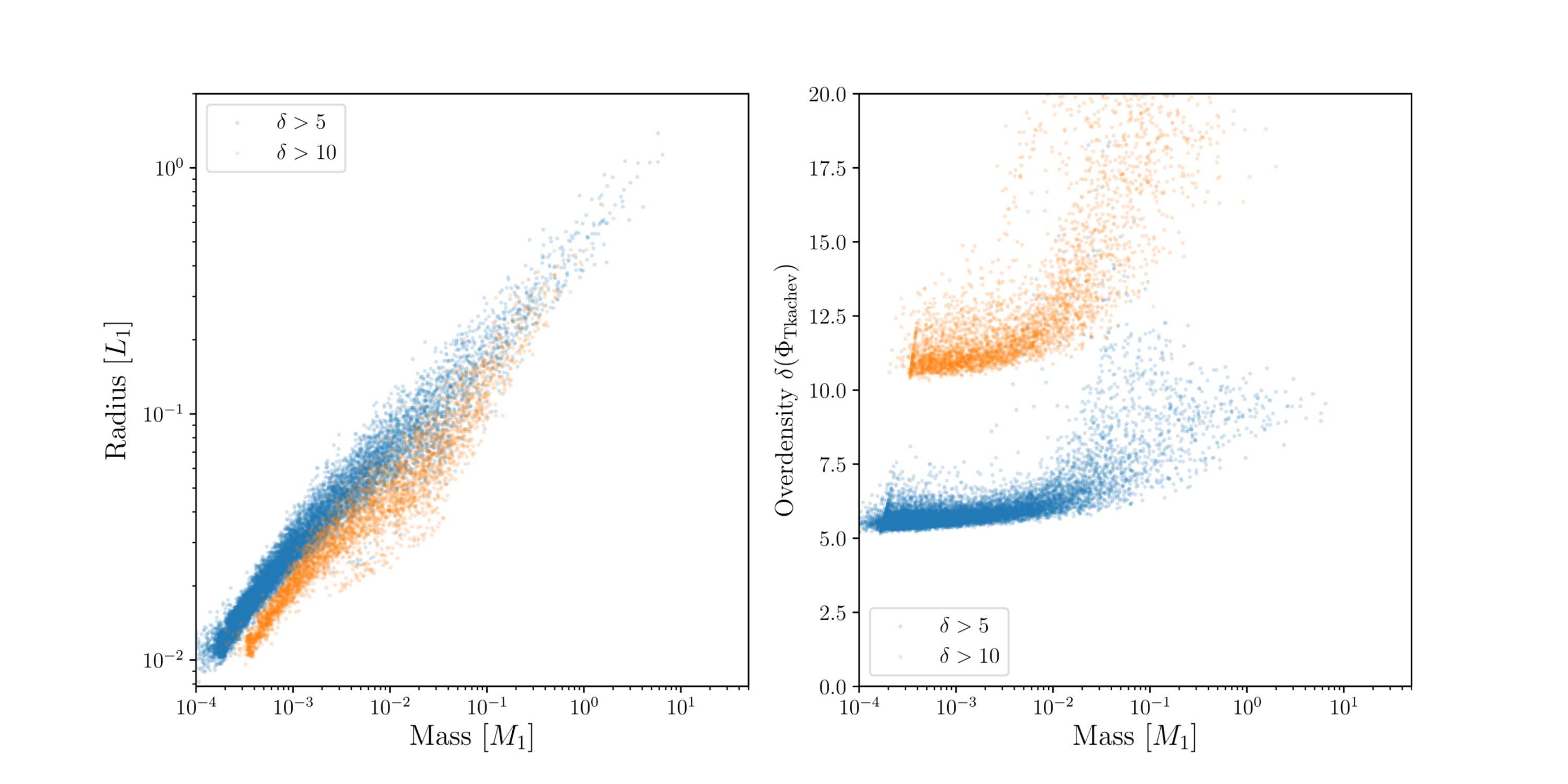}
\caption{On the left, distribution of minicluster seed masses and radii for $\delta+1\ge5$ and $\delta+1\ge10$. The radii increase with the third root of the mass roughly. On the right, distribution of the
average overdensity within a minicluster seed with its mass. High mass seeds typically have moderate average overdensities of $\mathcal{O}(10)$.}
\label{fig: mc-mass-radius-overdensity}
\end{center}
\end{figure}

Since the density fluctuations in our simulations are generally not spherical, defining the radius of a minicluster seed is not obvious. Here we quote it as the radius of a spherical shell, centred at the
seed's centre of mass and enclosing 50~\% of its total mass, given in units of $L_1$. We show how that radius is distributed with seed mass in the left panel of Fig.~\ref{fig: mc-mass-radius-overdensity}. As
one would expect, the radius roughly scales with the third root of the mass. Thus the minicluster seeds at the high-mass end of the distribution in Fig.~\ref{fig: mc-mass-hist} have radii comparable to $L_1$,
further confirming the picture that most of the mass in large-mass seeds resides in a rather extended region. In Fig.~\ref{fig: mc-mass-radius-overdensity} (right) we also show the average overdensity of all points within a minicluster seed. In a model of spherical collapse this quantity determines the time of collapse and the radius and density of a virialised object. Especially low mass objects have an average overdensity very close to the threshold. The scatter increases towards larger overdensities and larger masses, which also imply larger seed radii. Interestingly, the scatter is bigger in minicluster seeds with a large threshold further confirming that moderately over-dense regions contribute the most to a minicluster seed's mass and also to its average overdensity. High mass minicluster seeds with $M_\mathrm{mc}\sim M_1$ and a large average overdensity $\mathcal{O}(100)$, which would produce the strongest signals in lensing experiments, are mighty rare objects: all high mass seeds ($M_\mathrm{mc} \gtrsim M_1$), which we find for a threshold $\delta\ge5$, have average overdensities of at most $\mathcal{O}(10)$.

\begin{figure}[tbp]
\begin{center}
\includegraphics[width=\textwidth/2]{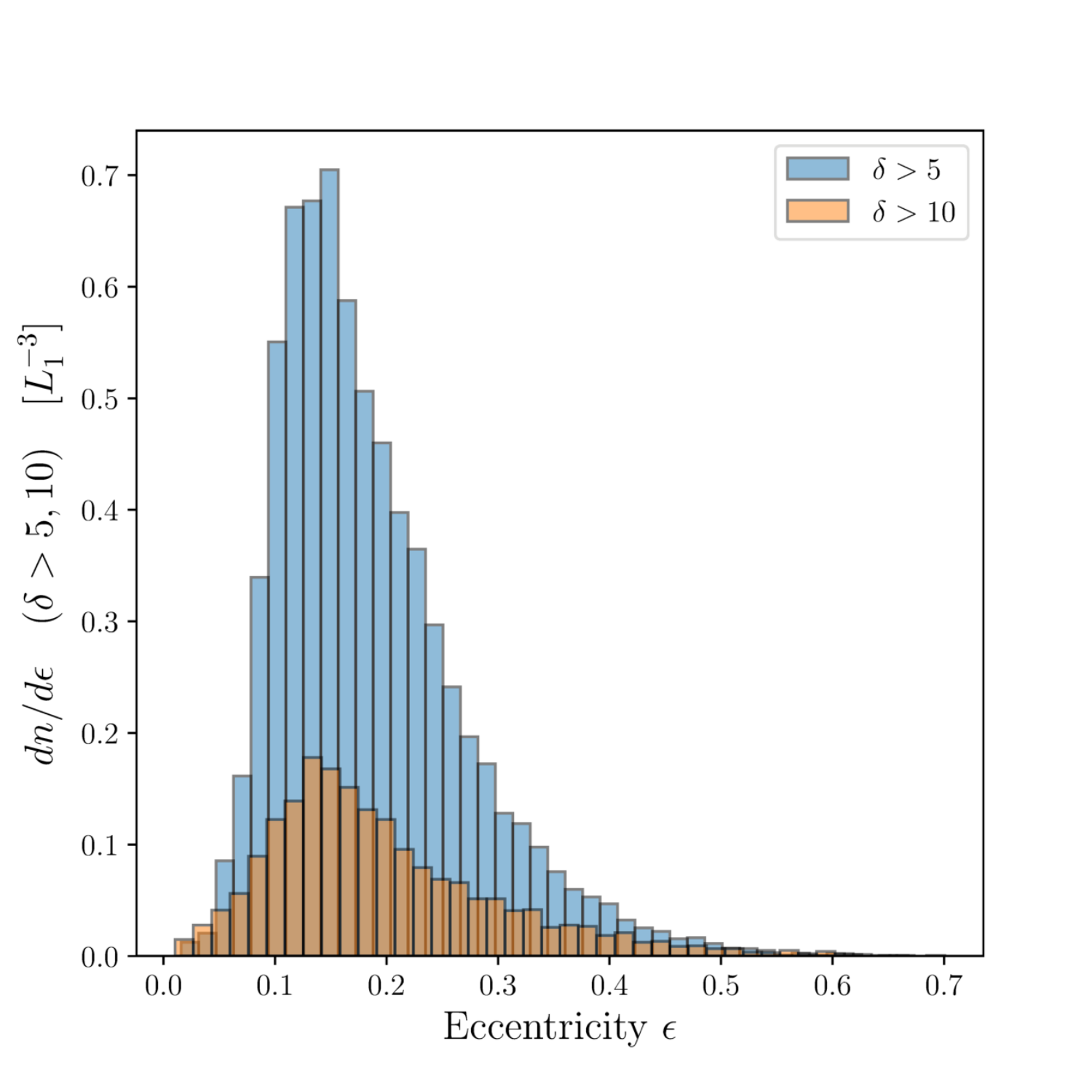}
\caption{Distribution of minicluster seeds as as function of the eccentricity parameter $\epsilon$.}
\label{fig: mc-eccentricity}
\end{center}
\end{figure}

Previous analysis found that shapes of axion density fluctuations created in the post inflation scenario are rather spherical \cite{Kolb:1994xc}. To quantify the sphericity of the fluctuations in our simulation
we define the ellipticity $\epsilon$ as
\begin{equation}
\epsilon = \frac{I_3 - I_1}{I_3 + I_1}\,,
\end{equation}
where $I_3$ and $I_1$ are the largest and the smallest eigenvalues of the momentum of inertia, respectively. For comparison, a sphere would have $\epsilon=0$ and a homogeneous ellipsoid with the larger axis
$40~\%$ bigger than the smaller would correspond to $\epsilon=0.2$. We show in Fig.~\ref{fig: mc-eccentricity} how $\epsilon$ is distributed over the minicluster seeds in our simulations. For a threshold of
$\delta\ge5$ only very few regions of high sphericity are encountered: the distribution in $\epsilon$ is peaked between $0.1$ and $0.2$ with a tail towards larger eccentricities. Increasing the threshold,
connected regions become somewhat more spherical, the tail towards large eccentricities flattens and more low-eccentricity regions appear. However the peak of the distribution remains at roughly the same
position. This picture is further confirmed when the connected regions are visualised directly. Especially those regions, which contain large masses are very amorphous and have irregular shapes, see Fig.
\ref{densityslices} and Fig.~\ref{picsWKB}. For lower-mass minicluster seeds the shapes become more compact, some of them arising from axiton cores, however, as Fig.~\ref{fig: mc-eccentricity} indicates, many of them remain elongate.

As stressed before, our analysis doesn't not account for the proximity of neighbouring regions and for the density field between individual minicluster seeds. It is not guaranteed that each minicluster seed will form an individual, virialised minicluster after the gravitational collapse of the distribution. Rather, a hierarchical collapse of smaller structures within larger ones and a complicated system of mergers and fragmentations is expected, given the complicated, irregular structure of the density field. These complicated density patterns inevitably emerge from the full dynamics of the axion field, i.e. strings, domain walls and axitons, considered in our simulations. Thus, we believe there is no benefit in pushing the analysis of minicluster seeds to further detail. Of observational interest are the mass function and radii of virialised objects after they have participated in the hierarchical formation of larger halos. To model these properties correctly N-body simulations of the gravitational evolution of the density field are needed. We will present such simulations in a separate publication.

\section{Summary, discussion and conclusions}\label{conclusions}

In this paper we present numerical simulations of the evolution of the axion field around the QCD epoch in the post-inflation scenario. Our study is dedicated to the analysis of the final density contrast of the simulation to shed light on the formation of axion miniclusters.  

There are several problems associated with this kind of simulations. First, the random nature of the initial conditions in the post-inflation scenario leads to the formation of topological structures, more exactly global strings and domain walls. The topological structures necessitate a UV completion of the theory. As done throughout the literature, we therefore assume that the axion arises as the phase of a complex scalar field with a spontaneously broken PQ-symmetry. There is a large hierarchy between the axion and the saxion mass, which sets the tension and thickness of cosmic strings. At the moment simulations with physical values for the axion to saxion mass ratio are not achievable, as they would require extremely fine grids, of $\mathcal{O}(10^{57})$ points (!), to capture all dynamics correctly. Instead, we adopt the viewpoint that the final axion yield is sensitive only to the logarithm of the string tension and that, therefore, the physical picture should be close to what we can achieve in our simulations. We also advocate the use of PRS or ``fat'' strings, as these increase the dynamical range of the simulations and overall are closer to the physical picture. Our viewpoint was argued in many previous works and is further supported by the fact that our simulations, which use the largest volumes employed up to date (up to $8192^3$ points), seem to point into the same direction.

The second issue is connected to the saxion to axion mass ratio as well. Since the axion mass increases in time with a fast power, our simulations inevitably reach the point where both masses become comparable, resulting in the unphysical destruction of strings and domain walls. This problem was solved using a mixed approach: on the one hand, we keep the string tension large enough to ensure we could reach the times when topological structures disappear naturally. Then we switch to an axion-only simulation, where the saxion mass is explicitly sent to infinity and decouples.

Third, during the late times of our axion-only simulation we observe the development of axitons -- pseudo breathers, which are constantly flashing and re-collapsing, thereby emitting mildly relativistic axions. Axitons are expected to diffuse away when the axion mass saturates to its zero-temperature value, however, due to the limited dynamical range of our simulations, we can not reach such late times. Although the number density of axitons increases with time, our study suggests that only axion emission from axitons at early times contributes significantly to the axion density field, at least at the large and intermediate scales of our simulation we are most interested in. This assumption is one of the largest uncertainties in our study, besides the smallness of the axion to saxion mass ratio and will be reviewed in a future study dedicated specially to axitons. For now our approach is to ensure that we capture the dynamics of the first axitons correctly, subsequently we evolve the axion field for some period without self interactions to allow remaining axitons to diffuse away and all high-momentum modes to reach the non-relativistic regime. This final phase is computed analytically in the WKB approximations and allows us to arrive at a density contrast, which is frozen until gravitational instability around matter radiation equality transforms our fluctuations into axion miniclusters.

The consistency of our results across several volumes, lattice spacings and string tension parameters $\lambda_{\prs}$ gives us enough confidence to trust our results at the qualitative level. Despite the complicated dynamics, involving a large hierarchy of scales, we believe we capture all major effects accurately and that our simulations are the most physical ones, concerning the dynamics of the axion field in the post-inflation scenario, up to date.

Comparing our results with those of previous works, we confirm the existence of a scaling regime in the evolution of strings. We correctly reproduce the peak announcing the annihilation of the string network and our largest simulation shown no sign of unphysical domain wall destruction. The spectrum of the axion number follows a power law $n(k) \propto k^{-3.5}$, verifying the previous notion that the energy is dominated by the UV, but most axions are generated at low momenta. A more detailed study of the axion spectrum produced from string decay and its extrapolation to physical string tensions is left for a future paper, however, our results seem to be largely consistent with previous works.

The object of our study is the distribution of the axion dark matter at the end of our simulation and how it defines the initial conditions for the later formation of axion miniclusters. Due to the fast increasing axion mass, axions become non-relativistic very fast. Almost instantly, most of the axion distribution freezes out around $\ct\sim 1.5-2.5$. The resulting distribution preserves many features of the string-wall scaling regime: frozen, roughly spherical waves that emanated from the collapse of the last string loops, as well as more homogeneous, larger patches that have a similar value of the axion field (misalignment angle). The axion self-interactions implied by the $-\cos\theta$ potential play an important role in this epoch. They are attractive and tend to enhance the density of lumps in interference patterns and regions of large misalignment angle. They also cause the only obvious changes in this late part of the axion field evolution, the wave collapse of dense regions into long-lived pseudo-breathers, oscillons that are usually called axitons. In the cores of axitons, the axion field reaches large values where the attractive self-interaction saturates. Axitons suffer violent oscillations and emit bursts of spherical axion waves, the first of which become also frozen in the axion distribution and are noticeable at later times. The characteristic size of an axiton core is the inverse axion mass $1/m_A$ and therefore, the cores shrinks and so does the wavelength of the emitted axions, which are typically moderately relativistic.

We have first studied the power spectrum of density fluctuations, see Fig.~\ref{dimvar},~\ref{dimvar_summary},~\ref{axievol} and \ref{pspnsp}. After the scaling regime, the large scale part converges well to the expected shape. The dimensionless variance $\Delta_k^2$ rises as $k^3$ (white noise) at large scales and flattens at the mode related to the size of the causal horizon at the time when axions become non-relativistic. We find this comoving momentum to be $k\sim 3/L_1$, see Fig.~\ref{dimvar_summary}, which is larger than previously assumed in the literature. We think that this is due to two reasons. One is the existence of cosmic strings, which are boundaries where the axion field takes all values from 0 to $2\pi$. Strings are very dense and have a large inertia and thus the axion field takes some time to drag them to collapse. Consequently, its correlation length will be smaller that the horizon size. Second, our simulations are performed in very large boxes where we can simulate many relativistic modes. Although the spectrum is red, is only just so. Some radiation is able to free-stream and suppress the power at large scales, which therefore shifts the flattening $k$ to higher momenta. These observations are supported by Fig.~\ref{dimvar_summary}, where we show that increasing the string tension (which increases the string inertia, and was mostly accomplished by decreasing the lattice spacing at the same time) tends to suppress the power at small scales. When $\Delta_k^2$ becomes flat at wave numbers larger than $k\sim 3/L_1$ its value does not exceed $1$.

We have compared this power spectrum with the recent semi-analytical estimate of~\cite{Enander:2017ogx}, see Fig.~\ref{enander}. In that reference, only the axion field is considered and self-interactions are neglected by approximating the potential as $\chi\theta/^22$. Thus, their final spectrum is dictated by the initial conditions, taken as white noise with a cut-off around $L_1$ and normalised to match $\langle \theta^2\rangle =\pi^2/3$. An obvious limitation of this approach is that the shape of the axion spectrum and the choice of cut-off affect the normalisation. Compared with their assumptions, our simulations output an axion spectrum with a higher cut-off and a very soft decline $\propto k^{-3.5}$, see Fig.~\ref{spectrumPlot}. Therefore, although the basic features are very similar in order of magnitude, our power spectrum of density fluctuations is pushed to higher-momenta, i.e. smaller scales, and has much more structure at small scales, $k\gg 3/L_1$. The final density maps show significant correlations and small non-Gaussianities. These are related to the initial conditions, which include cosmic strings and domain walls, and to axion-self interactions. Indeed, in Fig.~\ref{pspnsp} we have build a power-spectrum from our axion spectrum by assuming uncorrelated axion modes and, although the result agrees quantitatively with the full numerical result, there are $\mathcal{O}(1)$ differences.

In summary, the typical $\sim L_1$ axion density fluctuations associated with axion minicluster seeds, tend to be smaller than previously thought. Their overdensities, although of the order of $\mathcal{O}(1)$, are moderate. Most importantly, the dimensionless variance is relatively large at intermediate and small scales. We do not find an exponential cut-off but a soft power law that follows the spectrum of axions. Therefore, there is a lot of structure below the $\mathcal{O}(L_1)$ length scales. We expect both isolated small minicluster seeds and substructure inside the typical $\mathcal{O}(L_1)$ minicluster.

In earnest, at the smallest scales, axion self-interactions and axitons in particular murk the interpretation of the power spectrum and do not allow our results to be completely trustworthy. Axitons do not contain a large number of axions (they are dense but very small), but nevertheless they do dominate the power spectrum at scales related to the axion mass at a particular time. They imprint a very particular peak in the variance $\Delta_k^2$ that shifts to smaller scales as time evolves, see Fig.~\ref{dimvar}. We have shown that the position of the peak corresponds to the axion mass (Fig.~\ref{spectrumPeak}) as one expects from the emission of semi-relativistic axions with very large amplitude from a small source. We have also found that the peak height increases in time, see Fig.~\ref{spectrumPeak}, and we interpret this as an increase of the number of axitons. We have attempted to correlate the peak height with the axiton number with a simple model, with meagre success, see Fig.~\ref{figAnumber}, but there are many reasons to believe that the correlation will be more complicated. We have followed the evolution of the power spectrum in time, showing that mode amplitudes decrease slowly or are frozen until the axiton peak reaches them, they increase significantly when the axion mass becomes comparable to the mode, and they slowly decrease later, see Fig.~\ref{axievol}. The decrease can be interpreted as the dispersion of the axion waves emitted by the roughly point-like axitons and the compression of the axiton cores. It seems that modes become frozen slowly after the axiton peak has gone well over them. We have devoted Sec.~\ref{sectionaxitons} to understand the main features of axitons and their impact on the power spectrum. There is a critical index $\n=2$ for the time-increase of the axion mass ($m^2_A\propto \ct^{\n}$), above which axitons will tend to re-collapse after emitting axion bursts and they are therefore persisting. For, $\n<2$ we should find the well known pseudo-breathers or oscillons, which are known to evaporate after a few bursts. The time scale for the self-interactions to produce a collapse increases in time, and the most likely moment for dense regions to suffer wave collapse is shortly after they become non-relativistic. The collapse time strongly depends on the initial amplitude, too, so that only regions with large initial misalignment angle at are expected to suffer wave collapse. These two effects imply that the position of axitons will be closely correlated. Indeed, we see in our simulations that axiton cores clump in clusters, some of them in what we called axiton chains, see Fig.~\ref{axichain}. Similar effects were observed in the pioneering literature~\cite{Kolb:1993hw}. We have argued that axions emitted from late axiton cores are expected to be largely irrelevant for the density fluctuations. This is due to two effects: The first is that axions emitted at late times are more
relativistic than early ones, and thus diffuse much farther when released from the core, as shown in Fig.~\ref{axitodel}. Second, the core volume decreases faster than its energy density so cores can radiate comparatively less axions. Since the large scales of the axion field are essentially frozen by the end of our simulations and the only active spots are small axitons and their recently emitted waves, we have devised a somewhat drastic procedure to obtain a better picture of the final distribution of axion dark matter. We follow the axion evolution as late as we can without severely compromising the spatial resolution of axitons, typically $\ct\sim 4.5$. Second, at this late time, we neglect self-interactions and evolve the field in momentum space using a WKB approximation. This evolution disperses very efficiently the axiton waves and cores (which have already reached very small sizes) and reveals the tail of the large scale spectrum. The tail decreases smoothly following the expectation from the relatively red tilted axion-spectrum, see Fig.~\ref{pspnsp}. The WKB evolution preserves the large and intermediate scales with the many features from the first axiton generations like fossil rings and chains. Unfortunately, the procedure implies a certain degree of arbitrariness, as different ending times and WKB durations imply small changes of the final spectrum. Moreover, non-linearities at small scales are largely erased by the WKB procedure, while physically some of them are expected to survive. In summary, we are confident that the resulting power spectrum of fluctuations is a good representation of the axion dark matter distribution at large, intermediate and small scales, except for the small scales in those rare regions where an overabundance of axitons is present. We note that our simulations are the finest-resolution simulations of the axion-field ever performed in the deep-non-linear epoch. 

An alternative to deal with small scales and axitons would have been to cut the growth of the axion mass at an earlier time, so that axitons become pseudo-breathers and disperse within our simulation time scale, as done in~\cite{Kolb:1993hw} and \cite{buschmann}. We have argued that the most relevant axiton features are due to the earlier (bigger) axitons, so by making axitons bigger for a longer time this procedure might enhance the non-linearities at intermediate scales. Nevertheless, we think that a comparison of both approaches would be very interesting and relevant to understand the axion dark matter distribution at the smallest scales.

A final note of caution is in order. The non-linearities around the axiton cores change the axion number by fusing many non-relativist axions into relativistic ones, i.e. decreasing axion number. At the same time, large field axion oscillations (happening in the core) redshift slower than ordinary axion dark matter, which has the opposite effect. We only detect a very small drift of the axion number (obtained by integrating the axion spectrum) after the strings and walls have disappeared so this phenomenon is negligible during our simulations. The effect of a single axiton on the axion number must decrease with time, because the axiton core volume shrinks very fast. However, we have detected a very large growth of the number of axiton cores, and if this trend continues it is conceivable that the global axion number might
be sensibly affected, changing the axion dark matter yield! We observe that axitons appear mostly in nearby groups where the axion field was large, $\theta(\ct\sim 1)\sim \pi$ and we think that only these superdense regions will be affected. If the net axiton effect was to decrease axion number, we expect that this can only imply a moderate, maybe up to a $\mathcal{O}(20\%)$ decrease. We note, however, that the axiton dynamics can be strongly affected by discretisation effects. For instance we have seen that the axiton peak height depends on our lattice spacing, see Fig.~\ref{axievol}. In our opinion, these speculations highly motivate further work on axion dark matter non-linearities.
 
Let us come back to the large and intermediate scales where we think we understand the axion dark matter fluctuations well. We have analysed the statistical properties of the density contrast's Fourier modes and showed that the distribution of their modulus is Gaussian to a large extent, see. Fig.~\ref{statis_distri}. Non-Gaussianities can be detected by computing higher momenta of the distribution and appear at the percent level. We think the small-$k$ region would benefit from more statistics, but the effect appears to be significant overall. We conclude here building the axion dark matter field as a Gaussian density field is not a very bad first approximation. However, the approximation misses the phase correlations that arise from the string-wall network and from the axitons. Therefore, we advise to use a simulation, like the ones presented here, to determine the axion dark matter distribution before studying the gravitational collapse and axion miniclusters.

Finally, we have studied the dark matter distribution of our simulations in position space to directly challenge some claims in the literature about the size and density of minicluster seeds. In many works it has been assumed that the axion field can be taken to consist of uncorrelated, approximately homogeneous patches of size $\sim L_1$. A distribution of overdensities of these patches was presented in Fig.~2 of~\cite{Kolb:1995bu}, which could be used to compute the radius and mass of a typical virialised minicluster assuming spherical collapse. The distribution shown in~\cite{Kolb:1995bu} appears to be simply the distribution of density contrasts of the different \emph{points} of the simulation (actually the cumulative distribution is shown). The analogous distribution for our simulations has been presented in Fig.~\ref{dpddelta} and agrees qualitatively with~\cite{Kolb:1995bu}, at least at small values of the contrast $\delta$. However, Fig.~2 of~\cite{Kolb:1995bu} is not the distribution of overdensities of $\mathcal{O}(L_1)$ minicluster seeds! We find that ultra-dense points in our simulation tend to group in small clusters, not in very large $\mathcal{O}(L_1)$ regions. This can be seen already from the dimension-less variance at $k<1/L_1$, which is sensibly smaller than 1. Indeed, we have estimated the overdensity of a region of size $\sigma$ by integrating the power spectrum and showed our results in Fig.~\ref{varianceseed} as a function of $\sigma$. For
$\sigma\sim 1/L_1$, $\langle \delta^2_\sigma\rangle$ is sensibly smaller than one. This result assumes  Gaussian distributed modes and should fail particularly in the tail of the distribution, but not very much
in the variance shown. 

We finally analysed minicluster seeds by a simply friends-of-friends clustering algorithm applied to the points of our final density grids. We have build minicluster seeds by associating points above a certain density threshold. If we set a small density threshold $\delta_t\sim 1,2$ the connected regions are very large and contain a lot of mass but they are \emph{quite amorphous, i.e largely non-spherical}. Our method is best suited to study the high density minicluster seeds, precisely those that can be worst represented by the Gaussian statistics implied by Fig.~\ref{varianceseed}. The mass and radius of seeds obtained by two different thresholds of $\delta_t=5,10$ are shown in Fig.~\ref{fig: mc-mass-hist} and \ref{fig: mc-mass-radius-overdensity}. Note that all the points analysed for $\delta_t>10$ are of course included in the $\delta_t>5$ analysis. Most of the large density objects are extremely small and have small masses $\ll M_1$ (the average mass inside $L_1^3$). Looking at Fig.~\ref{fig: mc-mass-hist}, we find that the abundance of minicluster seeds of mass $\sim M_1$ and overdensity $\delta>5$ is only $\mathcal{O}(0.01/L_1^3)$ and with overdensity $\delta>10$ about an order of magnitude smaller. Fig.~\ref{fig: mc-mass-radius-overdensity} shows that the largest seeds tend to be more massive (left) and that the contrast tends to be larger in the most massive seeds (only by a small amount, i.e. $\delta\sim 10$ for $\delta_t>5$). We also see that the maximum contrast decreases with increasing mass for masses $>0.01 M_1$. All these findings agree with the conclusions drawn by studying the power-spectrum. In particular, we emphasise once more the huge number of small seeds with masses $<0.01M_1$ and large overdensities. We have also studied the sphericity of high density seeds to check how the hypothesis of spherical collapse applies for them. Fig.~\ref{fig: mc-eccentricity} shows that even this high-density seeds retain eccentricities of $\sim 0.2$ on average, which corresponds to $\mathcal{O}(1)$ distortions from sphericity.  

The picture drawn from our simulations is thus one of a huge amount of small size and small mass miniclusters seeds. Large miniclusters will probably tend to have a sizeable number of smaller objects inside.
In a future publication we will use the results of these simulations to numerical evolve the gravitational collapse of axion miniclusters as the next step to study their very spectacular phenomenology, were
they to survive until today and comprise a large fraction of the cold dark matter of the Universe.

\section{Acknowledgements}

In the already few years invested on this project, we have enjoyed, learnt much and got support and ideas during discussions with many dear colleagues, I. Tkachev, J. Niemeyer, K. Saikawa, C. O'Hare, G. Raffelt,
A. Ringwald, I. G. Irastorza, D. Marsh, G. Sigl, G. Dvali and G. Moore amongst many others. JR is supported by the Ramon y Cajal Fellowship 2012-10597, the grant FPA2015-65745-P (MINECO/FEDER), the EU through the ITN ``Elusives"
H2020-MSCA-ITN-2015/674896 and the Deutsche Forschungsgemeinschaft under grant SFB-1258 as a Mercator Fellow. AV is supported by the U.S. National Science Foundation under grant PHY14-14614. JS receives
funding/support from the European Union's Horizon 2020 research and innovation programme under the Marie Sklodowska-Curie grant agreement No 674896. The numerical work was mostly done in the supercomputers
Draco and Cobra of the Max Planck computer data facility (MPCDF). Much of the development was also done in the Caesaraugusta node of the Spanish supercomputing network and on TACC Stampede 2 and PSC Bridges
supercomputers under XSEDE allocation PHY170045. Our numerical calculations make extensive use of the excellent Fast Fourier Transform library (FFTw3)~\cite{FFTW05} of M.~Frigo and S.~Johnson. 
\appendix

\section{Time scale for non-relativist axion lump evolution}
Starting with \eqref{eom11} with $R_{\ct\ct}$ neglected and expanding the sine
\be
\ctheta_{\ct\ct} -\nabla^2 \ctheta + m_\ctheta^2 \(\ctheta -\frac{1}{6\ct^2}\ctheta^3+...\) = 0.
\ee
with $m_\ctheta = \ct^{\n+2}$.
We introduce the non-relativistic ansatz,
\be
\ctheta(\ct, \veca x) = \frac{\nrc(\ct, \veca x)}{\sqrt{2 m_\ctheta}}e^{-i W_0(\ct)} + \text{h.c.}
\ee
(recall the phase integral \eqref{Wphase} at $\veca{ k}=0$, which gives $\partial_\ct W_0 = m_\ctheta$.) into the e.o.m. getting,
\be
\(- i \sqrt{2 m_\ctheta} \partial_\ct \nrc - \frac{1}{\sqrt{2 m_\ctheta}} \nabla^2 \nrc - \frac{3 \cmassTO}{6 \ct^2}\frac{\nrc |\nrc|^2}{(2\cmass)^{3/2}} + ... \)e^{-i W_0} + \text{h.c.} = 0
\ee
where the ellipsis stands for terms suppressed by larger powers of $1/\cmass$, terms $\partial_{\ct}^2\nrc$ and the high harmonic $\propto e^{-2 i W_0}$ whose effects on the fundamental oscillation one can
neglect. Equating to zero the parenthesis, one gets the usual Gross-Pitaevskii-Poisson equation,
\be
\label{gpeq}
i \partial_\ct \nrc = - \frac{ \nabla^2 \nrc}{2 m_\ctheta} - \frac{1}{8 \ct^2}\nrc |\nrc|^2.
\ee
Note that as time increases, both right-hand side terms become less relevant.  The first term gives us gradient pressure and tends to make $\nrc$ diffuse, homogenise through oscillations. The second term due to
self-interactions gives the negative pressure that drives the wave collapse instability.
If both terms can be neglected (the case at late times) the solution is $\nrc=\nrc (\veca x)$, a constant in time. Now, note that in the regime of our simulations $\cmass = \ct^{\n/2+1}$. If $\n/2+1>2\to \n>2$
the gradient pressure term decreases much faster than the self-interaction term and self-interactions dominate the evolution of $\nrc$, in agreement with our energetic estimates of Sec.~\ref{sectionaxitons}.
For $\n<2$ the gradient pressure will eventually win, even if the instability develops due to a large initial value of $\nrc$. Compared with the discussion of Sec.~\ref{sectionaxitons} this equation allows to
get a quantitative feeling of the dynamics of the collapse as well. As long as the gradient pressure is negligible, the time scale for the collapse should be $\sim 8\ct^2/|\nrc|^2$.

We can be even more exact by comparing \eqref{gpeq} with the equation used in \cite{Levkov:2016rkk} in a non-expanding background with constant axion mass. Their mass $m$ becomes our conformal mass, $\cmass$,
and their self interaction term $g_4^2$ corresponds to $1/\cmass \ct^2$. Reference \cite{Levkov:2016rkk} finds a self-similar solution, which should be locally valid also for a expanding Universe, so we can
read off the evolution of $\nrc$ at the centre of our lump from the time derivative of their equation (8). This gives \eqref{wavecollapseestimate}, which agrees very good with the above estimate. Just note that
$|\nrc|^2\sim \cmass \ctheta^2\sim  \ct^{\n/2+1} \Theta^2 \ct^2$, which we have labelled $\bar\Theta$ for simplicity.

\footnotesize

\bibliographystyle{utphys}
\bibliography{axionDM}

\providecommand{\href}[2]{#2}\begingroup\raggedright\begin{thebibliography}{10}

\bibitem{Peccei:1977ur}
R.~Peccei and H.~R. Quinn, ``{Constraints imposed by CP conservation in the
  presence of instantons},''
\href{http://dx.doi.org/10.1103/PhysRevD.16.1791}{{\em Phys.Rev.} {\bfseries
  D16} (1977) 1791--1797}.

\bibitem{Peccei:1977hh}
R.~Peccei and H.~R. Quinn, ``{CP conservation in the presence of instantons},''
\href{http://dx.doi.org/10.1103/PhysRevLett.38.1440}{{\em Phys.Rev.Lett.}
  {\bfseries 38} (1977) 1440--1443}.

\bibitem{Wilczek:1977pj}
F.~Wilczek, ``{Problem of strong P and T invariance in the presence of
  instantons},''
\href{http://dx.doi.org/10.1103/PhysRevLett.40.279}{{\em Phys.Rev.Lett.}
  {\bfseries 40} (1978) 279--282}.

\bibitem{Weinberg:1977ma}
S.~Weinberg, ``{A new light boson?},''
\href{http://dx.doi.org/10.1103/PhysRevLett.40.223}{{\em Phys.Rev.Lett.}
  {\bfseries 40} (1978) 223--226}.

\bibitem{Peccei:2006as}
R.~Peccei, ``{The Strong CP problem and axions},''
  \href{http://dx.doi.org/10.1007/978-3-540-73518-2_1}{{\em Lect.Notes Phys.}
  {\bfseries 741} (2008) 3--17},
\href{http://arxiv.org/abs/hep-ph/0607268}{{\ttfamily arXiv:hep-ph/0607268
  [hep-ph]}}.

\bibitem{Raffelt:1996wa}
G.~G. Raffelt, {\em {Stars as laboratories for fundamental physics}}.
\newblock Cambridge U. Press, 1996.
\newblock
\url{http://wwwth.mpp.mpg.de/members/raffelt/mypapers/199613.pdf}.
\newblock

\bibitem{Viaux:2013lha}
N.~Viaux, M.~Catelan, P.~B. Stetson, G.~Raffelt, J.~Redondo, {\em et~al.},
  ``{Neutrino and axion bounds from the globular cluster M5 (NGC 5904)},''
  \href{http://dx.doi.org/10.1103/PhysRevLett.111.231301}{{\em Phys.Rev.Lett.}
  {\bfseries 111} (2013) 231301},
\href{http://arxiv.org/abs/1311.1669}{{\ttfamily arXiv:1311.1669
  [astro-ph.SR]}}.

\bibitem{Bertolami:2014wua}
M.~M. Miller~Bertolami, B.~E. Melendez, L.~G. Althaus, and J.~Isern,
  ``{Revisiting the axion bounds from the Galactic white dwarf luminosity
  function},'' \href{http://dx.doi.org/10.1088/1475-7516/2014/10/069}{{\em
  JCAP} {\bfseries 1410} no.~10, (2014) 069},
\href{http://arxiv.org/abs/1406.7712}{{\ttfamily arXiv:1406.7712 [hep-ph]}}.

\bibitem{Ayala:2014pea}
A.~Ayala, I.~Dom{\'\i}nguez, M.~Giannotti, A.~Mirizzi, and O.~Straniero,
  ``{Revisiting the bound on axion-photon coupling from Globular Clusters},''
  \href{http://dx.doi.org/10.1103/PhysRevLett.113.191302}{{\em Phys. Rev.
  Lett.} {\bfseries 113} no.~19, (2014) 191302},
\href{http://arxiv.org/abs/1406.6053}{{\ttfamily arXiv:1406.6053
  [astro-ph.SR]}}.

\bibitem{Chang:2018rso}
J.~H. Chang, R.~Essig, and S.~D. McDermott, ``Supernova 1987a constraints on
  sub-gev dark sectors, millicharged particles, the qcd axion, and an
  axion-like particle,'' \href{http://dx.doi.org/10.1007/JHEP09(2018)051}{{\em
  Journal of High Energy Physics} {\bfseries 2018} no.~9, (Sep, 2018) 51},
  \href{http://arxiv.org/abs/1803.00993}{{\ttfamily arXiv:1803.00993
  [hep-ph]}}.
\url{https://doi.org/10.1007/JHEP09(2018)051}.

\bibitem{Dvali:2017ruz}
G.~Dvali and S.~Zell, ``{Classicality and Quantum Break-Time for Cosmic
  Axions},'' {\em Journal of Cosmology and Astroparticle Physics} {\bfseries
  2018} no.~07, (2018) 064, \href{http://arxiv.org/abs/1710.00835}{{\ttfamily
  arXiv:1710.00835 [hep-ph]}}.
\url{http://stacks.iop.org/1475-7516/2018/i=07/a=064}.

\bibitem{Preskill:1982cy}
J.~Preskill, M.~B. Wise, and F.~Wilczek, ``{Cosmology of the Invisible
  Axion},''
\href{http://dx.doi.org/10.1016/0370-2693(83)90637-8}{{\em Phys. Lett.}
  {\bfseries 120B} (1983) 127--132}.

\bibitem{Abbott:1982af}
L.~F. Abbott and P.~Sikivie, ``{A Cosmological Bound on the Invisible Axion},''
\href{http://dx.doi.org/10.1016/0370-2693(83)90638-X}{{\em Phys. Lett.}
  {\bfseries 120B} (1983) 133--136}.

\bibitem{Dine:1982ah}
M.~Dine and W.~Fischler, ``{The Not So Harmless Axion},''
\href{http://dx.doi.org/10.1016/0370-2693(83)90639-1}{{\em Phys. Lett.}
  {\bfseries 120B} (1983) 137--141}.

\bibitem{diCortona:2015ldu}
G.~Grilli~di Cortona, E.~Hardy, J.~Pardo~Vega, and G.~Villadoro, ``{The QCD
  axion, precisely},'' \href{http://dx.doi.org/10.1007/JHEP01(2016)034}{{\em
  JHEP} {\bfseries 01} (2016) 034},
\href{http://arxiv.org/abs/1511.02867}{{\ttfamily arXiv:1511.02867 [hep-ph]}}.

\bibitem{Borsanyi:2016ksw}
S.~Borsanyi {\em et~al.}, ``{Calculation of the axion mass based on
  high-temperature lattice quantum chromodynamics},''
  \href{http://dx.doi.org/10.1038/nature20115}{{\em Nature} {\bfseries 539}
  no.~7627, (2016) 69--71},
\href{http://arxiv.org/abs/1606.07494}{{\ttfamily arXiv:1606.07494 [hep-lat]}}.

\bibitem{Irastorza:2018dyq}
I.~G. Irastorza and J.~Redondo, ``{New experimental approaches in the search
  for axion-like particles},''
  \href{http://dx.doi.org/https://doi.org/10.1016/j.ppnp.2018.05.003}{{\em
  Progress in Particle and Nuclear Physics} {\bfseries 102} (2018) 89 -- 159},
  \href{http://arxiv.org/abs/1801.08127}{{\ttfamily arXiv:1801.08127
  [hep-ph]}}.
\url{http://www.sciencedirect.com/science/article/pii/S014664101830036X}.

\bibitem{Sikivie:1983ip}
P.~Sikivie, ``{Experimental tests of the invisible axion},''
\href{http://dx.doi.org/10.1103/PhysRevLett.51.1415,
  10.1103/PhysRevLett.51.1415}{{\em Phys.Rev.Lett.} {\bfseries 51} (1983)
  1415}.

\bibitem{Hogan:1988mp}
C.~Hogan and M.~Rees, ``{Axion Miniclusters},''
\href{http://dx.doi.org/10.1016/0370-2693(88)91655-3}{{\em Phys.Lett.}
  {\bfseries B205} (1988) 228--230}.

\bibitem{Kolb:1993zz}
E.~W. Kolb and I.~I. Tkachev, ``{Axion miniclusters and Bose stars},''
  \href{http://dx.doi.org/10.1103/PhysRevLett.71.3051}{{\em Phys.Rev.Lett.}
  {\bfseries 71} (1993) 3051--3054},
\href{http://arxiv.org/abs/hep-ph/9303313}{{\ttfamily arXiv:hep-ph/9303313
  [hep-ph]}}.

\bibitem{Kolb:1993hw}
E.~W. Kolb and I.~I. Tkachev, ``{Nonlinear axion dynamics and formation of
  cosmological pseudosolitons},''
  \href{http://dx.doi.org/10.1103/PhysRevD.49.5040}{{\em Phys.Rev.} {\bfseries
  D49} (1994) 5040--5051},
\href{http://arxiv.org/abs/astro-ph/9311037}{{\ttfamily arXiv:astro-ph/9311037
  [astro-ph]}}.

\bibitem{Zurek:2006sy}
K.~M. Zurek, C.~J. Hogan, and T.~R. Quinn, ``{Astrophysical Effects of Scalar
  Dark Matter Miniclusters},''
  \href{http://dx.doi.org/10.1103/PhysRevD.75.043511}{{\em Phys.Rev.}
  {\bfseries D75} (2007) 043511},
\href{http://arxiv.org/abs/astro-ph/0607341}{{\ttfamily arXiv:astro-ph/0607341
  [astro-ph]}}.

\bibitem{Kolb:1994fi}
E.~W. Kolb and I.~I. Tkachev, ``{Large amplitude isothermal fluctuations and
  high density dark matter clumps},''
  \href{http://dx.doi.org/10.1103/PhysRevD.50.769}{{\em Phys. Rev.} {\bfseries
  D50} (1994) 769--773},
\href{http://arxiv.org/abs/astro-ph/9403011}{{\ttfamily arXiv:astro-ph/9403011
  [astro-ph]}}.

\bibitem{Nelson:2018via}
A.~E. Nelson and H.~Xiao, ``{Axion Cosmology with Early Matter Domination},''
  \href{http://dx.doi.org/10.1103/PhysRevD.98.063516}{{\em Phys. Rev.}
  {\bfseries D98} no.~6, (2018) 063516},
\href{http://arxiv.org/abs/1807.07176}{{\ttfamily arXiv:1807.07176
  [astro-ph.CO]}}.

\bibitem{Visinelli:2018wza}
L.~Visinelli and J.~Redondo, ``{Axion Miniclusters in Modified Cosmological
  Histories},''
\href{http://arxiv.org/abs/1808.01879}{{\ttfamily arXiv:1808.01879
  [astro-ph.CO]}}.

\bibitem{TheMADMAXWorkingGroup:2016hpc}
{\bfseries MADMAX Working Group} Collaboration, A.~Caldwell, G.~Dvali,
  B.~Majorovits, A.~Millar, G.~Raffelt, J.~Redondo, O.~Reimann, F.~Simon, and
  F.~Steffen, ``{Dielectric Haloscopes: A New Way to Detect Axion Dark
  Matter},'' \href{http://dx.doi.org/10.1103/PhysRevLett.118.091801}{{\em Phys.
  Rev. Lett.} {\bfseries 118} no.~9, (2017) 091801},
\href{http://arxiv.org/abs/1611.05865}{{\ttfamily arXiv:1611.05865
  [physics.ins-det]}}.

\bibitem{Millar:2016cjp}
A.~J. Millar, G.~G. Raffelt, J.~Redondo, and F.~D. Steffen, ``{Dielectric
  Haloscopes to Search for Axion Dark Matter: Theoretical Foundations},''
  \href{http://dx.doi.org/10.1088/1475-7516/2017/01/061}{{\em JCAP} {\bfseries
  1701} no.~01, (2017) 061},
\href{http://arxiv.org/abs/1612.07057}{{\ttfamily arXiv:1612.07057 [hep-ph]}}.

\bibitem{Horns:2012jf}
D.~Horns, J.~Jaeckel, A.~Lindner, A.~Lobanov, J.~Redondo, {\em et~al.},
  ``{Searching for WISPy Cold Dark Matter with a Dish Antenna},''
  \href{http://dx.doi.org/10.1088/1475-7516/2013/04/016}{{\em JCAP} {\bfseries
  1304} (2013) 016},
\href{http://arxiv.org/abs/1212.2970}{{\ttfamily arXiv:1212.2970}}.

\bibitem{Tkachev:2014dpa}
I.~I. Tkachev, ``{Fast Radio Bursts and Axion Miniclusters},''
  \href{http://dx.doi.org/10.1134/S0021364015010154}{{\em JETP Lett.}
  {\bfseries 101} no.~1, (2015) 1--6},
  \href{http://arxiv.org/abs/1411.3900}{{\ttfamily arXiv:1411.3900
  [astro-ph.HE]}}.
[Pisma Zh. Eksp. Teor. Fiz.101,no.1,3(2015)].

\bibitem{Pshirkov:2016bjr}
M.~S. Pshirkov, ``{May axion clusters be sources of fast radio bursts?},''
  \href{http://dx.doi.org/10.1142/S0218271817500687}{{\em Int. J. Mod. Phys.}
  {\bfseries D26} no.~07, (2017) 1750068},
\href{http://arxiv.org/abs/1609.09658}{{\ttfamily arXiv:1609.09658
  [astro-ph.HE]}}.

\bibitem{Ruffini:1969qy}
R.~Ruffini and S.~Bonazzola, ``{Systems of selfgravitating particles in general
  relativity and the concept of an equation of state},''
\href{http://dx.doi.org/10.1103/PhysRev.187.1767}{{\em Phys. Rev.} {\bfseries
  187} (1969) 1767--1783}.

\bibitem{Kaup:1968zz}
D.~J. Kaup, ``{Klein-Gordon Geon},''
\href{http://dx.doi.org/10.1103/PhysRev.172.1331}{{\em Phys. Rev.} {\bfseries
  172} (1968) 1331--1342}.

\bibitem{Chavanis:2011zi}
P.-H. Chavanis, ``{Mass-radius relation of Newtonian self-gravitating
  Bose-Einstein condensates with short-range interactions: I. Analytical
  results},'' \href{http://dx.doi.org/10.1103/PhysRevD.84.043531}{{\em Phys.
  Rev.} {\bfseries D84} (2011) 043531},
\href{http://arxiv.org/abs/1103.2050}{{\ttfamily arXiv:1103.2050
  [astro-ph.CO]}}.

\bibitem{Chavanis:2011zm}
P.~H. Chavanis and L.~Delfini, ``{Mass-radius relation of Newtonian
  self-gravitating Bose-Einstein condensates with short-range interactions: II.
  Numerical results},''
  \href{http://dx.doi.org/10.1103/PhysRevD.84.043532}{{\em Phys. Rev.}
  {\bfseries D84} (2011) 043532},
\href{http://arxiv.org/abs/1103.2054}{{\ttfamily arXiv:1103.2054
  [astro-ph.CO]}}.

\bibitem{Chavanis:2016dab}
P.-H. Chavanis, ``{Collapse of a self-gravitating Bose-Einstein condensate with
  attractive self-interaction},''
  \href{http://dx.doi.org/10.1103/PhysRevD.94.083007}{{\em Phys. Rev.}
  {\bfseries D94} no.~8, (2016) 083007},
\href{http://arxiv.org/abs/1604.05904}{{\ttfamily arXiv:1604.05904
  [astro-ph.CO]}}.

\bibitem{Chavanis:2017loo}
P.-H. Chavanis, ``{Phase transitions between dilute and dense axion stars},''
  \href{http://dx.doi.org/10.1103/PhysRevD.98.023009}{{\em Phys. Rev.}
  {\bfseries D98} no.~2, (2018) 023009},
\href{http://arxiv.org/abs/1710.06268}{{\ttfamily arXiv:1710.06268 [gr-qc]}}.

\bibitem{Levkov:2018kau}
D.~G. Levkov, A.~G. Panin, and I.~I. Tkachev, ``Gravitational bose-einstein
  condensation in the kinetic regime,''
  \href{http://dx.doi.org/10.1103/PhysRevLett.121.151301}{{\em Phys. Rev.
  Lett.} {\bfseries 121} (Oct, 2018) 151301},
  \href{http://arxiv.org/abs/1804.05857}{{\ttfamily arXiv:1804.05857
  [astro-ph.CO]}}.
\url{https://link.aps.org/doi/10.1103/PhysRevLett.121.151301}.

\bibitem{Veltmaat:2018dfz}
J.~Veltmaat, J.~C. Niemeyer, and B.~Schwabe, ``{Formation and structure of
  ultralight bosonic dark matter halos},''
  \href{http://dx.doi.org/10.1103/PhysRevD.98.043509}{{\em Phys. Rev. D}
  {\bfseries 98} (Aug, 2018) 043509},
  \href{http://arxiv.org/abs/1804.09647}{{\ttfamily arXiv:1804.09647
  [astro-ph.CO]}}.
\url{https://link.aps.org/doi/10.1103/PhysRevD.98.043509}.

\bibitem{Bai:2016wpg}
Y.~Bai, V.~Barger, and J.~Berger, ``{Hydrogen Axion Star: Metallic Hydrogen
  Bound to a QCD Axion BEC},''
  \href{http://dx.doi.org/10.1007/JHEP12(2016)127}{{\em JHEP} {\bfseries 12}
  (2016) 127},
\href{http://arxiv.org/abs/1612.00438}{{\ttfamily arXiv:1612.00438 [hep-ph]}}.

\bibitem{Bai:2017feq}
Y.~Bai and Y.~Hamada, ``{Detecting Axion Stars with Radio Telescopes},''
  \href{http://dx.doi.org/https://doi.org/10.1016/j.physletb.2018.03.070}{{\em
  Physics Letters B} {\bfseries 781} (2018) 187 -- 194},
  \href{http://arxiv.org/abs/1709.10516}{{\ttfamily arXiv:1709.10516
  [astro-ph.HE]}}.
\url{http://www.sciencedirect.com/science/article/pii/S0370269318302661}.

\bibitem{Eby:2017xaw}
J.~Eby, M.~Leembruggen, J.~Leeney, P.~Suranyi, and L.~C.~R. Wijewardhana,
  ``{Collisions of Dark Matter Axion Stars with Astrophysical Sources},''
  \href{http://dx.doi.org/10.1007/JHEP04(2017)099}{{\em JHEP} {\bfseries 04}
  (2017) 099},
\href{http://arxiv.org/abs/1701.01476}{{\ttfamily arXiv:1701.01476
  [astro-ph.CO]}}.

\bibitem{Hertzberg:2018zte}
M.~P. Hertzberg and E.~D. Schiappacasse, ``{Dark Matter Axion Clump Resonance
  of Photons},'' {\em Journal of Cosmology and Astroparticle Physics}
  {\bfseries 2018} no.~11, (2018) 004,
  \href{http://arxiv.org/abs/1805.00430}{{\ttfamily arXiv:1805.00430
  [hep-ph]}}.
\url{http://stacks.iop.org/1475-7516/2018/i=11/a=004}.

\bibitem{Braaten:2015eeu}
E.~Braaten, A.~Mohapatra, and H.~Zhang, ``{Dense Axion Stars},''
  \href{http://dx.doi.org/10.1103/PhysRevLett.117.121801}{{\em Phys. Rev.
  Lett.} {\bfseries 117} no.~12, (2016) 121801},
\href{http://arxiv.org/abs/1512.00108}{{\ttfamily arXiv:1512.00108 [hep-ph]}}.

\bibitem{Iwazaki:2014wka}
A.~Iwazaki, ``{Axion Stars and Fast Radio Bursts},''
  \href{http://dx.doi.org/10.1103/PhysRevD.91.023008}{{\em Phys. Rev. D}
  {\bfseries 91} (Jan, 2015) 023008},
  \href{http://arxiv.org/abs/1410.4323}{{\ttfamily arXiv:1410.4323 [hep-ph]}}.
\url{https://link.aps.org/doi/10.1103/PhysRevD.91.023008}.

\bibitem{Raby:2016deh}
S.~Raby, ``{Axion star collisions with Neutron stars and Fast Radio Bursts},''
  \href{http://dx.doi.org/10.1103/PhysRevD.94.103004}{{\em Phys. Rev.}
  {\bfseries D94} no.~10, (2016) 103004},
\href{http://arxiv.org/abs/1609.01694}{{\ttfamily arXiv:1609.01694 [hep-ph]}}.

\bibitem{Visinelli:2017ooc}
L.~Visinelli, S.~Baum, J.~Redondo, K.~Freese, and F.~Wilczek, ``{Dilute and
  dense axion stars},''
  \href{http://dx.doi.org/10.1016/j.physletb.2017.12.010}{{\em Phys. Lett.}
  {\bfseries B777} (2018) 64--72},
\href{http://arxiv.org/abs/1710.08910}{{\ttfamily arXiv:1710.08910
  [astro-ph.CO]}}.

\bibitem{Eby:2016cnq}
J.~Eby, M.~Leembruggen, P.~Suranyi, and L.~C.~R. Wijewardhana, ``{Collapse of
  Axion Stars},'' \href{http://dx.doi.org/10.1007/JHEP12(2016)066}{{\em JHEP}
  {\bfseries 12} (2016) 066},
\href{http://arxiv.org/abs/1608.06911}{{\ttfamily arXiv:1608.06911
  [astro-ph.CO]}}.

\bibitem{Eby:2017xrr}
J.~Eby, M.~Leembruggen, P.~Suranyi, and L.~C.~R. Wijewardhana, ``{QCD Axion
  Star Collapse with the Chiral Potential},''
  \href{http://dx.doi.org/10.1007/JHEP06(2017)014}{{\em JHEP} {\bfseries 06}
  (2017) 014},
\href{http://arxiv.org/abs/1702.05504}{{\ttfamily arXiv:1702.05504 [hep-ph]}}.

\bibitem{Levkov:2016rkk}
D.~G. Levkov, A.~G. Panin, and I.~I. Tkachev, ``{Relativistic axions from
  collapsing Bose stars},''
  \href{http://dx.doi.org/10.1103/PhysRevLett.118.011301}{{\em Phys. Rev.
  Lett.} {\bfseries 118} no.~1, (2017) 011301},
\href{http://arxiv.org/abs/1609.03611}{{\ttfamily arXiv:1609.03611
  [astro-ph.CO]}}.

\bibitem{Kolb:1995bu}
E.~W. Kolb and I.~I. Tkachev, ``{Femtolensing and picolensing by axion
  miniclusters},'' \href{http://dx.doi.org/10.1086/309962}{{\em Astrophys.J.}
  {\bfseries 460} (1996) L25--L28},
\href{http://arxiv.org/abs/astro-ph/9510043}{{\ttfamily arXiv:astro-ph/9510043
  [astro-ph]}}.

\bibitem{Fairbairn:2017sil}
M.~Fairbairn, D.~J.~E. Marsh, J.~Quevillon, and S.~Rozier, ``{Structure
  formation and microlensing with axion miniclusters},''
  \href{http://dx.doi.org/10.1103/PhysRevD.97.083502}{{\em Phys. Rev.}
  {\bfseries D97} no.~8, (2018) 083502},
\href{http://arxiv.org/abs/1707.03310}{{\ttfamily arXiv:1707.03310
  [astro-ph.CO]}}.

\bibitem{Fairbairn:2017dmf}
M.~Fairbairn, D.~J.~E. Marsh, and J.~Quevillon, ``{Searching for the QCD Axion
  with Gravitational Microlensing},''
  \href{http://dx.doi.org/10.1103/PhysRevLett.119.021101}{{\em Phys. Rev.
  Lett.} {\bfseries 119} no.~2, (2017) 021101},
\href{http://arxiv.org/abs/1701.04787}{{\ttfamily arXiv:1701.04787
  [astro-ph.CO]}}.

\bibitem{Katz:2018zrn}
A.~Katz, J.~Kopp, S.~Sibiryakov, and W.~Xue, ``{Femtolensing by Dark Matter
  Revisited},'' \href{http://dx.doi.org/10.1088/1475-7516/2018/12/005}{{\em
  JCAP} {\bfseries 1812} (2018) 005},
\href{http://arxiv.org/abs/1807.11495}{{\ttfamily arXiv:1807.11495
  [astro-ph.CO]}}.

\bibitem{Niikura:2017zjd}
H.~Niikura, M.~Takada, N.~Yasuda, R.~H. Lupton, T.~Sumi, S.~More, A.~More,
  M.~Oguri, and M.~Chiba, ``{Microlensing constraints on primordial black holes
  with the Subaru/HSC Andromeda observation},''
\href{http://arxiv.org/abs/1701.02151}{{\ttfamily arXiv:1701.02151
  [astro-ph.CO]}}.

\bibitem{Inomata:2017vxo}
K.~Inomata, M.~Kawasaki, K.~Mukaida, and T.~T. Yanagida, ``{Double inflation as
  a single origin of primordial black holes for all dark matter and LIGO
  observations},'' \href{http://dx.doi.org/10.1103/PhysRevD.97.043514}{{\em
  Phys. Rev.} {\bfseries D97} no.~4, (2018) 043514},
\href{http://arxiv.org/abs/1711.06129}{{\ttfamily arXiv:1711.06129
  [astro-ph.CO]}}.

\bibitem{Kibble:1980mv}
T.~W.~B. Kibble, ``{Some Implications of a Cosmological Phase Transition},''
\href{http://dx.doi.org/10.1016/0370-1573(80)90091-5}{{\em Phys. Rept.}
  {\bfseries 67} (1980) 183}.

\bibitem{Kim:1979if}
J.~E. Kim, ``{Weak Interaction Singlet and Strong CP Invariance},''
\href{http://dx.doi.org/10.1103/PhysRevLett.43.103}{{\em Phys. Rev. Lett.}
  {\bfseries 43} (1979) 103}.

\bibitem{Shifman:1979if}
M.~A. Shifman, A.~I. Vainshtein, and V.~I. Zakharov, ``{Can Confinement Ensure
  Natural CP Invariance of Strong Interactions?},''
\href{http://dx.doi.org/10.1016/0550-3213(80)90209-6}{{\em Nucl. Phys.}
  {\bfseries B166} (1980) 493--506}.

\bibitem{Hardy:2016mns}
E.~Hardy, ``{Miniclusters in the Axiverse},''
  \href{http://dx.doi.org/10.1007/JHEP02(2017)046}{{\em JHEP} {\bfseries 02}
  (2017) 046},
\href{http://arxiv.org/abs/1609.00208}{{\ttfamily arXiv:1609.00208 [hep-ph]}}.

\bibitem{Fleury:2015aca}
L.~Fleury and G.~D. Moore, ``{Axion dark matter: strings and their cores},''
  \href{http://dx.doi.org/10.1088/1475-7516/2016/01/004}{{\em JCAP} {\bfseries
  1601} (2016) 004},
\href{http://arxiv.org/abs/1509.00026}{{\ttfamily arXiv:1509.00026 [hep-ph]}}.

\bibitem{Klaer:2017ond}
V.~B. Klaer and G.~D. Moore, ``{The dark-matter axion mass},''
  \href{http://dx.doi.org/10.1088/1475-7516/2017/11/049}{{\em JCAP} {\bfseries
  1711} no.~11, (2017) 049},
\href{http://arxiv.org/abs/1708.07521}{{\ttfamily arXiv:1708.07521 [hep-ph]}}.

\bibitem{Hiramatsu:2012gg}
T.~Hiramatsu, M.~Kawasaki, K.~Saikawa, and T.~Sekiguchi, ``{Production of dark
  matter axions from collapse of string-wall systems},''
  \href{http://dx.doi.org/10.1103/PhysRevD.86.089902,
  10.1103/PhysRevD.85.105020}{{\em Phys.Rev.} {\bfseries D85} (2012) 105020},
\href{http://arxiv.org/abs/1202.5851}{{\ttfamily arXiv:1202.5851 [hep-ph]}}.

\bibitem{Kawasaki:2014sqa}
M.~Kawasaki, K.~Saikawa, and T.~Sekiguchi, ``{Axion dark matter from
  topological defects},''
  \href{http://dx.doi.org/10.1103/PhysRevD.91.065014}{{\em Phys. Rev. D}
  {\bfseries 91} (Mar, 2015) 065014},
  \href{http://arxiv.org/abs/1412.0789}{{\ttfamily arXiv:1412.0789 [hep-ph]}}.
\url{https://link.aps.org/doi/10.1103/PhysRevD.91.065014}.

\bibitem{Kawasaki:2018aa}
M.~Kawasaki, T.~Sekiguchi, M.~Yamaguchi, and J.~Yokoyama, ``{Long-term dynamics
  of cosmological axion strings},''
  \href{http://dx.doi.org/10.1093/ptep/pty098}{{\em Progress of Theoretical and
  Experimental Physics} no.~9, (2018) 091E01},
  \href{http://arxiv.org/abs/1806.05566}{{\ttfamily arXiv:1806.05566
  [hep-ph]}}.
\url{http://dx.doi.org/10.1093/ptep/pty098}.

\bibitem{Gorghetto:2018myk}
M.~Gorghetto, E.~Hardy, and G.~Villadoro, ``{Axions from Strings: the
  Attractive Solution},'' \href{http://dx.doi.org/10.1007/JHEP07(2018)151}{{\em
  Journal of High Energy Physics} {\bfseries 2018} no.~7, (Jul, 2018) 151},
  \href{http://arxiv.org/abs/1806.04677}{{\ttfamily arXiv:1806.04677
  [hep-ph]}}.
\url{https://doi.org/10.1007/JHEP07(2018)151}.

\bibitem{Press:1989yh}
W.~H. Press, B.~S. Ryden, and D.~N. Spergel, ``{Dynamical Evolution of Domain
  Walls in an Expanding Universe},''
\href{http://dx.doi.org/10.1086/168151}{{\em Astrophys. J.} {\bfseries 347}
  (1989) 590--604}.

\bibitem{Moore:2001px}
J.~N. Moore, E.~P.~S. Shellard, and C.~J. A.~P. Martins, ``{On the evolution of
  Abelian-Higgs string networks},''
  \href{http://dx.doi.org/10.1103/PhysRevD.65.023503}{{\em Phys. Rev.}
  {\bfseries D65} (2002) 023503},
\href{http://arxiv.org/abs/hep-ph/0107171}{{\ttfamily arXiv:hep-ph/0107171
  [hep-ph]}}.

\bibitem{Bonati:2015vqz}
C.~Bonati, M.~D'Elia, M.~Mariti, G.~Martinelli, M.~Mesiti, F.~Negro,
  F.~Sanfilippo, and G.~Villadoro, ``{Axion phenomenology and
  $\theta$-dependence from $N_f = 2+1$ lattice QCD},''
  \href{http://dx.doi.org/10.1007/JHEP03(2016)155}{{\em JHEP} {\bfseries 03}
  (2016) 155},
\href{http://arxiv.org/abs/1512.06746}{{\ttfamily arXiv:1512.06746 [hep-lat]}}.

\bibitem{Asaka:1998xa}
T.~Asaka and M.~Yamaguchi, ``{Hadronic axion model in gauge mediated
  supersymmetry breaking and cosmology of saxion},''
  \href{http://dx.doi.org/10.1103/PhysRevD.59.125003}{{\em Phys. Rev.}
  {\bfseries D59} (1999) 125003},
\href{http://arxiv.org/abs/hep-ph/9811451}{{\ttfamily arXiv:hep-ph/9811451
  [hep-ph]}}.

\bibitem{Yamaguchi:1998gx}
M.~Yamaguchi, M.~Kawasaki, and J.~Yokoyama, ``{Evolution of axionic strings and
  spectrum of axions radiated from them},''
  \href{http://dx.doi.org/10.1103/PhysRevLett.82.4578}{{\em Phys. Rev. Lett.}
  {\bfseries 82} (1999) 4578--4581},
\href{http://arxiv.org/abs/hep-ph/9811311}{{\ttfamily arXiv:hep-ph/9811311
  [hep-ph]}}.

\bibitem{Yamaguchi:1999yp}
M.~Yamaguchi, ``{Scaling property of the global string in the radiation
  dominated universe},''
  \href{http://dx.doi.org/10.1103/PhysRevD.60.103511}{{\em Phys. Rev.}
  {\bfseries D60} (1999) 103511},
\href{http://arxiv.org/abs/hep-ph/9907506}{{\ttfamily arXiv:hep-ph/9907506
  [hep-ph]}}.

\bibitem{Davis:1985pt}
R.~L. Davis, ``{Goldstone Bosons in String Models of Galaxy Formation},''
\href{http://dx.doi.org/10.1103/PhysRevD.32.3172}{{\em Phys. Rev.} {\bfseries
  D32} (1985) 3172}.

\bibitem{Hiramatsu:2010yu}
T.~Hiramatsu, M.~Kawasaki, T.~Sekiguchi, M.~Yamaguchi, and J.~Yokoyama,
  ``{Improved estimation of radiated axions from cosmological axionic
  strings},'' \href{http://dx.doi.org/10.1103/PhysRevD.83.123531}{{\em Phys.
  Rev.} {\bfseries D83} (2011) 123531},
\href{http://arxiv.org/abs/1012.5502}{{\ttfamily arXiv:1012.5502 [hep-ph]}}.

\bibitem{RedondoPatras2018}
{J. Redondo}, ``{The dark-matter axion mass}.'' Talk at the 14th Patras
  Axion-WIMP-WISP Workshop.

\bibitem{Klaer:2017qhr}
V.~B. Klaer and G.~D. Moore, ``{How to simulate global cosmic strings with
  large string tension},''
  \href{http://dx.doi.org/10.1088/1475-7516/2017/10/043}{{\em JCAP} {\bfseries
  1710} (2017) 043},
\href{http://arxiv.org/abs/1707.05566}{{\ttfamily arXiv:1707.05566 [hep-ph]}}.

\bibitem{Vilenkin:2000jqa}
A.~Vilenkin and E.~P.~S. Shellard, {\em {Cosmic Strings and Other Topological
  Defects}}.
\newblock Cambridge University Press, 2000.
\newblock
\url{http://www.cambridge.org/mw/academic/subjects/physics/theoretical-physics-and-mathematical-physics/cosmic-strings-and-other-topological-defects?format=PB}.
\newblock

\bibitem{Zhitnitsky:2018mav}
A.~Zhitnitsky, ``{Solar Flares and the Axion Quark Nugget Dark Matter Model},''
  \href{http://dx.doi.org/https://doi.org/10.1016/j.dark.2018.08.001}{{\em
  Physics of the Dark Universe} {\bfseries 22} (2018) 1 -- 15},
  \href{http://arxiv.org/abs/1801.01509}{{\ttfamily arXiv:1801.01509
  [astro-ph.SR]}}.
\url{http://www.sciencedirect.com/science/article/pii/S2212686418301006}.

\bibitem{Liang:2016tqc}
X.~Liang and A.~Zhitnitsky, ``{Axion field and the quark nugget's formation at
  the QCD phase transition},''
  \href{http://dx.doi.org/10.1103/PhysRevD.94.083502}{{\em Phys. Rev.}
  {\bfseries D94} no.~8, (2016) 083502},
\href{http://arxiv.org/abs/1606.00435}{{\ttfamily arXiv:1606.00435 [hep-ph]}}.

\bibitem{Ge:2017idw}
S.~Ge, X.~Liang, and A.~Zhitnitsky, ``{Cosmological Axion and Quark Nugget Dark
  Matter Model},'' \href{http://dx.doi.org/10.1103/PhysRevD.97.043008}{{\em
  Phys. Rev. D} {\bfseries 97} (Feb, 2018) 043008},
  \href{http://arxiv.org/abs/1711.06271}{{\ttfamily arXiv:1711.06271
  [hep-ph]}}.
\url{https://link.aps.org/doi/10.1103/PhysRevD.97.043008}.

\bibitem{Kolb:1994xc}
E.~W. Kolb and I.~I. Tkachev, {\em Axitons},
  \href{http://dx.doi.org/10.1007/978-1-4615-1855-6_6}{pp.~95--112}.
\newblock Springer US, Boston, MA, 1995.
\newblock \url{https://doi.org/10.1007/978-1-4615-1855-6_6}.

\bibitem{Omelyan2002}
I.~P. Omelyan, I.~Mryglod, and R.~Folk, ``{New optimized algorithms for
  molecular dynamics simulations},'' {\em Condensed Matter Physics} {\bfseries
  5} no.~3(31), (2002) 369--390.

\bibitem{McLachlan1992}
R.~I. McLachlan and P.~Atela, ``The accuracy of symplectic integrators,'' {\em
  Nonlinearity} {\bfseries 5} no.~2, (1992) 541--562.
  \url{http://stacks.iop.org/0951-7715/5/i=2/a=011}.

\bibitem{Yamaguchi:2002zv}
M.~Yamaguchi and J.~Yokoyama, ``{Lagrangian evolution of global strings},''
  \href{http://dx.doi.org/10.1103/PhysRevD.66.121303}{{\em Phys. Rev.}
  {\bfseries D66} (2002) 121303},
\href{http://arxiv.org/abs/hep-ph/0205308}{{\ttfamily arXiv:hep-ph/0205308
  [hep-ph]}}.

\bibitem{Yamaguchi:2002sh}
M.~Yamaguchi and J.~Yokoyama, ``{Quantitative evolution of global strings from
  the Lagrangian view point},''
  \href{http://dx.doi.org/10.1103/PhysRevD.67.103514}{{\em Phys. Rev.}
  {\bfseries D67} (2003) 103514},
\href{http://arxiv.org/abs/hep-ph/0210343}{{\ttfamily arXiv:hep-ph/0210343
  [hep-ph]}}.

\bibitem{Inpreparation}
J.~Redondo and A.~Vaquero, ``{In preparation},''.

\bibitem{Ballesteros:2016xej}
G.~Ballesteros, J.~Redondo, A.~Ringwald, and C.~Tamarit, ``{Standard
  Model---axion---seesaw---Higgs portal inflation. Five problems of particle
  physics and cosmology solved in one stroke},''
  \href{http://dx.doi.org/10.1088/1475-7516/2017/08/001}{{\em JCAP} {\bfseries
  1708} no.~08, (2017) 001},
\href{http://arxiv.org/abs/1610.01639}{{\ttfamily arXiv:1610.01639 [hep-ph]}}.

\bibitem{Micha:2004bv}
R.~Micha and I.~I. Tkachev, ``{Turbulent thermalization},''
  \href{http://dx.doi.org/10.1103/PhysRevD.70.043538}{{\em Phys. Rev.}
  {\bfseries D70} (2004) 043538},
\href{http://arxiv.org/abs/hep-ph/0403101}{{\ttfamily arXiv:hep-ph/0403101
  [hep-ph]}}.

\bibitem{Enander:2017ogx}
J.~Enander, A.~Pargner, and T.~Schwetz, ``{Axion minicluster power spectrum and
  mass function},'' \href{http://dx.doi.org/10.1088/1475-7516/2017/12/038}{{\em
  JCAP} {\bfseries 1712} no.~12, (2017) 038},
\href{http://arxiv.org/abs/1708.04466}{{\ttfamily arXiv:1708.04466
  [astro-ph.CO]}}.

\bibitem{Schiappacasse:2017ham}
E.~D. Schiappacasse and M.~P. Hertzberg, ``{Analysis of Dark Matter Axion
  Clumps with Spherical Symmetry},''
  \href{http://dx.doi.org/10.1088/1475-7516/2018/03/E01,
  10.1088/1475-7516/2018/01/037}{{\em JCAP} {\bfseries 1801} (2018) 037},
  \href{http://arxiv.org/abs/1710.04729}{{\ttfamily arXiv:1710.04729
  [hep-ph]}}.
[Erratum: JCAP1803,no.03,E01(2018)].

\bibitem{Bonati:2013tt}
C.~Bonati, M.~D'Elia, H.~Panagopoulos, and E.~Vicari, ``{Change of ??
  Dependence in 4D SU(N) Gauge Theories Across the Deconfinement Transition},''
  \href{http://dx.doi.org/10.1103/PhysRevLett.110.252003}{{\em Phys.Rev.Lett.}
  {\bfseries 110} no.~25, (2013) 252003},
\href{http://arxiv.org/abs/1301.7640}{{\ttfamily arXiv:1301.7640 [hep-lat]}}.

\bibitem{Braaten:2016dlp}
E.~Braaten, A.~Mohapatra, and H.~Zhang, ``{Emission of Photons and Relativistic
  Axions from Axion Stars},''
  \href{http://dx.doi.org/10.1103/PhysRevD.96.031901}{{\em Phys. Rev.}
  {\bfseries D96} no.~3, (2017) 031901},
\href{http://arxiv.org/abs/1609.05182}{{\ttfamily arXiv:1609.05182 [hep-ph]}}.

\bibitem{Salmi:2012ta}
P.~Salmi and M.~Hindmarsh, ``{Radiation and Relaxation of Oscillons},''
  \href{http://dx.doi.org/10.1103/PhysRevD.85.085033}{{\em Phys. Rev.}
  {\bfseries D85} (2012) 085033},
\href{http://arxiv.org/abs/1201.1934}{{\ttfamily arXiv:1201.1934 [hep-th]}}.

\bibitem{DiVecchia:1980yfw}
P.~Di~Vecchia and G.~Veneziano, ``{Chiral Dynamics in the Large n Limit},''
\href{http://dx.doi.org/10.1016/0550-3213(80)90370-3}{{\em Nucl. Phys.}
  {\bfseries B171} (1980) 253--272}.

\bibitem{buschmann}
{M. Buschmann and B. Safdi and J. Foster}, ``{Numerical Simulation of the Axion
  Field through the QCD Phase Transition }.'' Talk/poster at the 14th Patras
  Workshop on Axions, WIMPs and WISPs.

\bibitem{scikit-learn}
F.~Pedregosa, G.~Varoquaux, A.~Gramfort, V.~Michel, B.~Thirion, O.~Grisel,
  M.~Blondel, P.~Prettenhofer, R.~Weiss, V.~Dubourg, J.~Vanderplas, A.~Passos,
  D.~Cournapeau, M.~Brucher, M.~Perrot, and E.~Duchesnay, ``Scikit-learn:
  Machine learning in {P}ython,'' {\em Journal of Machine Learning Research}
  {\bfseries 12} (2011) 2825--2830.

\bibitem{FFTW05}
M.~Frigo and S.~G. Johnson, ``The design and implementation of {FFTW3},'' {\em
  Proceedings of the IEEE} {\bfseries 93} no.~2, (2005) 216--231. Special issue
  on ``Program Generation, Optimization, and Platform Adaptation''.

\end{thebibliography}\endgroup
\end{document}